\documentclass[prc,twocolumn,showpacs,superscriptaddress]{revtex4-1}
\usepackage{dsfont}
\usepackage{graphicx,amsmath}
\def\be{\begin{eqnarray}}
\def\ee{\end{eqnarray}}
\def\bc{\begin{center}}
\def\ec{\end{center}}

\def\rmF{{\rm F}}

\def\half{{\textstyle \frac12}}

\newcommand{\lsim}{\stackrel{\scriptstyle <}{\phantom{}_{\sim}}}
\newcommand{\gsim}{\stackrel{\scriptstyle >}{\phantom{}_{\sim}}}

\begin{document}
\title{Vector-boson condensates, spin-triplet superfluidity of paired neutral and charged fermions,  and 3P$_2$ pairing of nucleons
}
\author{D. N. Voskresensky}
 \affiliation{ National Research Nuclear
    University (MEPhI), 115409 Moscow, Russia}
   \affiliation{ Joint Institute for Nuclear Research,
		Joliot-Curie street 6,
		141980 Dubna, Russia}
\begin{abstract}
After reminding of  properties of  the condensate of the complex scalar field in the external uniform magnetic field $H$
focus is made  on the study of phases of the complex neutral  vector boson fields
coupled with magnetic  field by the
Zeeman coupling and phases of the charged vector boson fields.
The systems  may behave  as nonmagnetic and ferromagnetic superfluids and ordinary and ferromagnetic superconductors.
Response of these superfluid and superconducting systems  occupying half  of space on the external uniform static magnetic field $H$ is thoroughly studied.
Then  the
spin-triplet pairing of
neutral fermions  at conserved spin is considered.  Novel phases  are found. In external magnetic field the phase with zero mean spin  proves to be unstable to the formation of a phase with a non-zero spin. For a certain parameter choice  ferromagnetic superfluid phases are formed already for $H=0$, characterized by an own  magnetic field $h$. For $H>H_{\rm cr 2}$ the spin-triplet pairing and  ferromagnetic superfluidity
continue to exist above the ``old'' phase transition critical temperature $T_{\rm cr}$.   Formation of domains  is discussed.
Next, spin-triplet pairing of charged fermions is studied.  Novel phases are found.
Then,  the  3P$_2$ $nn$ pairing in neutron star matter is studied.
Also a  3P$_2$ $pp$ pairing is considered.
Numerical estimates are performed in the BCS weak coupling limit and beyond it for the $3P_2$ $nn$ and $pp$ pairings, as well as for the 3S$_1$ $np$ pairing.

\end{abstract}
\date{\today}
 \maketitle

\section{Introduction}\label{Intro}

In the condensed matter physics the spin-ordered
pairing is known from the studies of the $^3$He liquid, heavy-fermion systems
like UPt$_3$, some Rb-and La-based superconductors and other
materials, see Refs.~\cite{Legett,SaulsAdv,Saxena,Saxena1,Saxena2,Saxena3,Hillier} for review.
A Josephson supercurrent through the strong ferromagnet CrO$_2$ was observed in
\cite{Keizer}, from which it was inferred that it is a spin triplet supercurrent.  A long-range supercurrent in Josephson junctions containing Co (a strong ferromagnetic material) was observed \cite{Khaire}
when one inserted thin layers of either PdNi or CuNi weakly ferromagnetic alloys between the Co and the two superconducting Nb electrodes.

In cold atomic Fermi gases strong magnetic dipolar interaction may cause pairing in the state with orbital angular momentum $L$, spin $S$ and total angular momentum $J$  equal to one, i.e in the $(2S+1)L_J$=3P$_1$~\cite{LiWu}.
Isotopes $^{161}$Dy and $^{163}$Dy
are the most magnetic fermionic  atoms with magnetic moments as high
as $10\mu_{e,\rm B}$, where $\mu_{e,\rm B}$ is the electron Bohr magneton. The lowest temperature reached in experiments ~\cite{Lev} for the spin-polarized $^{161}$Dy is a factor 0.2 below the Fermi temperature $\epsilon_{\rm F}$=300 nK. The co-trapped $^{162}$Dy cools to approximately critical temperature
for the Bose-Einstein condensation,  realizing a novel nearly quantum
degenerate dipolar Bose-Fermi gas mixture.
In some systems such as dilute Fermi gases the p-wave pairing may occur even in the case of a repulsive interaction~\cite{pwave-pairing,Kagan,Kagan1}. Conventional electron-phonon interactions induce triplet pairing in time-reversal 3$d$ Dirac semimetals, if magnetic impurities or exchange interaction are sufficiently strong, cf. \cite{Rosenstein:2015iza} and refs therein.    Very recently a paramagnetic Meissner effect in an Nb-Ho-Au structure was observed \cite{Bernardo}. In this system,  superconductivity  enhances  the  magnetic  signal  rather than expels   it.  Reference \cite{Espedal} demonstrated that by combining superconductors with spin-orbit coupled materials  the Meissner effect can be modulated by the orientation of an external magnetic field.

Since the magnetic field is the axial vector it is efficient pair
breaker for s- wave superconductors  but
it should not break pairs with parallel spins. The Zeeman coupling
of pairs with $S_z=\pm 1$, where $S_z$ is the spin projection on the quantization axis, is responsible for
this effect. For instance, the phase A of the p-wave spin-triplet pairing
in the $^3$He survives in the external magnetic field, which can also induce a  specific A$_1$-phase~\cite{AM1973,Levin}. The $^3$He-A$_1$ phase behaves as the magnetic superfluid in the external magnetic field. Superconductivity of the spin-triplet electron pairs in unconventional superconductors in external magnetic fields was extensively studied, see Ref.~\cite{VG1985,KR1998} and review   \cite{MineevSamokhin}.  Interaction of the vector order parameter with the magnetic field is introduced with the help of  the minimal coupling and  the Zeeman coupling.

In the description of the $^3$He the rotation field  was introduced in \cite{VolovikMineev84} with the help of the Galilean variable shift, similarly to that for the magnetic field. To the best of our knowledge a  possibility of the appearance of an own  magnetic and rotation fields in the whole volume of  fermion superfluids has not been considered.

The 1S$_0$ channel provides the largest    $nn$ and $pp$ attractive interactions at low densities in the neutron star matter. Thereby, A.B. Migdal suggested the Cooper pairing and superfluidity of neutrons in neutron stars in the 1S$_0$ state \cite{Migdal1959}.  With an increase of the baryon density the $NN$ interaction in the s-wave is weakened. The nucleon-nucleon ($NN$) phase shift in the 3P$_2$ partial wave becomes  the largest one among others at sufficiently high momenta owing to the strong spin-orbit $NN$ interaction in the vacuum that allows for the $nn$ pairing in 3P$_2$  state ~\cite{HGRR1970,Tamagaki1970,Takatsuka1971,Takatsuka:2004zq}. Therefore, in the neutron star interiors neutrons are supposed to be paired in the 1S$_0$ state at a low baryon density $n \leq (0.7-0.8) n_0$, where $n_0$ is the nuclear saturation density,  and in  $3P_2$ state  for $(3-4)n_0\geq n \geq 0.8 n_0$, cf. ~\cite{ST83} and recent works  \cite{Stein:2015bpa,Sedrakian:2018ydt}.
 However the value of the 3P$_2$ $nn$ gap
 in the neutron matter is poorly known and varies in various calculations
from tiny values $\lsim 10$~KeV up to values of the order of several MeV and may be more,
see Refs.~\cite{Takatsuka:2004zq,Schwenk:2003bc,Drischler:2016cpy,Ding:2016oxp,ChenLi,Dong,Zuo:2008zza}.
Uncertainties appear largely  due to a lack of knowledge of the efficiency of the three-body forces in a dense baryon matter \cite{Papakonstantinou:2017ewy}. Note that the cooling history of neutron stars is appropriately described in the nuclear medium cooling scenario within an ansatz that the 3P$_2$ pairing gap has only  a tiny value, cf.  \cite{Grigorian:2005fn,Grigorian:2016leu,Grigorian:2018bvg}.
Because of all these
uncertainties, and since microscopic calculations of the gap are
beyond the scope of the given work, we further consider the critical
temperature as an external phenomenological parameter varying in broad limits.

Mixing of 3P$_2$ and 3F$_2$ partial waves increases the value of the 3P$_2$ gap.  In some density interval the $nn$ 3PF$_2$ pairing may coexist with the $pp$ 1S$_0$ pairing. The 1S$_0$ channel is
most attractive for protons, as a consequence of their small
concentration in neutron stars but in the hyperon-enriched central regions
of sufficiently massive neutron stars proton concentration increases and protons can be paired also in the 3PF$_2$ state~\cite{Schulze}, as well as $\Lambda$ hyperons \cite{Raduta:2019rsk}.
Besides hyperons \cite{Glendenning:2001pe,Maslov:2015msa}, $\Delta (\frac{3}{2},\frac{3}{2})$ isobars may exist in  central regions of sufficiently massive neutron stars   \cite{Drago:2014oja,Kolomeitsev:2016ptu}. Pairing in the fermion systems of spin $3/2$ was recently discussed in \cite{Venderbos2016}.
Moreover,  a developed pion condensate may exist in  the central regions of sufficiently massive neutron stars. In the presence of the developed pion condensate only one Fermi sea of a mixture of the baryon quasiparticles consisted of neutrons, protons and $\Delta$ isobars is filled \cite{Campbell:1974qt,Campbell:1974qu,Baym:1975tm,Voskresensky:1982vd,Migdal:1990vm} and thereby they  can be paired  in the 3S$_1$ state.

The phases of the 3P$_2$ $nn$ pairing  were studied in~\cite{SS1978,SSt81,Yasui:2019unp,VS84,VS} within the BCS weak coupling approximation, when  the ground state corresponds to the symmetric (magnetically neutral) phase.   The order parameter for the 3P$_2$ $nn$ pairing  is the 3x3 matrix. The Ginzburg-Landau free-energy functional is ordinary considered as the expansion in the order parameter up to 4-th power. However  the 6-th order term calculated within  the BCS approximation proves to be negative \cite{SS1978} causing a possibility of the first-order phase transitions in the system.  Recently the Ginzburg-Landau free-energy functional was expanded in the order parameter up to 8-th power and coefficients of expansion were found in the BCS approximation \cite{Yasui:2019unp}. A particular role of the Zeeman and gradient terms, which  are of our key interest here, was not studied, cf. ~\cite{SSt81}.

Magnetic fields in ordinary pulsars, like  the Crub pulsar, reach values  $\sim (10^{12}-10^{13})$ Gs at their surfaces.
At the surface of magnetars magnetic fields may reach values $\gsim 10^{15}$Gs. In the interior, the magnetic field might be even stronger (up to $\sim 10^{18}$Gs) depending on the assumed (still badly known) mechanism of magnetic field formation~\cite{Ghosh}.
Still stronger magnetic fields appear in non-central heavy ion collisions. The first estimate of the value of the magnetic field in heavy-ion collisions  performed in Ref. ~\cite{VA1980}  argued for the presence of the magnetic fields of the order of $\sim 10^{17}\div 10^{18}$G  at collision energies $\sim $ GeV/A. Subsequent  calculations ~\cite{IST} demonstrated that typical values of magnetic fields may reach $\sim (10^{17}-- 10^{19})$Gs in heavy-ion collision experiments from GSI to LHC energies. Thus, the coupling of a spin-triplet order parameter to a magnetic field might be of  importance for the description of nuclear systems prepared in peripheral heavy-ion collisions.

For low densities the 3S$_1$ channel provides the largest  attractive interaction  for the $np$ pairing in the  isospin-symmetrical matter. With increasing density the 3D$_2$ channel becomes most attractive, cf.  \cite{Sedrakian:2018ydt}.
 One of the hypothesis for the explanation of the level
structure  of super-deformed (rotated) nuclei is the spin-triplet pairing~\cite{Shapiro,Shapiro1,Shapiro2}. Spin-triplet pairing in  $N=Z$ nuclei with $A>140$ may be favored, since the spin-orbit force becomes vanishing \cite{Bertsch:2009xz}. In the vicinity of the proton drip in heavier nuclei the spin-triplet pairing also could potentially  become important.
The $^3$SD$_1$ spin-triplet $np$ pairing in nuclei was studied in \cite{Guo:2018lna}. The BCS calculations for the symmetric matter with the vacuum interactions predict the $np$ pairing gaps as large as $\simeq 12$ MeV. Even with the effect of the depletion of the Fermi surface taken into account, one estimates the $np$  pairing gap in maximum to be as high as $\simeq 4$ MeV. Reference ~\cite{Cederwal} studying the level structure of $^{92}$Pd  found signals of the spin aligned $np$-paired state with $J=9$ and $L\neq 0$.

There exist millisecond pulsars, being fast-rotating neutron stars with the angular rotation  frequencies  as high as $\sim 10^4$ Hz, see Ref.~\cite{Ghosh}. The rotation frequency of the fireball  in ultrarelativistic heavy-ion collisions
at the freezeout \cite{Shuryak}  is estimated as $\sim 10^{22}$ Hz. For low energies the
spectator pieces in heavy-ion collisions can rotate at a still larger frequency  ($> 10^{22}$ Hz).
In a sense, the rotation acts in a neutral system similar to a magnetic field in a charged system. Thereby, description of the behavior of the spin-triplet condensates in the magnetic and the rotation fields is an important issue.

Another phenomenon, which might be relevant to our study, is a condensation of the charged $\rho$ mesons in a dense isospin-asymmetric baryon matter \cite{Voskresensky:1997ub,Kolomeitsev:2004ff,Kolomeitsev:2017gli}. The $\rho$ mesons, being bosons with the spin and isotopic spin equal to one, are described by the vector--isospin-vector field $\rho^a_\nu$, where $a=1,2,3$ is the isospin index and $\mu =0,1,2,3$ is the Lorentz index.
In the quantum field theory relevant phenomena are condensations of non-abelian charged $\rho$ and $W$ bosons in super-strong magnetic fields $\gsim 10^{19}$G in vacuum, see
Ref.~\cite{Chernodub:2010qx,Chernodub,Olesen}. Presence of strong magnetic fields in neutron star interiors would promote the charged $\rho$ meson condensation \cite{Mallick:2014faa}. Gluons become massive in the hot quark-gluon plasma and  may form condensates at some conditions.  Thereby a ferromagnetic superconductivity of the condensate of charged vector fields is another issue of our interest. The axial-vector--isospin-vector boson may also play a role in nuclear phenomena forming condensates at some conditions, cf.  \cite{Hashimoto:2014sha,MRho2019}. Finally, the order parameter in color superconductors has a matrix structure and a spin-triplet di-quark pairing is allowed in some cases \cite{Iida:2000ha,Buballa:2002wy,Pang:2010wk,Alford:2016dco}.

The Ginzburg-Landau description  is relevant not only below critical point for the order parameters but also for their long-range fluctuations within a fluctuation region above the critical point. The width of the fluctuation region is determined by the Ginzburg number Gi following the so called Ginzburg-Levanyuk criterion \cite{GS}. In the substances with a strong interaction between quasiparticles the Ginzburg number Gi$\sim 1$ and the fluctuation region should be  broad \cite{LarkinVarlamov}. For instance, the fluctuation region might be very broad for  the color-superconductors and for the proton pairing in neutron stars  \cite{Voskresensky:2004jp}.
Thereby the consideration of  a triplet paring correlations  above the critical point is also an important issue.

In this work we study nonmagnetic, diamagnetic, paramagnetic and ferromagnetic responses of
superfluid and superconductive  condensates of vector bosons and spin-triplet Cooper pairs.
 We start with a reminding  of superfluid and superconductive properties of the complex scalar field at the negative squared effective mass of the boson (in Sect. \ref{preliminary}) and then (in Sect. \ref{vectorboson}) we focus on the description of the complex   vector field of neutral and charged bosons at the conditions when their squared effective masses might be either negative or positive.  Influence of the external magnetic field is considered. Various nonmagnetic and ferromagnetic superfluid phases and nonmagnetic, superdiamagnetic and ferromagnetic superconducting phases will be studied. In  Sect.~\ref{vector} we perform a general analysis of the spin-triplet pairing of charge-neutral fermions with a magnetic
moment, interacting with the magnetic field by the Zeeman coupling. First, we assume that  spin-orbit forces are weak and spin of the pair is a good quantum number. The spin-triplet pairing is then described by a vector order parameter, as for a composed spin-one charge-neutral boson with an anomalous magnetic moment. Some of the phases  are characterized by the spin order parameter and a self-magnetization. In this sense we  deal with a ferromagnetic superfluidity.
  Note that  an another type of the   ferromagnetic superfluidity, when a  magnetization exists already in absence of the Cooper pairing and  remains in presence of the superfluidity, as it may occur in some uranium compounds, is not of our interest here, see \cite{Walker,Shopova}. In Sect. \ref{charged} we consider the spin-triplet pairing in charged fermion superconducting systems described by the vector order parameter.
In Sect.~\ref{nnpairing} focus is made on
the description of the 3P$_2$ $nn$ pairing in the neutron star matter. Various phases are found.
Some numerical evaluations  are performed in Sect.~\ref{BCS} for the 3P$_2$ $nn$ and $pp$ pairings and for the 3S$_1$ $np$ pairing and some
physical consequences of the ferromagnetic superfluidity and superconductivity for
neutron stars
and heavy-ion collisions are
specified.
In Sect.~\ref{conclusion} we formulate our conclusions.

Throughout the paper we use units $\hbar =c=1$, Lattin indices are $i=1,2,3$, Greek indices are Lorentz ones, $\mu =0,1,2,3$. For 3-vectors, where it does not cause a confusion,  we  use the ordinary 3-dimensional notations, $\vec{a}=(a_1,a_2,a_3)$. Summation over repeated indices is implied, if not presented explicitly.

\section{Preliminaries}\label{preliminary}

\subsection{Superfluidity and superconductivity of complex scalar fields}\label{superscalar}
\subsubsection{Lagrangian and equations of motion}
Consider the model described by the Lagrangian density
\begin{eqnarray}
L=D_\mu \phi D^\mu\phi^* -m_{\rm sc}^2 |\phi|^2 -\lambda|\phi|^4/2 -F_{\mu\nu}F^{\mu\nu}/(16\pi)\,,
\label{complL}
 \end{eqnarray}
$\phi =(\phi_1 -i\phi_2)/\sqrt{2}$ is the spin-zero complex field of a negatively charged boson,
$\phi_1$ and $\phi_2$ are real components, $\phi_{+} =\phi^*=(\phi_1 +i\phi_2)/\sqrt{2}$ is spin-zero complex field of a positively charged boson, $A_\mu$ is the electromagnetic field,
\begin{eqnarray}
F_{\mu\nu}=\partial_\mu A_\nu -\partial_\nu A_\mu\,,\quad
D_\mu =\partial_\mu +ieA_\mu -i\mu \delta_{\mu 0}\,,\label{long}
 \end{eqnarray}
 $e<0$ is the charge of the electron, $e^2=1/137$,
 $\mu$ is the  chemical potential of the negatively charged boson, e.g. in the neutron-star matter due  to reactions $n+e\to n+\pi^{-}$, $n\to p+\pi^{-}$ one gets $\mu_{\pi^-}=\mu_e=\mu_n -\mu_p$. The quantity $m_{\rm sc}^2 =m^2 +U$, where $m>0$ is the mass of the vector particle, $U$ is a scalar potential, which we assume to be zero in vacuum and heaving a negative value in the medium. We will  only use that in a deep potential,  $U<-m^2$, in the medium the quantity $m_{\rm sc}^2$ becomes negative.

Equations of motion are
\begin{eqnarray}
&D_\mu D^\mu \phi +m_{\rm sc}^2 \phi +\lambda |\phi|^2 \phi =0\,,\label{phich}\\
&\partial_\mu F^{\mu\nu}=-4\pi \delta L_{\phi}/\delta A_\nu =4\pi J^\nu \,,\label{Max}
\end{eqnarray}
with the 4-current-density  $$J^\nu = -ie\phi D^{*\nu} \phi^* +{\rm c.c.}\,,$$ which is conserved, $\partial_\nu J^\nu =0$, the abbreviation ${\rm c.c.}$ denotes complex conjugation, $L_{\phi}$ is the part of the Lagrangian density depending on $\phi$.

 For the case of the static field $\phi$ and the static magnetic field equations of motion render
\begin{eqnarray}
&(\nabla -i e\vec{A})^2 \phi -m_{\rm ef}^2 \phi -\lambda |\phi|^2 \phi =0\,, \label{phichst}\\
&\Delta \vec{A} = -4\pi \vec{J}\,,\,\, \vec{J}=ie(\phi\nabla\phi^*-\phi^*\nabla \phi) -2e^2\vec{A}|\phi|^2, \label{Cursc}
\end{eqnarray}
where $m_{\rm ef}^2 =m_{\rm sc}^2 -\mu^2\,$ has a sense of the squared effective mass term. It is used that $\mbox{div}\vec{A}=0$.

We introduce the Gibbs free-energy density  $G=F-\vec{M}\vec{H}-\vec{H}^2/8\pi$, $F$ is the free-energy density,
$\vec{M}=(\vec{h}-\vec{H})/4\pi $ is the density of the magnetization, $\vec{H}$ is the strength of the external uniform static magnetic field. Note that in the given paper we  use the definition of $G$,  which differs from the often used definition by the shift  on the constant value  $\vec{H}^2/8\pi$\,. Thus the  Gibbs free-energy density is
\begin{eqnarray}
G=|(\nabla -ie\vec{A}) \phi|^2  +m_{\rm ef}^2 |\phi|^2 +\frac{\lambda}{2} |\phi|^4 +
\frac{(\vec{h}-\vec{H})^2}{8\pi}\,,\label{Gsc}
\end{eqnarray}
where $\vec{h}=\mbox{curl} \vec{A}$.
The condensate of the charged boson field appears provided $m_{\rm ef}^2<0$ in a part of the space.

 A superfluid non-relativistic motion of the system with the velocity $\vec{v}$ is described with the help of the replacement $\vec{D} \to \vec{D} +i m_{\rm qp}\vec{v}$, where $m_{\rm qp}$ is a quasiparticle  mass-coefficient, which value is not of our interest at present.
 Replacing $\phi \to \phi e^{i\chi}$ we find the contribution to the density of the momentum of the system $\vec{J}_v = 2\nabla \chi |\phi|^2$,
 $\vec{v} =\nabla \chi$, $\vec{p}=m_{\rm qp}\vec{v}$ is the momentum of the particle of the superfluid.

\subsubsection{Neutral complex scalar field. Nonmagnetic superfluid phase}
Consider a complex scalar field, which does not interact with the electromagnetic field. Thus we put $e=0$. Then $\mu =0$ as well, and thereby $m^2_{\rm ef}=m^2_{\rm sc}$.

 Let us consider the half-space $x<0$ medium, where  $m_{\rm sc}^2 =  m^2_{0} <0$ is the constant, and the system
 is placed in the external uniform static magnetic field $\vec{H}$. For $x>0$,  $m_{\rm sc}^2 =  m^2>0$.
The specific interactions, which may  provide  inequality $m^2_{0} <0$, are not of our interest here.

 For $x\leq 0$  from Eqs. (\ref{phichst}), (\ref{Cursc}) putting there $e=0$ we obtain solutions
\begin{eqnarray}
\phi =f_0 \mbox{th} [(x-x_0)/(\sqrt{2}l_\phi)]\,,\quad  \vec{h}=\vec{h}_0=const
\,,\label{phisolneut}
\end{eqnarray}
$$ f_0 =\pm\sqrt{-m^2_0/\lambda}\,\theta (-m_{\rm 0}^2)\,,\quad l_\phi =1/{|m_{0}|}, \quad x_0=const\,,$$
$\theta (z)$ is the step function.
For $x\geq 0$ we  put $\phi =0$, $\vec{h}=\vec{H}$. It is possible to do provided $l_\phi\gg 1/m$, i.e., $|m_0|\ll m$, that we  assume  for simplicity. Such an approximation in the phase transition theory is usually called the Landau approximation.
From the boundary conditions for $x=0$ we get $x_0 =0$ and
$h_0 =H$.
Thus, we conclude that {\em the magnetic field and  condensate decouple.}

With these solutions we obtain the space-averaged Gibbs free-energy density
\begin{eqnarray}
\overline{G}= \frac{\int d^3 x G}{\int d^3 x} = -\frac{m_{0}^4}{2\lambda}\left(1-\frac{4\sqrt{2}}{3}\frac{l_\phi}{d_x}\right)\theta(-m_{\rm 0}^2)\,,
\label{phisolneutVec}
\end{eqnarray}
$d_x $ is the  length of the system in the $x$-direction. Note that for the semi-infinite matter $d_x \to \infty$
and surface-energy  term is vanishingly small. However   after the replacement $d_x \to  d_x/2$  Eq. (\ref{phisolneutVec}) holds also for the layer of a finite length provided $d_x \gg l_\phi$.

\subsubsection{Charged complex scalar field. Superdiamagnetic response, superconductivity and mixed Abrikosov state} Assume that $m_{\rm ef}^2 =m_{\rm sc}^2-\mu^2=m_{\rm ef,0}^2$ for $x<0$, with $m_{\rm ef,0}^2 =const <0$ and that $m_{\rm ef}^2 =m^2>0$ for $x>0$, and  the system is placed in the static uniform magnetic field parallel $z$.
For $1/l_\phi \gg eH l_h$, where $m_{\rm ef,0}^2$ now replaces the value $m_{\rm 0}^2$ in previous example,
$l_h$ is the penetration depth of the magnetic field in the medium, assuming  $l_\phi\gg 1/m$ from   Eq. (\ref{phichst})  we recover   solution (\ref{phisolneut}). With this solution at hand,   Eq. (\ref{Cursc}) in the gauge $\mbox{div} \vec{A}=0$ simplifies as
  \begin{eqnarray}
  \Delta \vec{A}- 8\pi e^2 |\phi|^2 \vec{A}=0\,, \quad x\leq 0\,.\label{Meis}
 \end{eqnarray}
For
$$l_h =1/ \sqrt{8\pi e^2 f_0^2} \gg l_\phi\,,$$  we may put $|\phi|^2 =f_0^2$ in (\ref{Meis}). The value $m_\gamma =1/l_h$
plays the role of the photon mass in the superconducting region, the quantity
$\kappa =\sqrt{l_h/l_\phi}$
 is  the  Ginzburg-Landau parameter. The inequality $1/l_\phi \gg eH l_h$ is rewritten as $H\ll H_{\rm cr}$, with the thermodynamical critical field
 $$H_{\rm cr} =\sqrt{4\pi }|m_{\rm ef}^2|/\sqrt{\lambda}\,.$$
 For $H\ll H_{\rm cr}$ the solution of Eq. (\ref{Meis}) is
 $$A_2 (x\leq 0)= Hl_h e^{x/l_h}\,,$$
where we used  the gauge $\vec{A}=(0, A_2 (x), 0)$ for $\vec{H}\parallel z$ and   the boundary conditions $A_2^{\prime} (x\to 0)=H$, $|A_2 (x)|<\infty$. This solution   demonstrates the
Meissner-Higgs effect of the repulsion of the magnetic field from the superconducting region. For $\vec{H}\parallel y$ a similar Meissner effect exists for $A_3 (x)$  with  $\vec{A}=(0, 0, A_3 (x))$.

The volume part of the space-averaged Gibbs free-energy density in the presence of the condensate, with the  magnetic field being repelled from the condensate matter (phase I: $\phi =f_0$, $\overline{h}=0$) is as follows:
$\overline{G}_{\rm I}=-\frac{m_{\rm ef}^4}{2\lambda}+\frac{H^2}{8\pi}\,.$
 The volume part of the averaged Gibbs free-energy density in the absence of the condensate, with the  magnetic field (phase  II: $\phi =0$,
 $\overline{h}=H$) is $\overline{G}_{\rm II}=0$. Thus for $H<H_{\rm cr}$ the condensate phase is energetically favorable, since $\overline{G}_{\rm I}<\overline{G}_{\rm II}$.

 For the Ginzburg-Landau parameter $\kappa \gg 1$ (actually it is sufficient to have $\kappa >1/\sqrt{2}$) in a range of the fields $H_{\rm c1}<H<H_{\rm c2}$  the Abrikosov mixed phase is formed.
 Already for $H>H_{\rm cr 1}$ (at $H_{\rm cr 1}<H_{\rm cr}$) the surface energy of the system  is decreased, if  there appear filament vortices of the normal phase. The typical transversal size of the normal filament vortex directed parallel $\vec{H}$  is $\sim l_\phi$, whereas the magnetic field decreases at the distance $\sim l_h$ in the transversal direction. Thus the Gibbs free
energy gain due to appearance of the single vortex is estimated as $\sim -\pi  l_h^2 d_z H^2/8\pi$ and the energy loss is $\sim \pi l^2_\phi d_z m_{\rm ef}^4/2\lambda$. Comparing the gain and loss contributions we see that the Gibbs free energy is indeed gained for $H<H_{c1} \sim H_c /\kappa$. For $H>H_{\rm cr 1}$ the vortices form the  triangular  lattice, which proves to be energetically more favorable compared to the quadratic lattice originally considered by Abrikosov, cf. \cite{Tilly-Tilly}. Thus for $H>H_{\rm cr 1}$ the  solution for the field $\phi$ should satisfy the  periodic boundary conditions.
Such a solution replaces the solution satisfying the boundary conditions for $x=0$ that we had for $H<H_{\rm cr 1}$. With subsequent  increase of $H$ the distance between vortices decreases, the condensate weakens and  vanishes for $H=H_{\rm cr 2}$.

For $H$ slightly  below $H_{\rm cr 2}$ the condensate field is weak and the equations of motion (\ref{phichst}), (\ref{Cursc})  can be linearized. Then the solution can be found analytically.
Eq. (\ref{phichst}) for $\vec{A}=(A_1 (y), A_2 (x), 0) $  renders
\begin{eqnarray}
 -(D_1^2 +D_2^2)\phi = -m_{\rm ef}^2 \phi\,.\label{aux}
 \end{eqnarray}
 With $\vec{A}\simeq (0, Hx,0)$, being the solution of the linearized Eq. (\ref{Cursc}),  we may rewrite Eq. (\ref{aux}) in the form
\begin{eqnarray}
-\frac{(\nabla -i e\vec{A})^2 \phi}{2m_{\rm aux}}  \simeq -\frac{m_{\rm ef}^2 \phi}{2m_{\rm aux}}  \, \end{eqnarray}
of the Schr\"odinger equation for the non-relativistic spin-less particle in the uniform magnetic field, where the quantity    $m_{\rm aux}$ is     an auxiliary mass coefficient.

 The energy in the ground state is
$E_{\rm min}=\frac{|m_{\rm ef}^2| }{2 m_{\rm aux}}= |e|H_{\rm cr 2}/2 m_{\rm aux}$, from where we find
\begin{eqnarray}
H_{\rm cr 2}=|m_{\rm ef}^2/e|=H_{\rm cr}\sqrt{2}\kappa\,.\label{Hcr2}
\end{eqnarray}

For the further ussage let us introduce the auxiliary condition
\begin{eqnarray}
D_i\phi_i  =0,  \,\,\, {\rm or}\,\,\, D\phi =0\,,\label{ansAx}
\end{eqnarray}
where $i=1,2$, $\phi_i = (\phi , -i \phi)$\,, $D=D_1 -i D_2$. Let us apply the operator $D=D_1 +i D_2$ to (\ref{ansAx}). Then we obtain equation
\begin{eqnarray}
(D_1^2 +D_2^2 -i[D_1,D_2]_{-})\phi =0, \,\,
 \end{eqnarray}
 with $[a,b]_{-}=ab-ba$, $i[D_1,D_2]_{-} \phi =e h_3 \phi\,$, $h_3 =\partial_1 A_2 -\partial_2 A_1\,$. Thus
 \begin{eqnarray}
 -(D_1^2 +D_2^2)\phi = -e h_3 \phi >0\,.\label{convEq}
 \end{eqnarray}
With $\vec{A}\simeq (0, H_{\rm c2} x,0)$ this equation is equivalent to  Eq. (\ref{aux}).
The solution has the form
$$\phi =\sum_{n=-\infty}^{\infty}C_n e^{ikny}\phi_n (x), \quad \phi_n (x)=e^{-(x-x_n)^2 /2l_{\phi}^2}\,,$$
where $x_n =nkl_{\phi}^2$, $C_{n+N}=C_n$, $k=|e^*| H_{\rm cr 2}x_0$, $N=1$ corresponds to the quadratic lattice, $N=2$, to the triangular one. This solution can be then used as the probe function to calculate the space-averaged Gibbs free-energy $\overline{G}$ within the mixed state and by variation of the free parameters to find its minimum.

In the toy model considered above we were not interested in specification of the  interactions, which
provide the inequality $m_{\rm sc}^2<0$ for the neutral system and inequality $m^2_{\rm ef}=m_{\rm sc}^2-\mu^2 <0$ for  the charged system.
 In the neutron-star matter there exists a fraction of protons,
and one can consider a possibility of the $\pi^-$ condensation, described by the negatively charged field $\phi =\phi_{\pi^-}$. Chemical potentials of particles fulfill equalities  $\mu_p =\mu_n -\mu_e$ and $\mu_{\pi^-}=\mu_e$. In the  approximation of the ideal pion gas the s-wave $\pi^-$ condensation would occur for $\mu_e (n)>m_\pi$, where $n$ is the baryon density. However it proves to be that the ideal gas approximation is hardly realized in a realistic problem due to the presence of the s-wave repulsive Weinberg-Tomozawa $\pi^- n$ interaction. The latter interaction
does not allow for the s-wave $\pi^-$ condensation up to high densities \cite{Migdal:1978az}.  The $\pi^-$ condensation with the field $\phi =f_0 e^{i\vec{k}_0\vec{r}}$ for $k_0\neq 0$ in neutron star matter may appear for $n>n_c^{\pi} \sim (1.5-3)n_0$ due to a strong p-wave $\pi N$ attraction \cite{Migdal:1978az,Migdal:1990vm}.  The condensate $\pi^-$ has properties of  an unconventional   superconductor of the second kind. In the external magnetic field for $H>H_{\rm cr 1}$ the vortices form the plane-layer structures rather then the filamentary structures and the value $H_{\rm cr 2}$
proves to be  very high \cite{VA1980,Migdal:1990vm}. Also, for $n>(n_c^{K^-} , n_c^{K^0}) \sim (2-4)n_0$ there may appear the s-wave \cite{Glendenning:2001pe,Kolomeitsev:2002pg} and p-wave \cite{Kolomeitsev:1995xz,Kolomeitsev:2002pg} antikaon condensates. The condensate $K^-$ has properties of  a superconductor and $\bar{K}^0$, of a superfluid.

\subsection{Zeeman coupling of neutral fermions and ferromagnetic state}

 In the quantum field theory in the famous Nambu-Jona-Lasinio (NJL) model \cite{Nambu:1961tp,Nambu:1961fr} the $\langle(\bar{\psi}\psi)^2\rangle$ self-interaction of quarks represents the squared chiral condensate $\bar{\psi}\psi$. Angular brackets denote averaging over the equilibrium state of the fermion medium. Reference \cite{Hashimoto:2014sha} considers a generalization of the NJL model with the spin-spin interaction term in the free-energy density  $b_s \langle(\bar{\psi}\vec{\gamma}\gamma_5 \psi)^2\rangle$ instead of $\langle(\bar{\psi}\psi)^2\rangle$ term in the original NJL model, $\gamma_i$, $\gamma_5$ are the Dirac matrices, the spin operator of the fermion is $\vec{S} =\frac{1}{2} \bar{\psi}\vec{\gamma}\gamma_5 \psi$, $i=1,2,3$. The spin-spin interaction for $b_s <0$  causes a spontaneous magnetization. Such an interaction appears also  in the model of the neutral massive  fermion field $\psi$ interacting with the own static magnetic field $\vec{h}=\mbox{curl} \vec{A}$ by the Zeeman coupling, $\vec{A}$ is the vector-potential of the magnetic field. The Lagrangian  density  is as follows
 $$L=\bar{\psi}(i\gamma^\nu \partial_\nu +i\gamma^0 \mu -m_{\rm F})\psi -U +\eta \vec{S}\vec{h}-\vec{h}^2/8\pi\,,$$
 $m_{\rm F}$ is the bare fermion mass, $\vec{\cal M}=\eta \vec{S}$ is the magnetic moment of the fermion,
 $U$ is a fermion interaction term not depending on $h$.

 The contribution to the Gibbs free-energy density dependent on $h$ is
 $$G_h =-\eta \langle(\bar{\psi}\vec{\gamma}\gamma_5 \psi)\vec{h}\rangle/2 +(\vec{h}-\vec{H})^2/8\pi\,.$$
 For $\langle\bar{\psi}{\gamma}^3\gamma_5 \psi\rangle \neq 0$ and $\langle\bar{\psi}{\gamma}^{1,2}\gamma_5 \psi\rangle =0$,
 minimizing $G_h$ in $h$ (let it be parallel $z$) one gets
 $$\vec{h} =\vec{H} +\vec{n}_3 \cdot 2\pi \eta \langle(\bar{\psi}{\gamma}^3\gamma_5 \psi)\rangle\,,$$
 where $\vec{n}_3 =(0,0,1)$,
  and
 $$G_h =-\pi \eta^2 \langle\bar{\psi}{\gamma}^3\gamma_5 \psi\rangle^2/2 +\eta \langle(\bar{\psi}{\gamma}_3\gamma_5 \psi)\rangle{H}_3/2\,.$$ The first term represents  a spin-spin interaction. For the  polarized spin state  this contribution to the free-energy density is negative (even for $H=0$). However  the positive Fermi gas energy term  for the polarized state is higher than that for the non-polarized state. For the fully polarized matter $\langle\bar{\psi}{\gamma}^3\gamma_5 \psi\rangle =n$, where  $n$ is the fermion density. Thus  the difference in the energy density for the fully spin-polarized matter and the non-polarized one for $H=0$, $T=0$ becomes $$E-E({h=0})=\frac{3^{5/3}\pi^{4/3}(2^{2/3}-1)n^{5/3}}{10 m_{\rm F}^*} -\frac{\pi \eta^2 n^2}{2}\,,$$
  $m_{\rm F}^*$ is the effective fermion mass resulting from interactions not dependent on $h$. Thus in this toy model the ferromagnetic state  becomes energetically favorable only for an abnormally   high  density $n>n_{\rm cr} =\frac{3^{5}\pi (2^{2/3}-1)^3}{125 m_{\rm F}^{*3} \eta^6}$, that is not realized for densities reachable in neutron stars. Only in an extremely high external magnetic field the neutron star matter could be fully polarized. Reference  \cite{Hashimoto:2014sha} additionally included the axial anomaly term,  a contribution of the axial-vector meson condensate and the neutral pion condensate. With these additional contributions the critical density, above which the neutron star matter can be polarized, is strongly diminished up to the values reachable in the most massive neutron stars.

Note also that   there exists a possibility of a
ferromagnetic transition in quark matter interacting with one-gluon-exchange interaction \cite{Tatsumi:1999ab}, similarly to the ferromagnetism in electron gas.  	
Spontaneous spin polarization due to the tensor self-energies in quark matter within the NJL model was considered in Ref. \cite{Maruyama:2017mqv}.

\section{
Complex vector boson fields. Ferromagnetic superfluidity and superconductivity
} \label{vectorboson}
\subsection{Lagrangian, equations of motion, Gibbs free energy}

Let ${\phi}^{(j)}_\nu =(\phi^{(1)}_\nu,\phi^{(2)}_\nu,\phi^{(3)}_\nu)$ is the field of a massive
vector-isospin-vector boson, such as $\rho$ meson, with $\phi^{(1)}_\nu,\phi^{(2)}_\nu,\phi^{(3)}_\nu$ as real quantities. Latin superscript  $(1),(2),(3)$ describes isospin, whereas the Greek index is as above the Lorentz index $0,1,2,3$.
Instead of real fields $\phi^{(1)}_\nu,\phi^{(2)}_\nu$ it is convenient to introduce complex fields $$\phi_\nu =(\phi^{(1)}_\nu -i \phi^{(1)}_\nu)/\sqrt{2}\,,\,\,\phi^*_\nu =(\phi^{(1)}_\nu +i \phi^{(1)}_\nu)/\sqrt{2}\,.$$

In our toy model we will for simplicity put $\phi^{(3)}_\nu =0$. Then we  deal with a simpler problem of the description of the complex vector field $\phi_\mu$.
 The interaction of $\phi_\mu $ with the electromagnetic field is described with the help of the long-derivative replacement $\partial_\mu \phi_\nu \to D_\mu \phi_\nu$, cf. (\ref{long}), and the Zeeman term. Then the Lagrangian density for the interacting $\phi$ and  the electromagnetic fields   renders
\begin{eqnarray}
L_{\phi,A}&=-\frac{F_{\mu\nu}F^{\mu\nu}}{16\pi}-\frac{\phi_{\mu\nu}\phi^{*\mu\nu}}{2}+m_{\rm sc}^2 \phi_\nu \phi^{*\nu}
 \label{LagrA}\\
&+L_{\phi\phi}
+i\eta F_{\mu\nu}\phi^{*\mu} \phi^{\nu}
\,,\nonumber
 \end{eqnarray}
 $\phi_{\mu\nu} =D_{\mu}\phi_\nu -D_\nu \phi_\mu$, as above $m_{\rm sc}^2$ is the squared bare mass shifted by an attractive scalar potential.

 The self-interaction term we take in the form
\begin{eqnarray}
 L_{\phi\phi}=-\Lambda [(\phi_\nu \phi^{*\nu})^2 + \xi_1 (\phi_\nu \phi^{\nu})(\phi^{*}_\mu \phi^{*\mu})]
 \,,\label{phiphisimp}
 \end{eqnarray}
  where $\Lambda$ is a positive coupling constant. Simplifying consideration  we shall employ $\xi_1=0$, if other is not mentioned.
 A  non-abelian form of the self-interaction  was used in \cite{Nielsen:1978rm,Ambjorn:1988fx} in the problem  of the instability of the $W$  boson vacuum  in a strong external magnetic field, in \cite{Voskresensky:1997ub,Kolomeitsev:2004ff,Kolomeitsev:2017gli} for  the description of the charged $\rho$ meson condensation in the dense isospin-asymmetric baryon matter and
  in \cite{Chernodub:2010qx,Chernodub},  for the  description of the instability of the  $\rho$ meson vacuum in a strong external magnetic field. At the condition $\phi^{(3)}_\nu =0$, that we use, results of those works and ours here coincide provided $\xi_1=-1$.

 The Zeeman coupling term, $L_{\rm Zeeman}=i\eta F_{\mu\nu}\phi^{*\mu} \phi^{\nu}$, describes the interaction of the spin of the complex vector field with the electromagnetic field.   In absence of the anomalous  magnetic moment, the magnetic moment  of the $\rho^-$ meson would be  ${\cal M}_\rho = \eta/m_\rho =e/m_\rho$, $e<0$. With inclusion of a contribution of the anomalous  magnetic moment, ${\cal M}_\rho\neq 2e/(2m_\rho)$. Reference \cite{Krutov:2018mbu} finds
${\cal M}_\rho \simeq 2.2 e/(2m_\rho)$, other existing  calculations give other values. An important circumstance here is only that  in general case $\eta \neq e$.

Note that in a realistic problem of the behavior of the $\rho$ meson in isospin-asymmetric nuclear matter one should  include $\phi_0^{(3)}$ component, the electromagnetic interaction of the charged $\rho$ fields, and the $\rho$ interaction with fermions and other mesons, e.g. with the $\sigma$ meson field, cf. \cite{Voskresensky:1997ub,Kolomeitsev:2004ff,Kolomeitsev:2017gli}.

Equations of motion for the fields $\phi^\nu$ render:
\begin{eqnarray}
&D^\mu D_\mu \phi^\nu - D^\nu D_\mu \phi^\mu  -i(e +\eta) F^{\mu\nu}\phi_\mu \label{motVec}\\
& +m_{\rm sc}^2 \phi^\nu -2\Lambda (\phi^{*}_\mu \phi^{\mu})\phi^\nu
 =0\,,\nonumber
 \end{eqnarray}
where we used the identity
\begin{eqnarray}
[D_\mu , D_\nu]_{-}  \psi =ieF_{\mu\nu}\psi\,,\label{iden}
 \end{eqnarray}
 and
\begin{eqnarray}
&\partial_\mu F^{\mu\nu}=4\pi J^{\nu}\,, \quad {\rm with}\nonumber\\
&J^{\nu}=ie D^\nu \phi_\mu \cdot \phi^{*\mu}-ie \phi^{*\mu}D_\mu \phi^\nu  +\rm c.c.\\
&-i(e+\eta)
\partial_\mu (\phi^{*\mu} \phi^{\nu}- \phi^{*\nu} \phi^{\mu})\,.\nonumber
 \end{eqnarray}
 Now the value $m^2_{\rm ef}=m_{\rm sc}^2
 -\mu_\phi^2$ has a sense of the squared effective mass of the complex vector field.

 From (\ref{motVec})   for $\eta =e$ neglecting $\sim \phi^3$ terms
 we recover ordinary Proca equation for the Bose particle with the spin one compatible with  the condition
 \begin{eqnarray}
 D_\mu \phi^\mu =0\,,\label{condDmu}
   \end{eqnarray}
 which is fulfilled identically   away from the sources of the electromagnetic field.
  To show this we apply the operator $D_\nu$ to the equation of motion (\ref{motVec})
  and  make use of the identity (\ref{iden}) and that away from the sources $J_\nu =0$.
   Contrary, the condition (\ref{condDmu}) is not necessarily compatible with the non-linear equation of motion (\ref{motVec}) and even with the  linear equation of motion at  $\eta \neq e$. Below (see discussion of Eq. (\ref{eqmotinh11ch})) we shall demonstrate a specific case,  when the condition (\ref{condDmu}) is compatible with the linear equations of motion for the charged field at $\eta \neq e$.

 For static vector fields $\phi^\nu =(0,\phi^i)$ and $A^\nu =(0,A^i)$
  the Gibbs free-energy density renders  (now in ordinary 3-dimensional notations)
\begin{eqnarray}
&G= m_{\rm ef}^2 |{\phi}_j|^2 +\Lambda ({\phi}_j{\phi}^{*}_j)^2
+\frac{(\vec{h}-\vec{H})^2}{8\pi}\label{Gn}\\&+|D_i \phi_j|^2 - D_j {\phi}_i D_i^{*} {\phi}_j^{*} +i\eta\epsilon_{jik}h_k\phi_j^{*}\phi_i
\,.\nonumber
\end{eqnarray}
The Zeeman coupling term $i\eta\epsilon_{jik}h_k\phi_j^{*}\phi_i$ describes the interaction of the spin density, $S_k \propto i\epsilon_{jik}\phi_j^{*}\phi_i$, with the static magnetic field $\vec{h}=\mbox{curl}\vec{A}$. The quantity $\vec{\cal{M}}=\eta \vec{S}$ is the magnetic moment, $\eta S^2={\cal{M}}_3 S_3$.

The identity (\ref{iden}) can be then written as
 \begin{eqnarray}
 i[D_i,D_j]_{-}=e\epsilon_{ijk}h_k\,,\label{idenSt}
 \end{eqnarray}
 where $D_j=(\nabla -ie\vec{A})_j$,  $\epsilon_{jkl}$ is the Levi-Civita tensor. With the identity (\ref{idenSt}) taken into account   equations of motion are simplified as
\begin{eqnarray}
&-D^2_i \phi_j
+D_j D_i \phi_i
+m_{\rm ef}^2 \phi_j \nonumber\\
&+2\Lambda  |{\phi}_i|^2 \phi_j
+{i}(e+\eta) F_{ji}\phi_i
=0\,, \label{phichstVEC}
\end{eqnarray}
and
\begin{eqnarray}
&\Delta \vec{A}=-4\pi \vec{J}\, \,\, {\rm at}\,\,\, \mbox{div}\vec{A}=0\,,\,{\rm with}\label{AJ}\\
&J_i =-ie\phi^{*}_j D_i {\phi}_j+ie{\phi}^{*}_j D_j\phi_i \nonumber\\
&+{\rm c.c.}+i(e+\eta) \nabla_j ({\phi}^{*}_j\phi_i -{\phi}^{*}_i\phi_j)\,.\nonumber
\end{eqnarray}
The condition (\ref{condDmu})
\begin{eqnarray}
D_j{\phi}_j=0\,,\label{CondA}
\end{eqnarray}
cf. Eq. (\ref{ansAx}) in case of the scalar field.

\subsection{Charge-neutral complex vector field}\label{NeutVec} Consider   model describing a complex vector field
coupled with the electromagnetic field by  the Zeeman coupling ($\eta\neq 0$) in absence of  the minimal coupling (for $e=0$).
Equations (\ref{Gn}) and (\ref{phichstVEC}) hold now for $e=0$.

\subsubsection{ Superfluidity in nonmagnetic phase A} The simplest choice is when  only one Lorentz component of the complex vector field is non-zero.  Label such a choice as the phase A. The spin in this state is zero. For $m^2_{\rm sc}>0$ there are no solutions in this case. One can consider three sub-phases: A$_1$  [${\phi}^\nu =(0, \phi_1 (x),0,0)$ ], A$_2$  [${\phi}^\nu =(0, 0, \phi_2 (x),0)$ ], and A$_3$  [${\phi}^\nu =(0, 0,0,\phi_3 (x))$ ].

In the case of the uniform matter placed in the external static uniform magnetic field  $\vec{H}$ all three sub-phases are allowed. The magnetic field and the condensate decouple: $\vec{h}=\vec{H}$, $|\phi|^2 = -\frac{m_{\rm sc}^2}{2\Lambda}\theta (-m_{\rm sc}^2)$.
The Gibbs free-energy density is $G_{\rm A}=-\frac{m_{\rm sc}^4}{4\Lambda}\,\theta (-m_{\rm sc}^2)$.

Let now the medium fills half-space $x<0$,  where $m^2_{\rm sc}=m_0^2<0$ is a constant, placed in the external static uniform magnetic field  $\vec{H}$. We will assume the vector boson field $\vec{\phi}$  and the internal magnetic field $\vec{h}=\mbox{curl} \vec{A}$ to be functions only of $x$ (using the symmetry arguments), satisfying the boundary conditions for $x=0$.

{\bf Sub-phase A$_1$}  is  not allowed. Indeed, then   the  boundary condition $\phi_1 (x=0)=0$ for the vector boson field  cannot be  satisfied due to the absence of  $\propto \partial_1$ gradient term in Eq. (\ref{phichstVEC}).     Also notice that the condition $\partial_i {\phi}_i =0$ is not fulfilled in this case although the latter condition should be satisfied at least in the single particle approximation, in absence of the term $\propto \eta$ and for $e=0$.

{\bf Sub-phase A$_2$.}
 Then the condition $\partial_i {\phi}_i =0$ is fulfilled and   $\phi_2$ and $h$ satisfy equations of motion that follow from the variation of  (\ref{Gn}) for $e=0$ in $\phi_2$ and $h$, cf. Eqs. (\ref{phichstVEC}), (\ref{AJ}),
\begin{eqnarray}
\partial_1^2 {\phi}_2 -m_{\rm sc}^2 {\phi}_2 -2\Lambda ({\phi}_2{\phi}_2^{*}){\phi}_2=0\,,\quad \vec{h}=\vec{H}\,.
\end{eqnarray}
Appropriate solution for the condensate field gets the form (\ref{phisolneut})   for $m_{\rm sc}^2 =m_0^2 <0$, $|m_0|\ll m$,  now with $\lambda =2\Lambda$. Then we  find
\begin{eqnarray}
G_{\rm A_2}=|\partial_1 \phi_2|^2 +m_{\rm sc}^2 |{\phi}_2|^2 +\Lambda ({\phi}_2{\phi}_2^{*})^2\,,
\end{eqnarray}
whereas the averaged Gibbs free-energy is given by (\ref{phisolneutVec}) (with $\lambda$ replaced by $2\Lambda$), i.e,
\begin{eqnarray}
\overline{G}_{\rm A_2}= \frac{\int d^3 x G}{\int d^3 x} = -\frac{m_{\rm 0}^4}{4\Lambda}\left(1-\frac{4\sqrt{2}}{3}\frac{l_\phi}{d_x}\right)\theta(-m_{\rm 0}^2)\,.
\label{phisolneutVecA}
\end{eqnarray}

{\bf Sub-phase A$_3$.} Similarly we could employ  the field ansatz ${\phi}^\nu =(0, 0, 0,\phi_3 (x))$  with the same results as for the sub-phase A$_2$, $\overline{G}_{\rm A_2}=\overline{G}_{\rm A_3}$.

\subsubsection{Ferromagnetic superfluidity in phase B}
Let $$\widetilde{\Lambda}=\Lambda -2\pi\eta^2\,$$ be positive.
Consider  the field ansatze, which we name the phase B: ${\phi}^{\nu} =(0, 0, \phi_1 (x),\phi_2 (x))$ (sub-phase B$_1$); ${\phi}^{\nu} =(0, \phi_1 (x),0,\phi_2 (x))$ (sub-phase B$_2$); and ${\phi}^\nu =(0, \phi_1 (x), \phi_2 (x),0)$ (sub-phase B$_3$) with $\phi_2 =-C_0 i\phi_1$, where  $C_0$ is  real coefficient. We further take $C_0=1$ for $\eta <0$ and $C_0=-1$ for $\eta>0$, as it follows from the minimization of the Gibbs free-energy. Also, for  convenience we introduce the new variable
$\widetilde{\psi}=\sqrt{2}\phi_1 (x)$. We will show that now classical solutions  may exist not only for   $m_{\rm ef,0}^2<0$ but in some cases also for $m_{\rm ef,0}^2>0$.

Already for the uniform matter  the free energy is different for the cases when the mean spin $\vec{S}$ is parallel  to the external uniform static magnetic field $\vec{H}$ and perpendicular to it.
For instance, for the sub-phase B$_1$ at $\vec{H}\parallel x$ using (\ref{Gn}) with  $\vec{A}=(0,0,Hy\mp 4\pi\eta|\widetilde{\psi}|^2 y)$ we obtain
 \begin{eqnarray}
&h_1 =H\mp 4\pi\eta|\widetilde{\psi}|^2=const\,,\,\, h_2=h_3=0\,,\,\,\nonumber\\
&|\widetilde{\psi}|^2=\frac{-m_{\rm sc}^2 \mp \eta H}{2\widetilde{\Lambda}}\theta ({-m_{\rm sc}^2 \mp \eta H})\,,\nonumber\\
&G_{\rm B_1}(\vec{H}\parallel x) =-\frac{(-m_{\rm sc}^2 \mp \eta H)^2}{4\widetilde{\Lambda}}\theta ({-m_{\rm sc}^2 \mp \eta H})\,.\nonumber
\end{eqnarray}
The upper and lower signs here correspond to two projections of the spin in the ground state for negative and positive $\eta$, respectively.
For $\vec{H}\parallel z$ we have $\vec{A}=(0,Hx,\mp 4\pi\eta|\widetilde{\psi}|^2y),$
 \begin{eqnarray}
&h_1 =\mp 4\pi\eta|\widetilde{\psi}|^2=const\,,\,\, h_2 =0\,, \,\,h_3 =H\,, \nonumber\\
&|\widetilde{\psi}|^2=\frac{-m_{\rm sc}^2}{2\widetilde{\Lambda}}\,\theta ({-m_{\rm sc}^2})\,,\,\,
G_{\rm B_1}(\vec{H}\parallel z)=-\frac{m_{\rm sc}^4}{4\widetilde{\Lambda}}\,\theta (\frac{-m_{\rm sc}^2}{4\widetilde{\Lambda}})\,.\nonumber
\end{eqnarray}
Similarly can be obtained solutions for the B$_2$ and B$_3$ sub-phases.   We  used that $\widetilde{\Lambda}>0$.
It is the case, e.g., for  hadrons since then $\Lambda \sim 1$ and $\eta \sim e$. Otherwise by the first order phase transition there may appear a novel  C-phase, see below.

Let now the medium fills half-space $x<0$,  where $m^2_{\rm sc}=m_0^2<0$ is a constant, placed in the external static uniform magnetic field  $\vec{H}$. Again consider the vector boson field $\vec{\phi}$  and the internal magnetic field $\vec{h}=\mbox{curl} \vec{A}$ to be functions only of $x$ (using  the symmetry arguments), satisfying the boundary conditions for $x=0$.

In the sub-phase B$_1$ at $\vec{H}\parallel x$ the volume contribution to the Gibbs free energy proves to be the same as for the B$_2$ sub-phase at $\vec{H}\parallel y$   but the surface energy is $\sqrt{2}$ larger. In the sub-phase B$_1$ at $\vec{H}\parallel y$ the volume contribution to the Gibbs free energy proves to be the same as for the B$_2$ sub-phase at $\vec{H}\parallel x$   but the surface energy is again $\sqrt{2}$ larger.


{\bf Sub-phase B$_2$.}  Then the own magnetic field  has the component  ${h}_2 (x)\neq 0$ due to the corresponding non-zero Zeeman term. The condition $\partial_i {\phi}_i = 0$ is not fulfilled with this field ansatz.
  From (\ref{Gn}) for $e=0$ we have in the given case:
 \begin{eqnarray}
&G_{\rm B_2} = \frac{1}{2}|\nabla_x \widetilde{\psi}|^2 +m_{\rm sc}^2|\widetilde{\psi}|^2 +\Lambda  |\widetilde{\psi}|^4\nonumber\\
&+\frac{(\vec{h}-\vec{H})^2}{8\pi} \pm \eta h_2  |\widetilde{\psi}|^2\,.\label{B1neut}
\end{eqnarray}
Following the minimization of the energy, for $\eta <0$ we should take the upper sign, and for $\eta >0$, the lower sign, that relates to the choice $\phi_2 =\mp i \phi_1$, respectively.

{\em Consider first $\vec{H}\parallel z$.}  Minimizing $G_{\rm B_2}$ in $h$ we obtain
  \begin{eqnarray}
 h_2 =\mp 4\pi \eta\,|\widetilde{\psi}(x)|^2\,, \quad h_3 =H\,,\label{h2}
 \end{eqnarray}
 $h_1 =0$. As we see, the field $\vec{h}(x)$ satisfies the necessary boundary condition $\vec{h}(0)=\vec{H}$, $\vec{A}=(0, Hx, \pm 4\pi\eta \int^x |\widetilde{\psi}|^2 dx)$.

 Equation of motion for the field $\widetilde{\psi}$  is given by
 \begin{eqnarray}
\frac{1}{2}\partial_1^2 {\widetilde{\psi}} -m_{\rm sc}^2 \widetilde{\psi} -2\Lambda |\widetilde{\psi}|^2\widetilde{\psi} \mp \eta h_2 \widetilde{\psi} =0\,.\label{solphi2}
\end{eqnarray}
Using (\ref{h2}) we find for $x\leq 0$:
 \begin{eqnarray}
 \widetilde{\psi} (x) =\pm \sqrt{\frac{-m_{\rm 0}^2}{2\widetilde{\Lambda}}}\,\theta(-m_{\rm 0}^2)\, \mbox{th} \frac{x-x_0}{\sqrt{2}{l}^{\rm B_2}_\phi}\,,\label{solphi2B2}
 \end{eqnarray}
${l}^{\rm B_2}_\phi ={l}_\phi/\sqrt{2}$,  and assuming  $|m_{\rm sc}|\ll m$ we put
  $x_0=0$ to satisfy the boundary condition  $\widetilde{\psi} (0)=0$.
  With these solutions we find
  \begin{eqnarray}\overline{G}_{\rm B_2} (\vec{H}\parallel z)=-\frac{m^4_{\rm 0} }{4\widetilde{\Lambda}}\left(1-\frac{4\sqrt{2}{l}^{\rm B_2}_\phi }{3d_x}\right)\theta(-m_{\rm 0}^2)\,.\label{PrelimB}
  \end{eqnarray}
  Thus at $\eta \neq 0$  comparing (\ref{phisolneutVecA}) and (\ref{PrelimB}) we see that for any value of $\vec{H}\parallel z$ the sub-phase B$_2$  is energetically preferable compared with the sub-phases A.

  {\em Let now $H\parallel x$.} We get  $\vec{A}=(0, 0, Hy \pm 4\pi\eta \int^x |\widetilde{\psi}|^2 dx)$,
  $h_1=H$, $h_2 =\mp 4\pi \eta |\widetilde{\psi}|^2$, $h_3 =0$ and recover Eq. (\ref{solphi2B2}), and (\ref{PrelimB}) now for $\overline{G}_{\rm B_2} (\vec{H}\parallel x)$.

  {\em Let now $H\parallel y$.} With $\vec{A}=(0, 0, -Hx\pm 4\pi\eta \int^x |\widetilde{\psi}|^2 dx)$,
  we obtain
  $$h_1 =0\,,\,\, h_2= H\mp 4\pi \eta|\widetilde{\psi}(x)|^2\,,\,\, h_3=0\,,$$ and
  \begin{eqnarray}
 \widetilde{\psi} (x) =\pm \sqrt{\frac{-m_{\rm 0}^2\mp \eta H}{2\widetilde{\Lambda}}}\,\theta(-m_{\rm 0}^2\mp \eta H) \,\mbox{th} \frac{x}{\sqrt{2}{l}^{\rm B_2}_\phi}\,,\label{solB1}
 \end{eqnarray}
   for $x\leq 0$, with
\begin{eqnarray}
&\overline{G}_{\rm B_2}(\vec{H}\parallel y) =-\frac{(-m^2_{\rm 0}\mp\eta H)^2 }{4\widetilde{\Lambda}}\left(1-\frac{4\sqrt{2}{l}^{\rm B_2}_\phi }{3d_x}\right)\label{GB1Hx}\\
 &\times\theta(-m_{\rm 0}^2\mp \eta H)\,.\nonumber
 \end{eqnarray}

For  $\vec{H}\parallel y$ at $m_0^2<0$ the condensate amplitude grows with increasing value $H$.  Thus  for $\vec{H}\parallel y$ the energy is gained  compared  to the case $\vec{H}\parallel x$ and $\vec{H}\parallel z$.
The sub-phase B$_2$  is a ferromagnetic phase, since even for $H=0$ there exists an own field $h_1\neq 0$.

The classical vector field (\ref{solB1}) is developed for $-m_{\rm 0}^2\mp \eta H>0$.
Thus, in this case {\em  the  condensation occurs not only for $m_{\rm 0}^2<0$ (for arbitrary $H$) but also   for
\begin{eqnarray} H>H_{\rm cr}^{\rm neut}=|m_{\rm 0}^2|/|\eta|\,,\quad {\rm at}\quad m_{\rm 0}^2>0\,.\label{Hneut}
\end{eqnarray}}

For $H\neq 0$  we found that  $\overline{G}_{\rm B_2}(\vec{H}\parallel y)< \overline{G}_{\rm B_2 }(\vec{H}\parallel x)=\overline{G}_{\rm B_2 }(\vec{H}\parallel z)\,.$

{\bf Sub-phase B$_3$.} The condition $\partial_i {\phi}_i = 0$ is not fulfilled with this field ansatz.
The Gibbs free-energy density renders:
\begin{eqnarray}
&G_{\rm B_3} = \frac{1}{2}|\nabla_x \widetilde{\psi}|^2 +m_{\rm sc}^2|\widetilde{\psi}|^2 +\Lambda  |\widetilde{\psi}|^4\\
&+\frac{(\vec{h}-\vec{H})^2}{8\pi} \pm \eta h_3  |\widetilde{\psi}|^2\,.\nonumber
\end{eqnarray}
 Equation of motion for $\widetilde{\psi}$ is as follows
\begin{eqnarray}
\frac{1}{2}\partial_1^2 {\widetilde{\psi}} -m_{\rm sc}^2 \widetilde{\psi} -2\Lambda (\widetilde{\psi}\widetilde{\psi}^{*})\widetilde{\psi} \mp \eta h_3\widetilde{\psi} =0\,.\label{solphi21}
\end{eqnarray}
{\em Let $\vec{H}\parallel z$.} With $\vec{A}=(0, Hx\mp 4\pi \eta \int^x |\widetilde{\psi}|^2 dx)$ we get
\begin{eqnarray}
h_3 =F_{12}=H \mp 4\pi \eta |\widetilde{\psi} (x)|^2\,.\label{solh}\end{eqnarray}

The appropriate solution of Eq. (\ref{solphi21}) with the boundary conditions $\widetilde{\psi} (x)\to 0$ for $x\to 0$ and $\widetilde{\psi} (x)\to \pm\sqrt{\frac{-m_{\rm 0}^2-\eta H}{2\widetilde{\Lambda}}}$ for $x\to -\infty$ coincides with Eq. (\ref{solB1}) and
\begin{eqnarray}
&\overline{G}_{\rm B_3}(\vec{H}\parallel z) = \overline{G}_{\rm B_2}(\vec{H}\parallel y)\,.\label{B3Hvector}
\end{eqnarray}
Moreover, for $H\neq 0$  we have $\overline{G}_{\rm B_3}(\vec{H}\parallel z)< \overline{G}_{\rm B_3 }(\vec{H}\parallel x)=\overline{G}_{\rm B_3 }(\vec{H}\parallel y)\,.$

As in case of sub-phase B$_2$ at $\vec{H}\parallel y$, for  the sub-phase B$_3$ at $\vec{H}\parallel z$ the classical vector field is developed for $-m_{\rm 0}^2\mp \eta H>0$.
Thus,  the  condensation occurs not only at $m_{\rm 0}^2<0$ for arbitrary $H$ but also  at $m_{\rm 0}^2>0$ for
$H>H_{\rm cr}^{\rm neut}=
|m_{\rm 0}^2|/|\eta|.$

{\bf Domains.} The difference in volume and surface energies of the sub-phases causes a possibility of existence of the domains for $H\neq 0$ and $H=0$ with different directions of the own magnetic  field $\vec{h}$ in each domain, which  may merge  in presence of the external  fields.

{\bf About choice of self-interaction.} With the self-interaction taken in the form (\ref{phiphisimp})  for $\xi_1 =0$, that  we have  used,  for $\vec{H}\parallel z$ the sub-phase B$_3$ proves to be energetically preferable  compared to the other allowed sub-phases A$_2$, A$_3$ and B$_2$. For $\xi_1 \neq 0$ the situation becomes more complicated. For example, for $\xi_1 =-1$  in the A-phase the repulsive self-interaction term vanishes, whereas in the B-phase the repulsive self-interaction term does not depend on the value $\xi_1$. Thereby, for $\xi_1 =-1$ the A phase becomes energetically favorable compared to the B phase at least for $H=0$. Similar problems will be considered in next Section on example of fermions with spin-triplet pairing.

\subsubsection{Ferromagnetic superfluidity in phase C}
For $\widetilde{\Lambda} =\Lambda - 2\pi \eta^2<0$  by the first order phase transition there may appear a novel  C-phase. Since the hadron-hadron coupling  $\Lambda \gg e^2$, at least for the $\rho$ mesons the C phase  is not realized.
For the triplet pairing the C phase is possible, we shall return to this question in Sect. \ref{vector}.

\subsection{Charged complex vector field}\label{ChVect}
Now let the complex vector field be charged and interacting with the electromagnetic field by the minimal  and the Zeeman couplings.
Consider  first   the charged static complex vector field with $m_{\rm ef,0}^2 =m^2 -\mu^2_{\phi}$ in half-space $x<0$,  placed in the external static uniform magnetic field $\vec{H}$.  In this case  fields $\vec{h}$ and $\phi_i $ depend only on $x$.

\subsubsection{Nonmagnetic and superdiamagnetic responses of various superfluid sub-phases A}
Solutions exist only  for $m_{\rm ef,0}^2 =m^2 -\mu^2_{\phi}<0$.

{\bf Sub-phase A$_1$} is not realized, as in case of the  neutral complex field considered in Section \ref{NeutVec},   since  the appropriate boundary conditions at $x=0$ cannot be fulfilled with the ansatz $\phi_i =(\phi_1 (x), 0,0)$. The condition $\partial_i \phi_i =0$ is also not satisfied, even for $\eta =e$ and for the linearized equation of motion, when it must be fulfilled.

{\bf  Sub-phase A$_2$.} For $\phi_i =(0, \phi_2 (x),0)$, taking $\vec{H}\parallel z$, $\vec{A}_{\rm ext}=(0, Hx,0)$,  $\vec{A} =(0, A_2 (x),0)$, from (\ref{Gn}) we  obtain
\begin{eqnarray}
G_{\rm A_2}(\vec{H}\parallel z) =|\partial_1 \phi_2|^2 +m_{\rm ef}^2 |\phi_2|^2 +\Lambda |\phi_2|^4 +
\frac{(\vec{h}-\vec{H})^2}{8\pi}.\nonumber
\end{eqnarray}
 Minimizing $\overline{G}_{\rm A_2}$ in $h$ we see that the magnetic field and the condensate decouple, and $\vec{h}=\vec{H}$. The resulting expression for $\overline{G}_{\rm A_2}$,
  \begin{eqnarray}
&\overline{G}_{\rm A_2}(\vec{H}\parallel z)= -\frac{m_{\rm ef,0}^4}{4\Lambda}\left(1-\frac{4\sqrt{2}\,\tilde{l}_\phi}{3d_x}\right)\theta (-m_{\rm ef,0}^2)\,,
\end{eqnarray}
  coincides with (\ref{phisolneutVec}) after the replacement $\lambda\to 2\Lambda$,  $m_0\to m_{\rm ef,0}$ and $l_\phi =1/|m_0|\to \tilde{l}_\phi =1/|m_{\rm ef,0}|$. For $\vec{H}\parallel z$ the sub-phase A$_2$ is a nonmagnetic phase.

For $\vec{H}\parallel y$ the Gibbs free-energy density takes the form
\begin{eqnarray}
&G_{\rm A_2}(\vec{H}\parallel y)=|\partial_1 \phi_2|^2 +e^2 A_3^2(x) |\phi_2|^2 +m_{\rm ef}^2 |\phi_2|^2 \nonumber \\
&+\Lambda |\phi_2|^4 +
\frac{(\vec{h}-\vec{H})^2}{8\pi}\,.
\end{eqnarray}
Comparison with (\ref{Gsc}) demonstrates that after the replacement $\lambda \to 2\Lambda$  the charged complex vector field
 is described completely the same as  the charged complex scalar field. Thus for low $H$ (for $H<H_{\rm cr 1}$) the magnetic field $h$ is repelled from the condensate region and
 \begin{eqnarray}
&\overline{G}_{\rm A_2}(\vec{H}\parallel y)\simeq \frac{H^2}{8\pi} -\frac{m_{\rm ef,0}^4}{4\Lambda}\left(1-\frac{4\sqrt{2}\,\tilde{l}_\phi}{3d_x}\right)\theta (-m_{\rm ef,0}^2).
\end{eqnarray}
 Thus the A$_2$ superconducting sub-phase for $\vec{H}\parallel y$ demonstrates a superdiamagnetic response on a weak external magnetic field, $\overline{h}=0$.  With an increase of $H$ in the interval $H_{\rm cr 1}<H<H_{\rm cr 2}$ there appears the Abrikosov mixed state of vortices alternating with the condensate, for $H=H_{\rm cr 2}$ the condensate disappears, and for $H>H_{\rm cr 2}$ the condensate does not exist.

 {\bf Sub-phase A$_3$.} For  $\phi_i =(0,0,\phi_3 (x))$ choosing $\vec{H}\parallel z$, $\vec{A}_{\rm ext}=(0, Hx,0)$, with $\vec{A} =(0, A_2 (x),0)$, i.e. with $\vec{h}\parallel z$, we are able to    satisfy the boundary condition $\vec{h}(x=0)=\vec{H}$.  The Gibbs free-energy density takes the form
\begin{eqnarray}
&G_{\rm A_3}(\vec{H}\parallel z)=|\partial_1 \phi_3|^2 +e^2 A_2^2(x) |\phi_3|^2 +m_{\rm ef}^2 |\phi_3|^2 \nonumber \\
&+\Lambda |\phi_3|^4 +
\frac{(\vec{h}-\vec{H})^2}{8\pi}\,.
\end{eqnarray}
Comparison with (\ref{Gsc}) demonstrates that after the replacement $\lambda \to 2\Lambda$  the charged complex vector field
 is described completely the same as  the charged complex scalar field. Thus for low $H$ (for $H<H_{\rm cr 1}$) the magnetic field $h$ is repelled from the condensate region and
 \begin{eqnarray}
&\overline{G}_{\rm A_3}(\vec{H}\parallel z)\simeq \frac{H^2}{8\pi} -\frac{m_{\rm ef,0}^4}{4\Lambda}\left(1-\frac{4\sqrt{2}\,\tilde{l}_\phi}{3d_x}\right)\theta (-m_{\rm ef,0}^2).
\end{eqnarray}
The sub-phase A$_3$ for a weak external magnetic field  $\vec{H}\parallel z$ is superdiamagnetic, and $\overline{G}_{\rm A_3}(\vec{H}\parallel z)=\overline{G}_{\rm A_2}(\vec{H}\parallel y)$.

 With an increase of $H$ in the interval $H_{\rm cr 1}<H<H_{\rm cr 2}$ there appears the Abrikosov mixed state of vortices alternating with the condensate and for $H>H_{\rm cr 2}$ the condensate disappears.

For $H\neq 0$ we find that  $\overline{G}_{\rm A_2}(\vec{H}\parallel z)<\overline{G}_{\rm A_3}(\vec{H}\parallel z)$. For $H\to 0$ both quantities coincide.

With $\vec{A}_{\rm ext}=(0, 0, Hy)$, we have $\vec{h}=\vec{H}$ and
\begin{eqnarray}
G_{\rm A_3}(\vec{H}\parallel x)=|\partial_1 \phi_3|^2  +m_{\rm ef}^2 |\phi_3|^2 +\Lambda |\phi_3|^4 \,.
\end{eqnarray}
Therefore  $\overline{G}_{\rm A_3}(\vec{H}\parallel x)=\overline{G}_{\rm A_2}(\vec{H}\parallel z)=\overline{G}_{\rm A_3}(\vec{H}\parallel y)$.

Thus most energetically preferable are the sub-phase A$_2$ for $\vec{H}\parallel z$ and the sub-phase A$_3$ for $\vec{H}\parallel x$.  In both cases the sub-phases are nonmagnetic and  the condensate of the charged vector field exists for arbitrary values of the external magnetic field.

\subsubsection{ Superconductivity in phase B}
We will show that, as in case of  the charge-neutral vector bosons,  classical solutions  may exist not only for   $m_{\rm ef,0}^2<0$, when the response on a weak external magnetic field is superdiamagnetic, but in presence of an overcritical external magnetic field also for $m_{\rm ef,0}^2>0$.

 {\bf{Sub-phase B$_3$.}}
 Let $\vec{H}\parallel z$ and employ $\vec{A}=(A_1(x,y),A_2(x,y),0)$, i.e., $\vec{h}\parallel z$.

Integrating by parts the gradient term in the Gibbs free-energy, using the identity (\ref{idenSt}) and retaining only the volume part in the Gibbs  free-energy we get:
\begin{align}
&\int d^3 x G_{\rm B_3}(\vec{H}\parallel z)=\int d^3 x \left[-\frac{1}{2}
\tilde{\psi}^{*}(D_1^2 +D_2^2)\tilde{\psi}\right]\label{gradc11}
\\
&+\int d^3 x \left[\frac{({h}_3 -{H})^2}{8\pi}+[m_{\rm ef}^2 +(\eta +\frac{e}{2}){h}_3]|\tilde{\psi}|^2 +\Lambda |\tilde{\psi}|^4 \right]\,,\nonumber
\end{align}
for $\eta <0$, $e<0$.
Varying the Gibbs free energy
 in $\widetilde{\psi}^{*}$
 we obtain equation of motion for the order parameter
 \begin{align}
&-
\frac{1}{2}
(D_1^2 +D_2^2)\widetilde{\psi}
+\left[m_{\rm ef}^2+(\eta +\frac{e}{2}){ h}_3\right]\widetilde{\psi}+2\Lambda  |\widetilde{\psi}|^2\widetilde{\psi} =0\,.\label{eqmotinh11ch}
\end{align}
Setting $e=0$ we recover Eq. (\ref{solphi2}). Choosing $A_1=0$ and varying (\ref{gradc11}) in $A_2$  we get
\begin{align}
\partial_1^2 A_2 =-4\pi J_2 =4\pi e^2 |\widetilde{\psi}|^2 A_2 -4\pi(\eta+\frac{e}{2}) \partial_1 |\widetilde{\psi}|^2\,,\label{J2}
\end{align}
cf. Eq. (\ref{AJ}) for the scalar charged bosons.
There are two typical lengths characterizing solutions of  these equations: $\tilde{l}_h =\sqrt{2}l_h$ characterizing the field $A_2(x)$ and   $\tilde{l}^{\rm B}_\phi =1/(\sqrt{2}|m_{\rm ef, 0}|)$, characterizing the field $\widetilde{\psi}(x)$, cf. quantities ${l}_h$ and ${l}_\phi$ introduced above.
We will see that there are two type of solutions of these  equations. One solution describes the Meissner screening effect, when the external magnetic field decreases on the length $\tilde{l}_h$ near the system boundary, whereas the condensate field reaches constant value for $-x >\tilde{l}_\phi$. In ordinary superconductors of the second kind this solution is realized for $H<H_{\rm cr 1}$.
Another type of solution describes periodic structures for $H_{\rm cr 1}<H<H_{\rm cr 2}$. Consider first a specifics of the Meissner effect in our case.
For $ -x\sim\tilde{l}_h  \gg \tilde{l}_\phi$, corresponding to the case $\kappa \gg 1$ that we consider, the term $4\pi (\eta +e/2) \partial_1 |\widetilde{\psi}|^2$ can be dropped and the
solution satisfying the boundary condition $h_3 (0) =H$ is $A_2(x)=H \tilde{l}_h e^{x/\tilde{l}_h}$. On the short distances $-x\sim 1/\tilde{l}_\phi$ from the surface the $y$ component of the vector-potential, $A_2 $, is a constant and the term $4\pi e^2 |\widetilde{\psi}|^2 A_2$ can be dropped for $H\ll 1/(\tilde{l}_\phi \tilde{l}_h)\sim H_{\rm cr}$. Then the solution (\ref{J2}), being valid for $-x\gsim \tilde{l}_h$, but
satisfying the appropriate boundary condition for $x=0$, $h_3 (0)=H$, $\widetilde{\psi}(0)=0$, renders
$$h_3 \simeq -4\pi (\frac{e}{2}+\eta)[|\widetilde{\psi}(x)|^2 -|\widetilde{\psi}(-\infty)|^2](1-e^{x/\tilde{l}_h}) +He^{x/\tilde{l}_h}\,.$$ This solution describes the screening Meissner effect.

For $H\ll H_{\rm cr}$ using estimate done above for the scalar charged field, we can replace $D_1^2 +D_2^2\to \partial_1^2 +\partial_2^2\to \partial_1^2$. The solution of Eq. (\ref{eqmotinh11ch}) then renders
$$\widetilde{\psi} (x) \simeq \pm f_0\,\theta(-m_{\rm ef, 0}^2)\, \mbox{th} \frac{x}{\sqrt{2}\,\tilde{l}^{\rm B}_\phi}\,,\quad f_0=\sqrt{\frac{-m_{\rm ef, 0}^2}{2{\Lambda}}}.$$
For the space-averaged Gibbs free-energy we obtain expression
\begin{eqnarray}
&\overline{G}_{\rm B_3}(\vec{H}\parallel z)\simeq \frac{H^2}{8\pi} -\frac{m_{\rm ef,0}^4}{4\Lambda}\left(1-\frac{4\sqrt{2}\,\widetilde{l}^{\rm B}_\phi}{3d_x}\right)\theta (-m_{\rm ef,0}^2).\label{B3ch}
\end{eqnarray}
We see that for $H\neq 0$, $$\overline{G}_{\rm A_2}(\vec{H}\parallel z)<
\overline{G}_{\rm B_3}(\vec{H}\parallel z)\,,$$
whereas
for $H\to 0$, due to a smaller surface energy contribution, for the system of the finite size we get $\overline{G}_{\rm B_3}(\vec{H}\parallel z)<\overline{G}_{\rm A_2}(\vec{H}\parallel z)$.

With increasing $H$ above the value $H_{\rm cr 1}$, there appears the Abrikosov lattice of vortices. For the ordinary metallic  superconductors and similarly for the case of the charged scalar field, with a subsequent increase of $H$ the condensate weakens and for $H=H_{\rm cr 2}$ it disappears. Assume  that  for $H$ near  the value $H_{\rm cr 2}$ the condensate is weak. Then we drop the non-linear term in Eq. (\ref{eqmotinh11ch}) and put $\vec{A}=(0, H_{\rm cr 2}x,0)$.   Thus, as for the case of the complex scalar field considered above at $H\simeq H_{\rm cr,2}$, we  find the solution satisfying  periodic boundary conditions. After dividing all terms in linearized  Eq. (\ref{eqmotinh11ch}) on an artificial mass coefficient the former equation acquires the form of the Schr\"odinger equation for the nonrelativistic particle in the uniform magnetic field $h_3=H$. The quantity
$$E_{\rm min}=-m_{\rm ef,0}^2 -(\eta +e/2)H = |e|H/2\,$$
is the minimal eigenvalue.
However, as we see, for  $m_{\rm ef,0}^2<0$, $\eta <0$, $e<0$ there is no solution of this equation and there is no  upper critical field $H_{\rm cr 2}$, at which the condensate vanishes with increasing $H$.

On the other hand the solution exists for $m_{\rm ef}^2>0$, $\eta <0$ at
\begin{align}
H> H_{\rm cr 2}=-m_{\rm ef}^2/\eta >0\,.  \label{Hc21}
\end{align}

Note that we did not use  the relation (\ref{CondA}).
 Now, using condition (\ref{CondA}) and the identity (\ref{idenSt}) we recower Eq. (\ref{convEq}), which coincides with the linearized Eq. (\ref{eqmotinh11ch})
at $h_3 =H_{\rm cr 2}$ for any $\eta <0$ at $m_{\rm ef}^2>0$.
Let   $H$ be slightly above $H_{\rm cr2}$. Then from (\ref{CondA}) we find that $\partial_1 |\phi_1|^2 =2e A_2 (x)|\phi_1|^2$. Setting this result in Eq. (\ref{J2})
 we obtain
 \begin{align}
\partial_1^2 A_2 +8\pi \eta e A_2  |\widetilde{\psi}|^2=0\,,\label{J2cond}
\end{align}
with the solution corresponding to the anti-screening effect, being  in accordance with our observation that the superconductivity of the charged vector bosons appears at $H>H_{\rm cr 2}$ for $m_{\rm ef}^2>0$, cf. statement of \cite{Olesen} that ``new superconductivity'' may anti-screen magnetic field.

Below I will demonstrate similarities and differences  in the description of the complex vector meson fields and  the spin-triplet pairing of fermions.

\section{Spin-triplet pairing in neutral fermion system described by
complex vector order parameter}\label{vector}

\subsection{Phenomenological Gibbs free energy density}\label{GL}

 A formalism  for description of the spin-triplet pairing in charged fermion systems,
where the non-zero spin of the Cooper pair might be considered as a conserved  quantum number, has been developed, cf.  ~\cite{VG1985,KR1998,MineevSamokhin} and refs. therein. In this Section  we employ  a similar formalism for the description of the spin-triplet pairing in neutral fermion systems, where the complex vector order parameter is coupled to the magnetic field by the Zeeman term. Novel phases will be found.

Consider pairing of identical fermions. Since the total wave function of the system of identical fermions is antisymmetric under their exchange, and the spin part  in the triplet state is symmetric,  the angular part behaves as $(-1)^L$ with odd $L$. To be specific let $L=1$.
 For the
description of the spin-triplet p-wave pairing of fermions the pairing gap is as follows
 \cite{VG1985}, $\hat{\Delta}(\vec{k})=\vec{\sigma}\vec{d}(\vec{k})i\sigma_2\,,$
where $\vec{d}(\vec{k})=-\vec{d}(-\vec{k})$ is an odd vector
function and ${\sigma}_j$ are the Pauli matrices, $j=1,2,3$.
If we considered pairing of nonidentical fermions, e.g. neutrons and protons, the  isospin quantum number should be taken into account, $S+L+T$ (spin plus orbital momentum plus isospin) should be odd. The $np$ 3S$_1$ phase shift is the largest among others at low nucleon-nucleon scattering energies.
Thus the  $np$ pairing in the 3S$_1$ channel is possible in the isospin-symmetric nuclear matter,
also described by the complex vector order parameter.

We present $\hat{\Delta}(\vec{k})=\psi_i \Phi_i (\vec{k})$,  where  $\Phi_i$ are
three basis functions. Let us postpone consideration of  the rotating systems (external rotation)  and also  disregard a possibility of an internal self-rotation. Thereby,
we  present the Gibbs
free-energy density  associated with the charge-neutral fermion
pairs paired in the  spin-triplet state  in  the form, cf. ~\cite{VG1985,KR1998,MineevSamokhin},
\begin{eqnarray}
\label{GLvector}
G &=&G_{\rm grad}^{\rm neut}+G_{\rm hom},\\
G_{\rm grad}^{\rm neut}&=& c_1 |\partial_i \psi_j|^2 + c_2 |\partial_i \psi_i|^2 + c_3 (\partial_i \psi_j)^* \partial_j \psi_i,\label{GLvector1}\nonumber\\
G_{\rm hom}&=&-a |\psi_i|^2 + b_1 (\psi_i\psi_i^*)^2 + b_2 (\psi_i\psi_i)(\psi^*_j\psi^*_j) \nonumber\\
&+& {\cal{M}} h_i iC\epsilon_{ijk}\psi_j^* \psi_k  +{(h_i -H_i)^2}/{(8\pi)}\nonumber\\
&+& b_3 \sum_j |\psi_j|^4+\{\gamma_k \psi_i\}^6\,,\label{GLvector2}\nonumber
 \end{eqnarray}
where  $\psi_i$ is the complex vector order parameter with indices
$i,j,k=1,2,3$ transformed as a vector indices under spin rotations, cf. Eq. (\ref{Gn}) introduced above for the complex vector boson fields.  The functional is symmetric under the U(1) phase transformations.
Coefficients $a$, $b_1$, $b_2$,   ${\cal{M}}$, $c_1$, $c_2$, $c_3$ are real quantities, and relations between $c_1$, $c_2$, $c_3$ should be such that the resulting surface term is positive. As we see at least should be
\begin{eqnarray}
c_1, c_2 \geq 0\,.
\end{eqnarray}

In the quantum field theory of the vector field, cf.  ~\cite{Nielsen:1978rm,Ambjorn:1988fx,Chernodub,Olesen} and Sect. \ref{vectorboson}, the gradient term is $\propto (D^\mu \phi^\nu -D^\nu \phi^\mu)^{*}(D_\mu \phi_\nu -D_\nu \phi_\mu)$, that corresponds to the choice $c_1=-c_3>0$, $c_2 =0$. In the BCS theory of clean materials one employs  \cite{MineevSamokhin} $c_1\simeq c_2\simeq c_3>0$. Reference \cite{KR1998} for the description of a new class of Ru-based superconductors uses the simplest choice   $c_2 =c_3 =0$, Ref. \cite{SaulsAdv} employs  also the choice $c_2=c_3\ll c_1\sim N(0)v_{\rm F}^2/(\pi^2 T_{\rm cr}^2)$ (E$_2$ model), $v_{\rm F}$ is the Fermi velocity. Using the most general gradient contribution consistent with the U(1) gauge symmetry and the rotational symmetry  Ref. \cite{Rosenstein:2015iza} calculated for the triplet
 superconductivity in 3D Dirac semimetals $c_3 =[u_L -u_T]/4$, $c_1=u_T/4$, $c_2 =0$,   $u_L =u_T/32$,  $u_T= \frac{7\zeta (3)N(0) v^2_{\rm F}}{15\pi^2 T_{\rm cr}^2}$,  i.e. $c_1\simeq -c_3$, $c_2=0$.  Bearing in mind these different possibilities, we further employ general expression not asking for any relations between $c_1$, $c_2$ and $c_3$. Reference \cite{Rosenstein:2015iza} also  derives  $b_1 = \frac{7\zeta(3)N(0)}{640 \pi^2 T_{\rm cr}^2}$ and $b_2=-b_1/3$. On the other hand the heat capacity measurements  performed for UPt$_3$
by several groups give $b_2/b_1 =(0.2-0.5)$, cf. \cite{SaulsAdv,Hasselbach}.

The quantity $h_i=\epsilon_{ijk}\partial A_k/\partial x_j$ is the actual value of the strength of the
 magnetic field, $\epsilon_{ijk}$, as above, is the Levi-Civita symbol. As in previous sections   $\vec{A}$ is the vector potential of the  magnetic field and $\vec{H}$ is the strength of the uniform external static magnetic field. Simplifying consideration we  neglect $\psi^2$-corrections  to the
$H^2$ magnetic energy  terms.

The term $\propto b_3 $   appears only in case of anisotropic systems. Thereby and for simplicity we further put $b_3 =0$, cf. \cite{MineevSamokhin,Rosenstein:2015iza}.
The  term $\{\gamma_k \psi_i\}^6$ in (\ref{GLvector}) symbolically means all possible combinations of the sixth order in the order parameter. For the sake of simplicity, where it does not lead to the  generation of instabilities, we  put $\gamma =0$.

Assuming that in absence of external fields for $\gamma =0$ we deal with the  second-order phase transition, we take
\begin{align}
a=\alpha_0 \varphi(t)\,,\quad t=(T_{\rm cr} -T)/T_{\rm cr}\,,
\label{aLand}
\end{align}
where the function $\varphi (t) = t +O(t^2)\,$ for small $t$,
$T_{\rm cr}$ has the sense of  the critical temperature of the pairing  transition for $H=0$, and
all the parameters $a_0 >0$, $b_1 >0$, and $b_2$, $b_3$, $c_1$, $c_2$, $c_3$ can be considered as $T$-independent constants for a small $t$. Also, simplifying  consideration  in this work we employ the mean-field theory. As is known, fluctuations of the order parameter prove to be significant in the vicinity of the critical point of the second-order phase transition, for $T$ near   $T_{\rm cr}$, cf. ~\cite{GS,LarkinVarlamov}. We will show that for certain sub-phases placed in external magnetic field the mean-field solutions may exist not only for $T<T_{\rm cr}$ but also for $T$ above $T_{\rm cr}$, i.e., below a higher value of the new critical temperature $T_{\rm cr}^{\rm H}$.    Thus, for mentioned sub-phases the fluctuation region is shifted to the vicinity of the critical temperature $T_{\rm cr}^{\rm H}$. As pointed out in Ref.~\cite{GS}, expansion in the order parameter is a primary
feature in the Landau  theory of phase-transitions, whereas an expansion in powers of $t$ is a secondary assumption valid for $T$ near $T_{\rm cr}$.
Therefore, at least for estimates, we may employ the functional (\ref{GLvector}) for  $T$ outside the vicinity of $T_{\rm cr}$ using $\varphi (T)=t$, $\varphi (T=0)=1$, cf.~\cite{LP1981}. Below, if not mentioned another,   to be specific we suppose that the external magnetic  field $\vec{H}$ is aligned parallel  to
$z$, i.e.  $\vec{H}_i=\delta_{i3}H$, although the behavior of the system described by the vector order parameter is sensitive to the choice of the direction of $H$ relatively the  surface, as we have demonstrated in previous section.

The mean spin density is carried by the order parameter
\begin{align}\label{avspin}
S_i= -iC\epsilon_{ijk}\psi_j^* \psi_k\,,
  \end{align}
where $C>0$ is a normalization constant.
 For $\vec{\psi}$ aligned along one of the axis 1, 2, 3 ($x$, $y$, or $z$) one has $\vec{S}=0$.

 Note that in case $b_2 =-b_1$ the self-interaction contribution to the Gibbs free-energy density,  $b_1 (\psi_i\psi_i^*)^2 + b_2 (\psi_i\psi_i)(\psi^*_j\psi^*_j)$, is reduced to the spin-spin interaction term $b_s S_i S_i$ with $b_s = b_1/C^2$ yielding the repulsion for $b_1 >0$, as in the Ginzburg-Landau treatment of superfluids described by a single order parameter, and the attraction for $b_1 <0$. For
$b_1 <0$ and $b_2=0$ the system is unstable.

 In difference with description of
magnetic superconductors performed in \cite{VG1985,KR1998,MineevSamokhin}, when dealing with neutral fermions we suppress
minimal coupling with the magnetic field but retain the Zeeman term assuming that neutral fermions under consideration
have magnetic moments.  The  orientation of the averaged spin related to the order parameter relatively the magnetic field depends on the sign of the magnetic moment of the pair.
The effective magnetic moment of the pair is
$\vec{\cal{M}}_{\rm pair}={\cal{M}}_{\rm pair} \vec{s}_{\rm pair}$, $\vec{s}_{\rm pair}$ is the spin of the pair.
Owing to the existence of the anomalous magnetic moment, the  neutron
pair with parallel spins gets the magnetic moment ${\cal{M}}_{nn}\simeq g_{nn} {\cal{M}}_{ N}$,
where ${\cal{M}}_N >0$ is the nucleon Bohr magneton, $g_{nn} =-2\cdot 1.91$ is the effective Lande factor. The proton pair has the magnetic moment ${\cal{M}}_{pp} \simeq  g_{pp} {\cal{M}}_N$ with $g_{pp} = 2\cdot 2.79$, ${\cal{M}}_N\simeq 3.15\cdot 10^{-18}$ MeV$/$G.  Note that the ratio of neutron to proton magnetic moments ${\cal{M}}_{nn}/{\cal{M}}_{pp} \simeq -0.68$ is close to the value $-2/3$ predicted by the valence quark model.
In the spin-orbit Fermi superfluids the role of the ${\cal{M}} h_i$ coefficient in the Zeeman term is played by the Rabi frequency \cite{PowellBaym}.
The volume-averaged Gibbs free-energy density $\overline{G}=\overline{F}-\overline{\vec{M}\vec{H}}$, where $\overline{F}$ is the averaged free-energy density,  $\vec{M}=(\vec{h}-\vec{H})/(4\pi)$ is the induced magnetization, $\overline{\vec{h}}=\vec{B}$  is the vector of the magnetic induction.

SO(3) symmetry is partially broken to its SO(2). Thereby, as in Ref. \cite{KR1998}, we may present
  \begin{align}\label{gensol}
  \vec{\psi}=f(\vec{n}\mbox{cos}\theta +i\vec{m}\mbox{sin}\theta)\,,
  \end{align}
  where $f$ is real and $\vec{n}$ and $\vec{m}$ are arbitrary unit vectors.  Let   $\phi$ is the angle between $\vec{n}$ and $\vec{m}$.   Then, for a uniform matter replacing (\ref{gensol}) in  (\ref{GLvector}) we find
 \begin{align}\label{GAgen}
 &G^{\rm hom}=-af^2 +\left[b_1+b_2\left(\mbox{cos}^2 (2\theta)+(\vec{n}\vec{m})^2
\mbox{sin}^2 (2\theta)\right)\right]f^4 \nonumber\\
&-C{\cal{M}} f^2\vec{h}[\vec{n}\times\vec{m}]\mbox{sin} (2\theta)+\frac{(\vec{h}-\vec{H})^2}{8\pi}+O(f^6)\,.
 \end{align}

Now we focus on the consideration of various phases in a  system of fermions with the 
spin-triplet pairing. First, consider the case when one can neglect contribution $\propto f^6$  formally setting  $\gamma =0$.
 Minimization in $h$ and $f$ yields
 \begin{align}
 \vec{h}=\vec{H}+4\pi C{\cal{M}} f^2[\vec{n}\times\vec{m}]\mbox{sin} (2\theta)\,,\label{hgen}
 \end{align}
\begin{align}
\label{fcos}
&f^2 = \frac{a+C{\cal{M}}\vec{H}[\vec{n}\times\vec{m}]\mbox{sin} (2\theta)}{2Y}\,\theta (f^2)\,,\\
&Y=b_1+
b_2 \left(\mbox{cos}^2 (2\theta) + (\vec{n}\vec{m})^2\mbox{sin}^2 (2\theta)\right)\nonumber\\
&-2\pi C^2{\cal{M}}^2 [\vec{n}\times\vec{m}]^2\mbox{sin}^2 (2\theta)\,.\label{Y}
\end{align}
Stable solution exists only for $Y>0$.
For $H=0$ the solution exists for $a>0$, $Y>0$.

With the solution (\ref{hgen}) -- (\ref{Y}) we get the Gibbs free-energy density
\begin{align}
 G^{\rm hom}= -\frac{\left[a+C{\cal{M}}\vec{H}[\vec{n}\times\vec{m}]\mbox{sin} (2\theta)\right]^2\theta(f^2)}{4Y}\,.\label{Gcos}
\end{align}

\subsection{Nonmagnetic superfluidity in  phase A}\label{DF}

\subsubsection{ Uniform matter}
  {\em The phase A} with zero   mean spin density (\ref{avspin}) corresponds to the choice: $\theta =0$. Then   $\vec{\psi}=f\vec{n}$, as it follows from (\ref{gensol}).

For $\theta =\phi =0$ in the stable phase A Eq. (\ref{GAgen}) simplifies as
  \begin{align}
\label{caseA}
G_{\rm A}^{\rm hom} =-a f^2 + (b_1 + b_2) f^4
\,.
\end{align}

 Eqs. (\ref{hgen}), (\ref{fcos}), (\ref{Gcos}) read
\begin{align}
\label{psiA}
f^2 &=f_0^2=\frac{a}{2(b_1 +b_2)} \theta(f^2_0)\,, \\ \label{hA}\vec{h}&=\vec{h}_0=H\,,
\\
\label{GA}
G_{\rm A}^{\rm hom} &=-\frac{a^2}{4(b_1 +b_2)}\theta(f_0^2)\,,
\end{align}
 for $T<T_{\rm cr}^{\rm A}\equiv T_{\rm cr}$ ($a>0$).   For $T>T_{\rm cr}$ we have $f =0$, $\vec{h}=0$, and $G_{\rm A}^{\rm hom} =0$. The gradient term (\ref{GLvector1}) is zero for the homogeneous solution.
In the critical point $G_{\rm A}^{\rm hom}=0$, $\partial G_{\rm A}^{\rm hom}/\partial T =0$ but $\partial^2 G_{\rm A}^{\rm hom}/\partial T^2 \neq 0$ that corresponds to the second-order phase transition at $T=T_{\rm cr}$.

Consider stability of the phase A respectively to the formation of a small spin density in the system for $H=0$.   Taking $|\theta |=\delta \theta \ll 1$ and allowing $\phi \neq 0$ in Eq. (\ref{Y}) we obtain
\begin{align}
\label{GcosB}
 &Y= b_1 +b_2 [1- 4(\delta\theta)^2(1- (\vec{n}\vec{m})^2)]\nonumber\\
 &-8\pi C^2{\cal{M}}^2[\vec{n}\times\vec{m}]^2(\delta\theta)^2\nonumber\\
 &=b_1+b_2 -4(\delta\theta)^2\sin^2\phi [b_2+
 2\pi C^2{\cal{M}}^2]
 \,,
\end{align}
that demonstrates stability of the phase A only provided
  \begin{align}
b_1+b_2>0 \,,
\label{stability-cond}
\end{align} (otherwise one should incorporate $\gamma\neq 0$ terms)
and for
 \begin{align}
  b_2  +2\pi C^2{\cal{M}}^2<0\label{b2}
   \end{align}
(otherwise the A-phase is unstable to the appearance of $\theta\neq 0$ and $\phi\neq 0$ in the ground state). Thus for $H=0$ the phase A is stable to appearance of a non-zero spin density in the system.
Note that for $b_2 =0$, that corresponds to $\xi_1 =0$ in the vector boson case considered in Sect. \ref{vectorboson}, the condition (\ref{b2}) is not fulfilled. In the vector boson case it was reflected in the fact that
for $\xi_1 =0$ in the B phase the Gibbs free energy is smaller than in the A phase.

\subsubsection{Sub-phases A$_1$, A$_2$, A$_3$. Gradient term.  Domains}

{\bf Sub-phases A$_1$, A$_2$, A$_3$.} Since $\vec{n}$ is fully characterized by its three projections,  we  may consider  three specific choices $\vec{n}=(1,0,0)$, $\vec{n}=(0,1,0)$, and $\vec{n}=(0,0,1)$: the A$_1$ sub-phase $(\psi_1 =\psi \neq 0,\,
 \psi_2 =\psi_3 =0)$, A$_2$ sub-phase $(\psi_2 =\psi\neq 0,\,
 \psi_1 =\psi_3 =0)$ and A$_3$ sub-phase $(\psi_3 =\psi \neq 0,\,
 \psi_1 =\psi_2 =0)$,
 which we have introduced in Sect. \ref{vectorboson}.
In the uniform neutral  superfluid  these states
are degenerate and correspond to the same Gibbs free energies.

{\bf Gradient term. Stability of A sub-phases.} We  focus now on the role of the gradient contribution to the free energy (\ref{GLvector}). Let the medium fills  the half-space $x<0$. Then $f=f(x)$ and  does not depend on $y$ and $z$ due to the uniformity of the system in these directions.
 The gradient contributions for sub-phases A$_1$, A$_2$ and A$_3$ are  different,
  \begin{align}
G^{\rm grad}_i= C_i (\partial_1 f)^2 \,, \quad i={\rm A}_1,{\rm A}_2,{\rm A}_3\,.\label{GgradA}
\end{align}
In the sub-phase A$_1$, $\psi_1 (x)\neq 0$, $\psi_2=\psi_3=0$, and  $C_{{\rm A}_1} = c_1+c_2+c_3$. For such a solution  $\mbox{div}\,\vec{\psi}\neq 0$.  In the sub-phases A$_2$ and A$_3$,
$C_{{\rm A}_2} =C_{{\rm A}_3}= c_1$. Here only $\psi_2 (x)\neq 0$ or $\psi_3(x)\neq 0$, respectively, and  the condition
$\mbox{div}\,\vec{\psi}=0$ is fulfilled.

Thus the stability conditions are
\begin{align}c_1 +c_2 +c_3 \geq 0\,,\quad c_1 \geq 0\,.\label{stabGrad}
\end{align}
Now let us check the stability of the phase A in presence of the gradient contribution to the Gibbs free energy   respectively the appearance of a small $\theta (x)$.
For $\vec{m}\parallel \vec{n}$ in Eq. (\ref{GgradA})  there appear extra terms
$(c_1 +c_2 +c_3)f^2(\partial_1 \theta)^2$ for $\vec{n}=(1,0,0)$ and $c_1 f^2(\partial_1 \theta)^2$
for $\vec{n}=(0,1,0)$ or $\vec{n}=(0,0,1)$.
For $\vec{m}\perp \vec{n}$, with $\vec{n}=(1,0,0)$ and $\vec{m}=(0,1,0)$ or $\vec{m}=(0,0,1)$  in Eq. (\ref{GgradA})   appears extra term $c_1 f^2(\partial_1 \theta)^2$. For  $\vec{n}=(0,1,0)$  for $\vec{m}=(0,0,1)$ there appears the term $c_1 f^2(\partial_1 \theta)^2$  and for $\vec{m}=(1,0,0)$, the term   $(c_1 +c_2+c_3) f^2(\partial_1 \theta)^2$.
As we see, in all these cases an increase of $\theta$ is energetically not profitable. Thus  the phase A is  stable respectively to the growth of  weak perturbations both in the uniform and the nonuniform matter.

Variation of the Gibbs free energy (\ref{GLvector}) in the field $f$
 yields  equations of motion
\begin{align}
C_i\partial_1^2 f + a f -2(b_1+b_2)f^3 =0\,,\label{eqpsgrad}
\end{align}
with the solutions satisfying the boundary condition $f (x=0)=0$,
 \begin{align}
f (x) =f_0 \mbox{th} \frac{x}{\sqrt{2}\,\xi_{\rm A_i}}\,,\quad \xi_{\rm A_i } =\sqrt{C_i/a}\,,\label{solpsiA}
\end{align}
$f_0$ is given by  Eq. (\ref{psiA}). Replacing (\ref{solpsiA}) in the expression for the Gibbs free energy, $\int d^3 x G ={\cal{G}}^{\rm vol}+{\cal{G}}^{\rm surf}$, we find that
 the surface contribution  is ${\cal{G}}^{\rm surf}_i\propto \xi_i S$, $S$ is the square in $y,z$ plane. ${\cal{G}}^{\rm surf}_i$ gets minimum  for the sub-phases A$_2$ and A$_3$, if
$0<c_1<c_1+c_2+c_3$, and for the sub-phase A$_1$, if  $c_1>c_1+c_2+c_3>0$.

 {\bf Domains.}
 Depending on how the system was prepared, it can consist of domains with different directions of the order parameter $\vec{\psi}$ in each domain.
 Due to the difference in the contributions to the surface energies in the longitudinal and transversal directions respectively the surface,  for a domain of a fixed volume it is profitable to become oblate or prolate in dependence on the sign of $c_2+c_3$.

For a slab of the sub-phase $A_1$ surrounded by  the matter in the  sub-phase $A_2$ due to the presence of the phase boundary  there appears a contribution to the surface energy, $\delta {\cal{G}}_{\rm A_1, A_2}^{\rm surf}={\cal{G}}^{\rm surf}_{\rm A_1} +{\cal{G}}^{\rm surf}_{\rm A_2}>0$.
However, as we have demonstrated, the solution  for the order parameter in the phase A characterized by a direction $\vec{n}$ is stable.
Thus, to melt the domain  should overcome the  energy barrier $\delta {\cal{G}}_{\rm A_1, A_2}^{\rm surf}.$
Necessary energy to overcome the barrier can be extracted, e.g., from  thermal fluctuations, or  from external  magnetic field, or for the  system subjected to the external rotation the required energy can be  taken from  the energy of the rotation.

Notice that   in difference with the case  $c_1 =-c_3\neq 0$, $c_2 = 0$ considered in Sect. \ref{vectorboson}, where the A$_1$ phase was not realized and the  sub-phases A$_2$ and A$_3$ had the same volume and surface energies,   here for $c_1 +c_2+c_3\neq 0$, $c_1 \neq 0$ and $c_1\neq c_1 +c_2+c_3$ all three sub-phases can be realized and the surface energy in the A$_1$ sub-phase differs from those in A$_2$ and A$_3$ sub-phases.

\subsection{Instability of A phase in external magnetic field, AH phase}\label{AH}

Above we have demonstrated stability of the  phase A (at zero mean spin density) to  formation of a nonzero spin state in absence of the external magnetic field.  Let us study stability of the ground state of the A-phase (conditions (\ref{stability-cond}), (\ref{b2}) are supposed to be fulfilled) respectively to the growth of $\theta$ and $\phi$, i.e. to the formation of a mean spin density in the system, for $H\neq 0$. Further  we consider energetically favorable cases, one corresponding to  $\vec{\cal{M}}\vec{ H} >0$ for ${\cal{M}} >0$ (for protons) and another for $\vec{\cal{M}}$  aligned antiparallel $z$ for ${\cal{M}} <0$ (for neutrons).
 Rewrite (\ref{Gcos}) as
\begin{align}
G^{\rm hom}_{\rm AH}= -\frac{[a+ C|{\cal{M}} H\zeta|]^2\theta (a+ C|{\cal{M}} H\zeta|)}{4[b_1 +{b}_2-\zeta^2 ({b}_2+2\pi C^2{\cal{M}}^2) ]}\,,
\end{align}
for ${\cal{M}} H >0$ with $\zeta =\mbox{sin}\phi\mbox{sin}(2\theta)>0$ and for ${\cal{M}} H <0$ with $\zeta =\mbox{sin}\phi\mbox{sin}(2\theta)<0$. The denominator is positive provided conditions  (\ref{stability-cond}), (\ref{b2}) are fulfilled.
As we can see, {\em for $H\neq 0$ the phase A proves to be unstable} in respect to production of a spin density, since it is energetically profitable to have $\zeta\neq 0$. Accordingly, cf. (\ref{hgen}), in presence of the external magnetic field the strength of the magnetic field becomes
$${h}={H}+4\pi C{\cal{M}} f^2 \zeta \,.$$

\subsubsection{ Paramagnetic response of superfluid in AH phase for $T<T_{\rm cr}$}
 For $a>0$, i.e. $T<T_{\rm cr}$, minimizing the Gibbs free-energy density in $\zeta =\mbox{sin}\phi\mbox{sin}(2\theta)$ we get at the extremum
\begin{align}
\zeta_m  =-\frac{C{\cal{M}} H (b_1+{b}_2)}{a ({b}_2 +2\pi C^2{\cal{M}}^2)}\,,\label{zetam}
\end{align}
 valid for $|\zeta_m| \leq 1$, with $\zeta_m \to 0$ for $H\to 0$.
  Note that with $\zeta=\zeta_m \neq 1$ we obtain $f^2>0$ in (\ref{fcos}) only for $a>0$, i.e. for $T<T_{\rm cr}$. Thus,  for $H\neq 0$ not all spins in the condensate are aligned in one direction  at  $T<T_{\rm cr}$. We deal with the novel phase, which we name  {\em the AH phase}
 when the conditions (\ref{stability-cond}), (\ref{b2}) are fulfilled but not all spins of the paired fermions   are aligned in one direction. For $b_2<0$ and $|C{\cal{M}}\zeta_m | H\ll a$ we find $h=H[1+2\pi C^2 {\cal{M}}^2/(|b_2|(b_1-|b_2|)]$. Also, from (\ref{zetam}) we find an additional  constraint,
 \begin{align}
 H\leq H_{\rm cr}^{\rm AH}(T<T_{\rm cr}) =\frac{a(|{b}_2|-2\pi C^2{\cal{M}}^2)}{|C{\cal{M}}| (b_1-|{b}_2|)}\label{HcrAH}
 \end{align}
  for $a>0$.

The Gibbs free-energy density in the ground state  for $T<T_{\rm cr}$ ($a>0$) can be presented as
\begin{align}
G^{\rm hom}_{\rm AH}\simeq -\frac{a^2}{4(b_1+{b}_2)}+\frac{C^2{\cal{M}}^2 H^2}{4({b}_2 +2\pi C^2{\cal{M}}^2)}
\,. \label{AHgibbs}
\end{align}
Although for $H\neq 0$ the resulting magnetic field $h\neq 0$, for $H\to 0$ we obtain $h\to 0$.

\subsubsection{ Instability  of AH phase for $T>T_{\rm cr}$. Transition to a ferromagnetic superfluid phase}
Now consider the case $a<0$, i.e.  $T>T_{\rm cr}$.
The actual critical temperature  is determined from the condition
$a+C{\cal{M}} H\zeta=0$ for ${\cal{M}} H>0$ and $\zeta =1$, and from $a-C{\cal{M}} H\zeta=0$ for ${\cal{M}} H<0$ and $\zeta =-1$.  For favorably aligned spins we obtain
\begin{align}
T_{\rm cr}^{\rm AH} =T_{\rm cr}(1 +|C{\cal{M}} H|/\alpha_0)>T_{\rm cr}\,,\quad {\rm for}\quad a<0\,.\label{TcrAH}
\end{align}
Thus the AH phase may exist not only for $T<T_{\rm cr}$ but also in the temperature interval $T_{\rm cr}<T<T_{\rm cr}^{\rm AH}$ and the critical temperature $T_{\rm cr}^{\rm AH}$ is increased with increasing $H$. In this respect the AH phase is similar to the A$_1$ phase of the $^3$He, cf. \cite{AM1973}.

The Gibbs free-energy density in the ground state  for $T_{\rm cr}<T<T_{\rm cr}^{\rm AH}$ is as follows
 \begin{align}
G^{\rm hom}_{\rm AH}= -\frac{[a+ |C{\cal{M}} H|]^2\,\theta (a+ |C{\cal{M}} H|)\,\theta(-a)}{4(b_1 -2\pi C^2{\cal{M}}^2) }\,.\label{bA}
\end{align}
As we see,  for $a<0$  still the condition $b_1 -2\pi C^2{\cal{M}}^2 >0$ should be satisfied for the stability of the phase.  In next sub-section \ref{BC} such a phase will be named the B phase.
Thus for $T_{\rm cr}<T<T_{\rm cr}^{\rm AH}$ the AH phase coincides with the B phase, if besides the conditions (\ref{stability-cond}), (\ref{b2}) also the condition $b_1 -2\pi C^2{\cal{M}}^2 >0$ is satisfied. For $T_{\rm cr}<T<T_{\rm cr}^{\rm AH}$ (for $a<0$) we put in
  Eq. (\ref{zetam}) $\zeta =1$ for $C{\cal{M}}>0$ and $\zeta =-1$ for $C{\cal{M}}<0$,
 that corresponds to the fact that all spins are aligned in one direction.
   We find that the condensate exists now for
  \begin{align}
  H>H_{\rm cr}^{\rm AH}(T_{\rm cr}<T<T_{\rm cr}^{\rm AH})=|a|/{|C\cal{M}}|\,.\label{HcrAH1}
 \end{align}

\subsection{Ferromagnetic superfluidity in phases B and C}\label{BC}

\subsubsection{Stability conditions}
We  name  {\em the phase B or C} the choice $\theta =\pi/4$, $\vec{n}\perp \vec{m}$, $H$ is arbitrary.
Setting  $\theta =\pi/4-\delta\theta$ in Eq. (\ref{Y}) we find
\begin{align}
&Y= b_1 +b_2 [4(\delta\theta)^2(1- (\vec{n}\vec{m})^2)+(\vec{n}\vec{m})^2]\nonumber\\
 &-2\pi C^2{\cal{M}}^2[\vec{n}\times\vec{m}]^2(1-4(\delta\theta)^2)\,
.
\label{GcosA}
\end{align}
We deal with {\em the phase B}, if
\begin{align}
b_1 -2\pi{\cal{M}}^2 C^2>0\,,
\label{conda}
\end{align}
and with {\em the phase C}, if
\begin{align}
b_1-2\pi {\cal{M}}^2 C^2 <0\,.\label{condC}
\end{align}
These conditions together with condition
\begin{align}
b_2+2\pi {\cal{M}}^2 C^2 >0\,\label{condB1}
\end{align}
replace the stability conditions (\ref{stability-cond}), (\ref{b2}), being fulfilled in case of the A phase.
Favorable direction of $\vec{H}$  is parallel to $[\vec{n}\times\vec{m}]$,
 as follows from (\ref{fcos}). This is in agreement with our observation done in previous section that
 the sub-phase B$_3$ with $\vec{H}\parallel z$ corresponds to the lowest Gibbs free energy.

For the phase C (at $b_1 -2\pi C^2{\cal{M}}^2<0$) one needs to include at least the 6-th order term ($\gamma \neq 0$) in the free energy. In order not to complicate  consideration we  further choose the simplest form of the $\{\gamma_i {\psi_i}\}^6$ term ($\gamma (\psi_i^{*}\psi_i)^3$) assuming $\gamma >0$. Note here that in the BCS weak coupling theory one  obtains $\gamma <0$ and expansion of the Gibbs free-energy should be continued up to the 8-th order \cite{Yasui:2019unp}.

For simplicity we   put $T_{\rm cr}^{\rm A}=T_{\rm cr}=T_{\rm cr}^{\rm B}$, on the other hand $T_{\rm cr}^{\rm C}\neq T_{\rm cr}$ since the phase transition to the phase C proves to be of the first order.

\subsubsection{Sub-phases B$_1$, B$_2$, B$_3$, C$_1$, C$_2$, C$_3$}
 In general we may consider three choices:
\begin{align}
& \psi_1 =0
\,,\,\,\,\psi_2 = \mp i\psi_3\equiv\frac{1}{\sqrt{2}}\widetilde{\psi}\,,\nonumber\\
&
\psi_2 =0\,, \,\,\,\psi_1 = \mp i\psi_3\equiv\frac{1}{\sqrt{2}}\widetilde{\psi}\,,\label{psi21}\\
&
\psi_3 =0\,, \,\,\,\psi_1 = \mp i\psi_2\equiv\frac{1}{\sqrt{2}}\widetilde{\psi}\,,\nonumber
\end{align}
for sub-phases B$_1$ (or C$_1$), B$_2$ (or C$_2$), and B$_3$ (or C$_3$), respectively.
In all these cases  $\psi_i\psi_i=0$.

With the simplest $\gamma (\psi_i^{*}\psi_i)^3$ term taken into account we have
\begin{align}
G^{\rm hom} & =-a |\widetilde{\psi}|^2 + b_1|\widetilde{\psi}|^4- C\vec{{\cal{M}}} \vec{h}|\widetilde{\psi}|^2
    +\gamma |\widetilde{\psi}|^6 \label{eqmot1}\\
    &+ \frac{(\vec{h}-\vec{H})^2}{8\pi} \,.\nonumber
\end{align}

The gradient contribution to the Gibbs free energy does not depend on $h$. Thus, varying (\ref{eqmot1}) in $h$
we obtain
\begin{align}
\vec{h}-\vec{H} = 4\pi  |\widetilde{\psi}|^2 C\vec{{\cal{M}}}\,.\label{hhom2hom}
\end{align}

In the B$_3$, C$_3$ sub-phases the averaged spin density and $\vec{h}$ are directed parallel or antiparallel $z$ and for $\vec{H}$ directed in $z$ we get  $h_{3}=H +4\pi  |\widetilde{\psi}|^2 C{\cal{M}}$.

In sub-phases  $B_1$ and C$_1$ the averaged spin density is directed parallel/antiparallel $x$, and
$\vec{H}$ directed in $z$ we have
\begin{align}
\label{hhom3}
{h}_{1} = 4\pi  |\widetilde{\psi}|^2 C{\cal{M}}\,, \quad {h}_{3} = H\,.
\end{align}
Similarly, in sub-phases B$_2$, C$_2$ the averaged spin density is directed parallel/antiparallel $y$.
Replacing (\ref{hhom2hom}) in (\ref{eqmot1})  we see that for $H\neq 0$ in sub-phases B$_3$, C$_3$ the energy density  is gained compared to sub-phases B$_1$, C$_1$ and B$_2$, C$_2$.
 In absence of $H$  the Gibbs free-energy density is the same for all the sub-phases B$_1$, B$_2$, B$_3$ and C$_1$, C$_2$, C$_3$, respectively. Thereby since (\ref{hhom3}) does not depend on $H$ the results for sub-phases B$_1$, B$_2$, and C$_1$, C$_2$, can be obtained from those for sub-phases B$_3$ and C$_3 $ by setting $H=0$.

\subsubsection{ Uniform solutions for phases B and C}
 For the uniform phases B and C  we find the solution
\begin{align}
|\widetilde{\psi}|^2 &=\widetilde{\psi}^2_0= -\frac{1}{3\gamma}(b_1 -2\pi C^2{\cal{M}}^2 )\nonumber\\
&\pm \frac{1}{3\gamma}\sqrt{(b_1-2\pi C^2{\cal{M}}^2)^2 + 3\gamma(a+C\vec{{\cal{M}}}\vec{H})}
\,.
\label{psihom1}
\end{align}
We need to retain the solution  corresponding to $|\widetilde{\psi}|^2 >0$. In case of the phase B it is the solution corresponding to the upper sign in (\ref{psihom1}) and in case of the phase C it is the solution corresponding to the lower sign.
From Eqs.~(\ref{eqmot1}), (\ref{hhom2hom}),  (\ref{psihom1}) we obtain
\begin{align}
G^{\rm hom}_{\rm{B},\rm{C}}=-\frac{\widetilde{\psi}^2_0}{3}\left[2(a+C\vec{{\cal{M}}}\vec{H})-(b_1-2\pi C^2{\cal{M}}^2)\widetilde{\psi}^2_0\right]\,.
\label{GB2}
\end{align}
We see that the energetically preferable direction of the spin is such that $\vec{{\cal{M}}}\vec{H}>0$.
Thus we may  replace $\vec{{\cal{M}}}\vec{H}$ to $|\vec{{\cal{M}}}\vec{H}|$.

Note that the ansatz $\psi_1 =\pm i\psi_2$ has been exploited previously in description of the unconventional superconductors, cf. \cite{KR1998,MineevSamokhin,Olesen},  but a possibility of appearance of an own magnetic field $h\neq 0$ was not considered.
Therefore \emph{phases B and C  are  novel magnetic phases: already in absence of the external magnetic field the matter in these phases  represents a ferromagnetic superfluid.}

 {\bf{ Uniform solutions for phase B.}}
Setting $\gamma \to 0$ in (\ref{psihom1}) with the plus-sign solution we find
\begin{align}
\widetilde{\psi}^2_0= \frac{a+|C\vec{{\cal{M}}}\vec{H}|}{2\,(b_1 - 2\pi C^2{\cal{M}}^2)}\,\theta\left(\frac{a+|C\vec{{\cal{M}}}\vec{H}|}{b_1 - 2\pi C^2{\cal{M}}^2}\right)\,,
\label{psihom}
\end{align}
for ${a+C\vec{{\cal{M}}}\vec{H}}>0$ provided the condition (\ref{conda}) is fulfilled. Using (\ref{psihom})  we obtain the own magnetic field $\vec{h}$:
\begin{align}
\vec{h}=\vec{H} +2\pi C\vec{{\cal{M}}}\frac{ (a+|C\vec{{\cal{M}}}\vec{H}|)}{b_1 -2\pi C^2{\cal{M}}^2}\,.
\label{hhom}
\end{align}
Chosing  ``$-$'' sign solution of Eq. (\ref{psihom1}) would lead to the positive value of $G$.

Replacing (\ref{psihom})  in (\ref{GB2}) we find the expression for
the Gibbs free-energy density
\begin{align}
G_{\rm B}^{\rm hom} =-\frac{(a+|C\vec{{\cal{M}}}\vec{H}|)^2}{4(b_1 -2\pi C^2{\cal{M}}^2)}\,\theta\left(\frac{a+|C\vec{{\cal{M}}}\vec{H}|}{b_1 - 2\pi C^2{\cal{M}}^2}\right)\,,
\label{GB1}
\end{align}
cf. Eq. (\ref{GB1Hx}) for the B$_1$ sub-phase for vector bosons  and Eq. (\ref{B3Hvector}) for B$_3$ sub-phase. For the B$_1$ sub-phase  here $\vec{{\cal{M}}}\vec{H}=0$ and for B$_3$ sub-phase $\vec{{\cal{M}}}\vec{H}=\pm{{\cal{M}}}{H}$.
Setting $H=0$ in (\ref{GB1}) we recover the Gibbs free-energies for the B$_1$ and B$_2$ sub-phases,
$$G_{\rm B_1,B_2}^{\rm hom} =G_{\rm B_3}^{\rm hom}(H=0)
\,,$$ cf. Eq. (\ref{PrelimB}) for neutral vector bosons. For $H=0$ in all three sub-phases B$_i$ there appears an internal magnetic field
\begin{align}
 \vec{h}(H=0) =\frac{2\pi a C\vec{{\cal{M}}}}{b_1 - 2\pi C^2{\cal{M}}^2}\,.
\label{hprop}
\end{align}
Thus, we found that  in sub-phases B$_i$ superfluidity ($\widetilde{\psi} \neq 0$)  coexists with ferromagnetism ($h(H=0)\neq 0$). With increasing $H$ the amplitude of the condensate grows.

 We see that in the presence of an external magnetic field  the sub-phase B$_3$, where $\vec{\cal{M}} \parallel z$ (for ${\cal{M}} >0$), becomes energetically preferable compared to  sub-phases B$_1$ and B$_2$. For ${\cal{M}} >0$ the preferable orientation of the averaged spin density $\vec{S}$ is parallel to $\vec{H}$. For ${\cal{M}} <0$ the preferable orientation of the averaged spin density $\vec{S}$ is antiparallel to $\vec{H}$.
Note that superfluidity may arise even in the state, where $\vec{{\cal{M}}}$ is antiparallel to $\vec{H}$ (for ${\cal{M}} >0$) provided $a-|{\cal{M}} H|>0$, however this state  corresponds to a higher Gibbs free energy than the state with $\vec{{\cal{M}}}$ parallel to $\vec{H}$.

{\em In the external magnetic field}, $H\neq 0$, the actual value of the critical temperature for the sub-phase B$_3$  is
\begin{align}
{T}_{\rm cr} ^{\rm B_3H}=T_{\rm cr} (1+|C{\cal{M}} H|/\alpha_0)\,,\label{newcrtem}
\end{align}
where $T_{\rm cr}$ is the critical temperature for $H=0$, provided one may use the parametrization $a =\alpha_0 t$ with  $t=(T_{\rm cr}-T)/T_{\rm cr}$.
Thus, for $H\neq 0$ ($\vec{H}\parallel z$) the sub-phase B$_3$ continues to exist above $T_{\rm cr}$ up to   $T={T}_{\rm cr}^{\rm B_3H}$. For
$T>{T}_{\rm cr}^{\rm B_3H}$ we have $\widetilde{\psi} =0$.
Notice that Eq. (\ref{GB1}) coincides with Eq. (\ref{bA}), which we have derived considering AH phase. However Eq. (\ref{GB1}) is  valid for all $T< T_{\rm cr}^{\rm B_3H}$ provided condition (\ref{condB1}) is fulfilled, whereas  Eq. (\ref{bA}) is valid for $T_{\rm cr}<T< T_{\rm cr}^{\rm AH}$ and at the condition (\ref{b2}) satisfied. For $T_{\rm cr}^{\rm B}=T_{\rm cr}^{\rm A}=T_{\rm cr}$, that we for simplicity postulated,  Eq. (\ref{newcrtem}) coincides with Eq. (\ref{TcrAH}).
We find that in the temperature interval $T_{\rm cr}<T<T_{\rm cr}^{\rm B_3H}$ the condensate exists now for
  \begin{align}
  H>H_{\rm cr}^{\rm BH}=|a|/{|C\cal{M}}|\,.\label{HcrBH1}
 \end{align}

For $H\gsim \alpha_0/|{\cal{M}}|$ the parametrization $a =\alpha_0  t$ might become invalid.
Using another  popular parametrization $\varphi =\ln (T/T_{\rm cr})$ in Eq. (\ref{aLand}) we  find
\begin{align}
{T}_{\rm cr}^{\rm B_3H} =T_{\rm cr} e^{C|{\cal{M}} H|/\alpha_0}\,.\label{newcrtem1}
\end{align}
Here we should notice that although expression (\ref{newcrtem1}) allows for ${T}_{\rm cr}^{\rm B_3H} \gg T_{\rm cr}$ the Ginzburg-Landau mean-field approach itself becomes  invalid for such temperatures.

 At the critical point $G_{\rm B_3}^{\rm hom}=0$, $\partial G_{\rm B_3}^{\rm hom}/\partial T =0$ but $\partial^2 G_{\rm B_3}^{\rm hom}/\partial T^2 \neq 0$ that, as in case of the phase A, corresponds to {\em{the second-order phase transition.}}

Note  that the quantity $b_1-2\pi C^2{\cal{M}}^2$ should not be too small. Otherwise, terms $\propto\gamma\psi^6$ must be taken into account.

Also note  that above we considered only  contributions to the Gibbs free-energy, which depend  on the pairing order parameter. However, the total Gibbs free-energy contains also a normal contribution of unpaired fermions. Owing to the normal term there appears a small paramagnetic contribution proportional to $h^2$. Simplifying consideration we disregarded this small correction term in our  calculations.

{\bf  Uniform solutions for phase C.} Now we assume that
conditions  (\ref{condC}) and (\ref{condB1}) are fulfilled.
The $\widetilde{\psi}^4$-term in the Gibbs free energy proves to be negative, and the problem should be reconsidered with taking into account  $\{\gamma_i {\psi_i}\}^6$ term,
which  provides stability (for $\gamma >0$).

Let us perform expansion of (\ref{psihom1}) in a small
$\gamma$. The minimum of the Gibbs free-energy density is realized for the choice of $``+"$ sign solution in Eq. (\ref{psihom1}).
Then from Eqs. (\ref{GB2})--(\ref{stabgrB}) we find
\begin{align}
\widetilde{\psi}^2_0 & \simeq\frac{2}{3\gamma}(2\pi C^2{\cal{M}}^2 -b_1)+\frac{(a+C\vec{{\cal{M}}}\vec{H})}{2(2\pi C^2{\cal{M}}^2 -b_1)}>0\,,\label{tildepsi2g}
\\
h &\simeq H+\frac{8\pi C{\cal{M}} }{3\gamma}(2\pi C^2{\cal{M}}^2 -b_1)\nonumber\\
&+\frac{2\pi C{\cal{M}} (a+C\vec{{\cal{M}}}\vec{H})}{(2\pi C^2{\cal{M}}^2 -b_1)}\,,\label{tildeh2g}
\\
G_{\rm C}^{\rm hom} & \simeq -\frac{4}{27\gamma^2}(2\pi C^2{\cal{M}}^2 -b_1)^3\,,\label{GB2g}
\end{align}
again with the energetically preferable direction of $\vec{{\cal{M}}}$ corresponding to $\vec{{\cal{M}}}\vec{H}>0$.
Expansion is valid for
\begin{align}
0<\gamma \ll (2\pi C^2{\cal{M}}^2 -b_1)^2/(a+|C{\cal{M}} H|)\, . \label{gamma}
\end{align}
The condensate amplitude grows with increasing $H$.
For the case ${H}\neq 0$ parallel $z$, which we consider, the sub-phase C$_3$ proves to be the most energetically profitable.
The results for C$_1$ and C$_2$ follow, if one puts $H=0$.

The value of the new critical temperature is determined by the condition of the vanishing of the square-root in Eq. (\ref{psihom1}),
\begin{eqnarray}
{T}_{\rm cr}^{\rm C_3H} =T_{\rm cr} \left[1
+\frac{(2\pi C^2{\cal{M}}^2-b_1)^2}{3\gamma \alpha_0}
+\frac{|{C\cal{M}} H|}{\alpha_0}\right],\label{TcC}
\end{eqnarray}
that holds provided the validity of the relation $a=\alpha_0  t$, $t=(T_{\rm cr}-T)/T_{\rm cr}$. Thus ${T}_{\rm cr}^{\rm CH}\geq {T}_{\rm cr}^{\rm C}$, where
\begin{eqnarray}
{T}_{\rm cr}^{\rm C} =T_{\rm cr} \left[1+\frac{(2\pi C^2{\cal{M}}^2-b_1)^2}{3\gamma \alpha_0}\right] > T_{\rm cr}.\label{TcC0}
\end{eqnarray}
 For the sub-phases $C_1$ and $C_2$ we have ${T}_{\rm cr}^{\rm C_2}(H=0)= {T}_{\rm cr}^{\rm C_1}(H=0) = {T}_{\rm cr}^{\rm C_3}(H=0)={T}_{\rm cr}^{\rm C}\,.$

 At the critical point $G_{\rm C}$ changes discontinuously, that corresponds to    {\em{the first order phase transition.}} Ferromagnetic superfluid solution (\ref{tildepsi2g}) -- (\ref{GB2g})
holds  provided conditions  (\ref{condC}), (\ref{condB1})  and (\ref{gamma})
are fulfilled.

\subsubsection{Gradient term. Domains}

For the system of a large but finite size already at $H=0$ the degeneracy of the sub-phases is  removed because of a difference in the gradient contributions in  the Gibbs free-energy density of various sub-phases. As in case of the phase A studied above, to be specific let us consider sample  filling the half-space $x<0$.  Then
  \begin{align}
G^{\rm grad}_i= C_i |\partial_1 \widetilde{\psi}|^2 \,, \,\,\, i={\rm B}_1 ({\rm C}_1),{\rm B}_2 ({\rm C}_2),{\rm B}_3 ({\rm C}_3)\,.\label{gradBC}
\end{align}
 For the sub-phases B$_3$ $({\rm C}_3)$ and B$_2$ $({\rm C}_2)$ the coefficient $C_{i} = c_1+(c_2  +c_3)/2$, and $\mbox{div}\,\vec{\psi}\neq 0$. For sub-phases   B$_1$ and ${\rm C}_1$:
$C_{i} = c_1$ and $\mbox{div}\,\vec{\psi}=0$. The stability conditions render
\begin{align}
c_1+c_2 /2 +c_3 /2> 0\,,\quad c_1 > 0\,.\label{stabgrB}
\end{align}

Consider  the phase B and put $\gamma =0$. Variation of the Gibbs free energy in fields, cf. (\ref{GLvector}),
 yields the equation of motion
\begin{align}
C_i\partial_1^2\widetilde{\psi} + (a+|C\vec{{\cal{M}}}\vec{H}|)\widetilde{\psi}-2(b_1 -2\pi C^2{\cal{M}}^2)|\widetilde{\psi}|^2\widetilde{\psi} =0\,,\label{eqpsgradB}
\end{align}
with the solution
satisfying the boundary condition $\widetilde{\psi} (x=0)=0$,
 \begin{align}
\widetilde{\psi} (x) =\widetilde{\psi}_0 \mbox{th} \frac{x}{\sqrt{2}\,\xi_{\rm B_i}}\,,\quad
\xi_{\rm B_i} =\sqrt{\frac{C_i}{a+|C\vec{{\cal{M}}}\vec{H}|}}\,, \label{solpsiB}
\end{align}
 instead of Eq. (\ref{solpsiA}) for the A-phase. The value $\widetilde{\psi}_0$ is determined by Eq. (\ref{psihom}).
At the fulfilled condition (\ref{conda}) 
the solution exists provided $a+|C\vec{{\cal{M}}}\vec{H}|>0$.

Replacing (\ref{solpsiB}) in the expression for the Gibbs free energy, $\int d^3 x G ={\cal{G}}^{\rm vol}+{\cal{G}}^{\rm surf}$, we find that
 the surface contribution  is ${\cal{G}}^{\rm surf}_i\propto \xi_i S$, $S$ is the square in $y,z$ plane. ${\cal{G}}^{\rm surf}_i$ gets minimum  for the sub-phases B$_3$ and B$_2$, if
$0<c_1+c_2/2 +c_3/2<c_1 $, and for the sub-phase B$_1$, if  $0<c_1<c_1+c_2/2 +c_3 /2 $.
Gradient terms do not contribute to the minimization of $G$ in $h$ and equation (\ref{hhom2hom}) continues to hold, from where using the boundary condition $\widetilde{\psi}(0)=0$
we find that $h(x\to 0)\to H$.

 The coordinate dependence of the field $\widetilde{\psi}$ in the phase C is    more involved, since  one needs to include at least  $\widetilde{\psi}^5$ term in the equation of motion to provide stability.

{\bf Domains.}
At the phase transition to phases B or C   there can be  formed  domains with different directions of $\vec{h}\parallel \vec{\cal{M}}$ and $\vec{\psi}$ in each domain.
As we have argued, when have considered the  sub-phases A,   an extra energy is needed to merge  the domains. For finite $T$  the required energy can be taken from thermal fluctuations. In presence of the external magnetic field or the external rotation  the extra energy can be taken from the energy of the magnetic and rotation fields, respectively.

\section{Spin-triplet pairing in charged fermion system described by complex
vector order parameter}\label{charged}
The spin-triplet pairing in the condensed matter, e.g. in systems with heavy fermions, is described by the vector order parameter at the effective charge of the pair $e_*=2e<0$, e.g. cf. \cite{VG1985,KR1998,MineevSamokhin}. In the nuclear systems the $np$ pairing in the 3S$_1$ channel is allowed for the case of the isospin-symmetric nuclear matter. The  3S$_1$ $np$ phase shift is the largest among others at low energies, cf. \cite{Sedrakian:2018ydt}. The $np$ pairing in the  3S$_1$ channel in absence of the spin-orbital interaction  is described by the vector order parameter at  $e_*=-e>0$.

\subsection{Gibbs free-energy density}\label{GFench}
For the description of the charged superconductors we may use Eqs. (\ref{GLvector})
for the Gibbs free-energy density ~\cite{KR1998,MineevSamokhin,Olesen} with $G_{\rm grad}^{\rm neut}$ replaced by $G_{\rm grad}^{\rm ch}$:
\begin{align}
G_{\rm grad}^{\rm ch}=c_1 |D_i \psi_j|^2 + c_2 |D_i \psi_i|^2 + c_3 (D_i \psi_j)^* D_j \psi_i\,,\label{GLvector3}
\end{align}
where $D_i =\partial_i -i e_{\rm *} A_i$, $A_i =(A_x, A_y, A_z)$, $e_{\rm *}$ is the charge of the fermion pair. The $D_i$ operators fulfill the relation for the commutator $i[D_i,D_j]_{-}=e_{\rm *}\epsilon_{ijk}h_k$, cf. Eq. (\ref{idenSt}) above. In  case when $\vec{h} =h\vec{n}_3$ with $\vec{n}_3 =(0,0,1)$ we have
\begin{align}
i[D_1,D_2]_{-}=e_{\rm *}h\,, \quad h=F_{12}=\partial_1 A_2 -\partial_2 A_1\,.\label{commut}
\end{align}

\subsection{Nonmagnetic phase A in the medium filling half of space placed  in uniform magnetic field}\label{Ach}
We deal with the phase A provided conditions (\ref{stability-cond}), (\ref{b2}) are fulfilled.
For this case a difference with the standard description of the superconductivity of spin-zero pairs is only in the specificity of the gradient terms. In absence of the external magnetic field    the description of the charged uniform system
 within the A phase remains the same as for the neutral system performed above. In presence of the external magnetic field the properties of the sub-phases A of a neutral spin-triplet superfluid and the charged one are different similarly to that we have demonstrated in previous section on example of the vector boson field.

Further consider a superconductor filling half of space $x<0$, placed in a homogeneous external magnetic field $\vec{H}$ parallel $z$, for $H>0$.  We may choose the gauge, where $\vec{A}$ has only one non-zero component $A_2(x)$ for $x<0$. We choose $\vec{A}_{\rm ext}=(0,Hx,0)$, satisfying the gauge condition $\mbox{div}\vec{A}_{\rm ext}=0$ and yielding $\mbox{rot} \vec{A}_{\rm ext}=\vec{H}$.

 Consider the phases A$_1$, A$_2$, A$_3$, which  are now not  degenerate.

 \subsubsection{ Sub-phase A$_1$ for $b_2 <0$} Consider first  phase A$_1$, where $\psi_1 =\psi (x)$ is real, $\psi_1 (x\to -\infty)\to \psi_0 =\sqrt{\frac{a}{2(b_1+b_2)}}$  for  $T<T_{\rm cr}^{\rm A}$ (to be specific we choose $``+"$ sign-solution), and allow for small perturbations of the fields $\psi_2 =-if_2 (x)$, $\psi_3 =-if_3 (x)$, and $A_2 =A_2(x)$, where $f_2$, $f_3$ are real quantities. The field $f_2$ is introduced to check stability of the A-phase in the presence of the external field $H$. Without  loss of the generality one may put $f_3 =0$.
 For simplicity assume that $A_2$ and $f_2$ are weak fields.

 The  gradient part of the Gibbs free-energy density
 can be presented as
 \begin{align}
&G_{\rm grad}^{\rm ch}=(c_1+c_2+c_3) (\partial_1 \psi)^2+c_1 e_{\rm *}^2 A_2^2(x)\psi^2 \nonumber\\
&+2c_3 e_{\rm *}A_2(x)\psi \partial_1 f_2+ c_1 (\partial_1 f_2)^2\,,\label{GgradAch}
\end{align}
written in quadratic approximation over the perturbative fields $A_2$ and $f_2$ and the derivatives $\partial_1$.
Stability conditions imply that $c_1+c_2+c_3 >0$, $c_1>0$.

Variation of the Gibbs free energy in the fields $\psi$, $A_2$, $f_2$ yields equations of motion
\begin{align}
(c_1+c_2+c_3)\partial_1^2\psi
+ a\psi-2(b_1+b_2)\psi^3 =0\,,\label{eqps}
\end{align}
\begin{align}
\partial_1^2 A_2 (x) -8\pi c_1 e_{\rm *}^2 \psi^2 A_2 (x)
+8\pi (C{\cal{M}}_3 -e_{\rm *} c_3)\psi \partial_1 f_2 =0\,,\label{cur}
\end{align}
\begin{align}
&c_1 \partial_1^2 f_2 +(c_3 e_{\rm *}-C{\cal{M}}_3)\psi  \partial_1 A_2 +c_3 e_{\rm *}A_2 \partial_1 \psi \nonumber\\
&+(a-2\psi^2 (b_1-b_2))f_2
=0\,,\label{f2}
\end{align}
written in linear approximation over perturbative fields.
The solution of Eq. (\ref{eqps}) for $a>0$ is given by   $\psi =\psi_0 \mbox{th} (x/\sqrt{2}\xi_{\rm A_1})$ with the coherence length $\xi_{\rm A_1} =\sqrt{(c_1+c_2+c_3)/a}$, cf. Eq. (\ref{solpsiA}).

In absence of the external magnetic field ($H=0$),  minimization in fields leads us to solutions (\ref{psiA}), (\ref{GA}) and $h=0$
for $T<T_{\rm cr}$  in the region $x<0$ everywhere except a surface layer.   In presence of a weak external magnetic field  there  exists complete Meissner effect.
We assume $d_{\rm A_1}/\xi_{\rm A_1}\gg 1$,
where $d_{\rm A_1}>0$ is the penetration depth for the magnetic field determined by Eq. (\ref{cur}).
Then  we may
 put $\psi=\psi_0 $ in Eqs. (\ref{cur}), (\ref{f2}).
In the theory of ordinary superconductors and for the case of the  charged scalar  bosons considered in Sect. \ref{preliminary} for $\vec{H}\parallel z$
the quantity $d_{\rm A_1}/\xi_{\rm A_1}$ is called the Ginzburg-Landau parameter, which value determines the behavior of the system. In the  case under consideration situation is a more involved.
 Explicit solution of Eq. (\ref{cur}) matched with that valid for $x\geq 0$ at the boundary $x=0$  is given by
 \begin{align}
A_2 (x)=Hd_{\rm A_1} e^{x/d_{\rm A_1}}\,.\label{solA}
\end{align}
 We search $f_2$ as
  \begin{align}
  f_2 (x)=D e^{x/d_{\rm A_1}},\label{f2a}
  \end{align}
  with a constant $D$. Since $D\neq 0$, to fulfill Eq. (\ref{f2}) for $A_2(x)\neq 0$, in case of the sub-phase A$_1$ there  appears a spin density in a surface layer. Dependence on $\psi (x)$ allows to fulfill the condition $f_2 (x=0)=0$. Since $\psi$ is dropping to zero on a scale $\xi_{\rm A_1}\ll d_{\rm A_1}$,  for $-x\sim d_{\rm A_1}\gg \xi_{\rm A_1}$ we may put $\psi=\psi_0$ in equation for $f_2$. Substituting (\ref{solA}) and (\ref{f2a}) in
  (\ref{cur}), (\ref{f2}) we find two solutions for $d_{\rm A_1}^{\pm}$:
\begin{align}
&1/d_{\rm A_1}^{2}=\psi_0^2 \left[\lambda_{\rm A_1} \pm \sqrt{\lambda_{\rm A_1}^2 +32\pi  e_{\rm *}^2 b_2}\right]\,,\label{sold}\\
&\lambda_{\rm A_1} =4\pi c_1 e_{\rm *}^2 -{2[b_2 +2\pi (C{\cal{M}}_3-c_3 e_{\rm *})^2]}/{c_1}\,.\nonumber
\end{align}
We should retain $+$-sign square root. Solution with other sign does not satisfy boundary condition $A_2^{'}(x=0)=H$.
The roots of Eq. (\ref{sold}) are positive (in accordance with the Meissner effect) for $$b_2 <0 \,\,\,{\rm at}\,\,\, -b_2-2\pi (C{\cal{M}}_3 -e_{\rm *} c_3)^2+2\pi c_1^2 e_{\rm *}^2 >0,$$ cf. condition (\ref{b2}) for neutral systems.
For $c_1
=\pm c_3$ the latter inequality is simplified as $-b_2-2\pi C^2{\cal{M}}^2 -4\pi c_1 e_{\rm *}C{\cal{M}} >0$. If the term $\propto e_{\rm *}$  is small compared to the term $\propto (-b_2-2\pi C^2{\cal{M}}^2)$, for $b_2 +2\pi C^2{\cal{M}}^2 <0$, the minimal among two  lengths, $d_{\rm A_1}^{\pm}$, becomes $d_{\rm A_1}^{+}\simeq \sqrt{c_1/(4(-b_2-2\pi C^2{\cal{M}}^2)\psi_0^2)}$.
A larger length then is $d_{\rm A_1}^{-}\simeq \sqrt{(1+ 2\pi C^2{\cal{M}}^2/b_2)/(8\pi e_{\rm *}^2 c_1\psi_0^2)}$.

To be specific let us further use that  $d_{\rm A_1}^{-}>_{\rm A_1}^{+}$. Then we may introduce the Ginzburg-Landau parameter as the ratio of the maximum among the lengths $d_{\rm A_1}^{-}$ and $d_{\rm A_1}^{+}$ to  $\xi_{\rm A_1}$, i.e.,
\begin{align}\kappa_{1,\rm A_1} =\frac{d_{\rm A_1}^{-}}{\xi_{\rm A_1}}=
\sqrt{\frac{(1+ 2\pi C^2{\cal{M}}^2/b_2)(b_1+b_2)}{4\pi e_{\rm *}^2 c_1 (c_1+c_2+c_3)}}\,.
 \end{align}
  Also, we further suppose that parameters are such that $\kappa_{1,\rm A_1}\gg 1$, cf. estimates performed below in Sect. \ref{BCS} in the BCS approximation.
 For $\kappa >1/\sqrt{2}$ the superconductor proves to be of the second-kind, cf. \cite{LP1981,Tilly-Tilly}, and with  increasing $H$ in the interval $H_{c1}^{\rm A_1}<H<H_{c2}^{\rm A_1}$ there appears a triangular Abrikosov lattice of vortices. The value $H_{c1}^{\rm A_1}\sim \frac{H_{\rm cr}}{\kappa_{1,\rm A_1}}$ is the lower critical field, such that for $H>H_{c1}^{\rm A_1}$ appearance of filament vortices  is  energetically profitable:
 $$H_{c1}^{\rm A_1}=\frac{H_{\rm cr}}{\sqrt{2}\kappa_{1,\rm A_1}}\,,\quad H_{\rm cr}=\sqrt{\frac{2\pi a^2}{b_1+b_2}}\,,$$
 where $H_{\rm cr}$ has a sense of the thermodynamical critical field, at which the Gibbs free energy of the phase with $\overline{h}=0$, $\psi =\psi_0$ coincides with that for $\overline{h}=H$, $\psi =0$. The over-line, as above, means averaging over the volume.

 To find the upper critical magnetic field one assumes $\psi $ to be tiny  and $A_2 \simeq Hx +O(\psi^2)$. As follows from Eq. (\ref{f2}), for fields nearby  $H_{c2}^{\rm A_1}$  the field $f_2$ is of the second-order smallness and can be dropped in equation for $\psi$. Then equation of motion for $\psi$ becomes
\begin{align}
&(c_1+c_2+c_3)\partial_{1}^2\psi +c_1 \widetilde{D}_{2}^2\psi
+ a\psi =0\,,\label{eqpsH2}
\end{align}
with $\widetilde{D}_{2}=\partial_2 -ie_{\rm *}Hx$, $f_2 =0$. From here
we find
 \begin{align}
H_{c2}^{\rm A_1}=\frac{a}{\sqrt{c_1(c_1+c_2+c_3)e_{\rm *}^2}}\equiv {H_{\rm cr}}{\sqrt{2}\kappa_{2,\rm A_1}}\,,
 \end{align}
for $a>0$,  being the upper critical field, at which the pairing is completely destroyed.
Here we introduced the quantity
\begin{align}
\kappa_{2,\rm A_1}
= \sqrt{\frac{b_1+b_2}{4\pi c_1 (c_1+c_2+c_3)e_{\rm *}^2 }}\,.
\end{align}
We see that $\kappa_{2,\rm A_1}\neq \kappa_{1,\rm A_1}$. For $b_2<0$ with above simplified estimate for $d_{\rm A_1}^{-}$, we find that $\kappa_{2,\rm A_1}> \kappa_{1,\rm A_1}$.

Recall that for $c_1+c_2+c_3 =0$ Eq. (\ref{eqps}) has no solution satisfying  appropriate boundary condition for $x=0$ and sub-phase A$_1$ is not realized, cf. discussion in Sect. \ref{vectorboson}.

 \subsubsection{ Instability of sub-phase A$_1$ for $b_2 >0$} For $b_2 >0$ one of the roots, $(d_{\rm A_1}^{+})^2$ or $(d_{\rm A_1}^{-})^2$, is negative that  means existence of the oscillating solution  corresponding to the penetration of the external magnetic field in the interior of the system.
Also, even in the absence of the external magnetic field  an own magnetic field $h$ is produced, as we will show.

Let us first put $H=0$ and search the fields in the form
$$A_2 (x) =h_0 k_0^{-1} \mbox{sin}(k_0x+\chi)\,, \quad f_2 =D\mbox{cos}(k_0x+\chi)\,,$$ with $h_0$ and $D$ being small constants and $\chi$ is a constant phase. Also assume that $1/k_0 \gg \xi_{\rm A_1}$. Then in Eqs. (\ref{cur}) and (\ref{f2})  we may put $\psi =\psi_0$. The spatially averaged Gibbs free energy becomes
\begin{align}
&\overline{G}^{\rm tot}_{\rm A_1}=-\frac{a^2}{4(b_1+b_2)}+\frac{c_1 e_{\rm *}^2\psi_0^2 h_0^2}{2k_0^2} +\frac{c_1 k_0^2 D^2}{2}+\frac{h^2_0}{16\pi}\nonumber\\
&+(C{\cal{M}}_3 -e_{\rm *} c_3)\psi_0 h_0 D -\frac{a D^2}{2} +(b_1-b_2)\psi_0^2 D^2 \,.\label{Ginsta1}
\end{align}
This expression can be rewritten as
\begin{align}
&\overline{G}^{\,\rm tot}_{\rm A_1}=-\frac{a^2}{4(b_1+b_2)}\label{Ginsta11}\\
&-\left[\frac{4\pi(C{\cal{M}}_3 -e_{\rm *} c_3)^2\psi_0^2}{1+8\pi c_1 e_{\rm *}^2\psi_0^2 /k_0^2}+2b_2\psi_0^2 -\frac{c_1 k_0^2}{2}\right]D^2\nonumber\\
&+\frac{1+8\pi c_1 e_{\rm *}^2\psi_0^2/ k_0^2}{16\pi }\left[h_0 +\frac{8\pi(C\mu_3 -e_{\rm *} c_3)\psi_0 D}{1+8\pi c_1 e_{\rm *}^2\psi_0^2/ k_0^2}\right]^2\,.\nonumber
\end{align}
Minimum of $\overline{G}^{\,\rm tot}_{\rm A_1}$ corresponds to
\begin{align}
h_0 =-\frac{8\pi(C{\cal{M}}_3 -e_{\rm *} c_3)\psi_0 D}{1+8\pi c_1 e_{\rm *}^2\psi_0^2/ k_0^2}\,.\label{hinsta11}
\end{align}
The occurrence of the oscillating fields is energetically profitable  provided
\begin{align}
\frac{4\pi(C{\cal{M}}_3 -e_{\rm *} c_3)^2\psi_0^2}{1+8\pi c_1 e_{\rm *}^2\psi_0^2 /k_0^2}+2b_2\psi_0^2 -\frac{c_1 k_0^2}{2}>0\,.\label{kinsta11}
\end{align}
This is so for $k_0$ varying in the range:
\begin{align}
\nu_{\mp}<k_0^2<\nu_{\pm}\,,\label{kineqinsta11}
\end{align}
the upper sign solution is here for $\nu =-\lambda_{\rm A_1} >0$ and lower sign one, for $\nu =-\lambda_{\rm A_1} <0$,
\begin{align}
\nu_{\pm}=(\nu \pm\sqrt{\nu^2+32\pi e_{\rm *}^2 b_2})\psi_0^2\,.\label{nuinsta11}
\end{align}
 Thus we have shown that for $b_2>0$  there exists an interval of values $k_0$ corresponding to the growing  fields $h$ and $f_2$. Thereby the linear approximation that we used becomes invalid. As we show below,  stable solutions then correspond to  the phases B or C.

\subsubsection{Sub-phase A$_2$} Consider the sub-phase A$_2$, where $\psi_2 =\psi (x)\neq 0$ is real. Assume that fields $A_2 (x)$ and   $\psi_1=-if_1(x)$ are small real quantities, and we assume $d_{\rm A{_2}}\gg \xi_{\rm A{_2}}$. Without  loss of the generality one may put $f_3 =0$.
Then
 the  gradient part of the Gibbs free-energy density in the quadratic approximation in perturbative fields can be presented as
 \begin{align}
&G_{\rm grad}^{\rm ch}=c_1 (\partial_1 \psi)^2+(c_1+c_2+c_3) [(\partial_1 f_1)^2 +e_{\rm *}^2 A_2^2(x)\psi^2] \nonumber\\
&+2c_3 e_{\rm *}A_2(x)\psi\partial_1 f_1\,.\label{GgradAch2}
\end{align}
As in case of the sub-phase A$_1$, the stability conditions imply that $c_1+c_2+c_3 >0$, $c_1>0$. Equations of motion for the  perturbative fields in the linear approximation become
\begin{align}
&c_1\partial_1^2\psi
+ a\psi-2(b_1+b_2)\psi^3 =0\,,\label{eqpsA2}
\end{align}
\begin{align}
&\partial_1^2 A_2 (x) -8\pi (c_1+c_2+c_3) e_{\rm *}^2 \psi^2 A_2 (x)\nonumber\\
&+8\pi (C{\cal{M}}_3 -e_{\rm *} c_3)\psi \partial_1 f_1 =0\,,\label{curA2}
\end{align}
\begin{align}
 &(c_1+c_2+c_3)\partial_1^2 f_1 +(c_3 e_{\rm *}-C{\cal{M}}_3)\psi\partial_1 A_2 \nonumber\\ &+c_3 e_{\rm *}A_2\partial_1\psi
 +(a-2\psi^2 (b_1-b_2))f_1
=0\,.\label{f2A2}
\end{align}
We may put $\psi =\psi_0$ in Eq. (\ref{f2A2}).
Solution of Eq. (\ref{eqpsA2}) reads $\psi =\psi_0 \mbox{th} [x/(\sqrt{2}\xi_{\rm A_2})]$ for $a>0$, and the coherence length $\xi_{\rm A_2} =\sqrt{c_1/a}$. Eq. (\ref{sold}) for the spectrum holds after the replacement
$c_1\leftrightarrow c_1+c_2+c_3$,
\begin{align}
&1/d_{\rm A_2}^{2}=\psi_0^2 \left[\lambda_{\rm A_2} \pm \sqrt{\lambda_{\rm A_2}^2 +32\pi  e_{\rm *}^2 b_2}\right]\,,\label{sold2}\\
&\lambda_{\rm A_2} =4\pi (c_1+c_2+c_3) e_{\rm *}^2 -\frac{2[b_2 +2\pi (C{\cal{M}}_3-c_3 e_{\rm *})^2]}{(c_1+c_2+c_3)}\,.\nonumber
\end{align}

We deal with the superconductor of the second kind for $d_{\rm A{_2}}\gg \xi_{\rm A{_2}}$. The value of the critical field $H_{c1}^{\rm A_2}= H_{\rm cr}/(\sqrt{2}\,\kappa_{1,\rm A_{2}})$, $\kappa_{1,\rm A_{2}}=d_{\rm A{_2}}/ \xi_{\rm A{_2}}$, with $d_{\rm A{_2}}$ corresponding to the maximum length among $d^{+}_{\rm A{_2}}$ and $d^{-}_{\rm A{_2}}$. Assume $c_1<c_1+c_2+c_3$. Then $\xi_{A_2}<\xi _{A_1}$. For $b_2<0$, assuming that the terms $\propto e^2_{\rm *}$ in
(\ref{sold2}) are small we find $d_{A_2}^{+}\simeq\sqrt{(c_1+c_2+c_3)/(4(-b_2-2\pi C^2{\cal{M}}^2)\psi_0^2)} >d_{A_1}^{+}\simeq\sqrt{c_1/(4(-b_2-2\pi C^2{\cal{M}}^2)\psi_0^2)} $.
 For $d_{A_2}^{-}$ we get
 $d_{\rm A_2}^{-}\simeq \sqrt{(1+ 2\pi C^2{\cal{M}}^2/b_2)/(8\pi e_{\rm *}^2 (c_1+c_2+c_3)\psi_0^2)}<d_{\rm A_1}^{-}$. Assuming that $d_{\rm A_2}^{-}>d_{\rm A_2}^{+}$ we find that the Ginzburg-Landau parameter related to the maximum among $d$-lengths is $\kappa_{\rm A_2}= \kappa_{\rm A_1}$. Also $\kappa_{2,\rm A_2}= \kappa_{2,\rm A_1}$ and $H_{c2}^{\rm A_2}=H_{c2}^{\rm A_1}$.

As the sub-phase A$_1$, the sub-phase A$_2$ proves to be unstable for $b_2>0$ in respect to the growing of the oscillating  fields $h$ and $f_1$. Equations (\ref{Ginsta1}) -- (\ref{nuinsta11}) continue to hold after the replacement $c_1\leftrightarrow c_1+c_2+c_3$.

In the particular case $c_1+c_2+c_3 =0$, the sub-phase A$_2$ for $H\parallel z$ is nonmagnetic, cf. discussion in Sect. \ref{vectorboson}.

\subsubsection{ Sub-phase A$_3$} Now consider the sub-phase A$_3$, where $\psi_3 =\psi (x)$.
In this case
 \begin{align}
G_{\rm grad}^{\rm ch}=c_1 |\partial_1 \psi|^2+c_1 e_{\rm *}^2 A_2^2(x)|\psi|^2 \,.\label{GgradAch3}
\end{align}
In the quadratic order in the perturbative fields $\psi_1 (x)$, $\psi_2(x)$ their contribution to the Gibbs free energy decouples with that for the fields $\psi_3 =\psi (x)$ and $A_2$.
The stability conditions imply that $c_1>0$. In this sub-phase we have
\begin{align}
&\xi_{\rm A_3}=\sqrt{c_1/a}\,,\,\,\,d_{\rm A_3}= 1/\sqrt{8\pi c_1 \psi_0^2 e_{\rm *}^2}\,,\nonumber\\
&\kappa_{1,\rm A_3}=\sqrt{\frac{b_1+b_2}{4\pi c_1^2 e_{\rm *}^2}}\,,\label{ksikappaA3}
\end{align}
from where it follows that $\kappa_{1,\rm A_3}=\kappa_{2,\rm A_3}$.
As above, we suppose that  $\kappa_{\rm A_3}\gg 1$ (although it is sufficient to have   $\kappa_{\rm A_3}>1/\sqrt{2}$).

Equations of motion for the fields  $\psi_1 =if_1$ and $\psi_2=if_2$ decouple in the linear approximation, e.g., we have
\begin{align}
(c_1+c_2+c_3) \partial^2_1 f_{1} +4b_2 \psi_0^2 f_{1} =0\,.
\end{align}
Equation for $f_2$ appears after the replacement $c_1+c_2+c_3\to c_1$. For $b_2 <0$ the energetically profitable solutions correspond to $f_1, f_2 =0$.

For $b_2 >0$ there are oscillating solutions indicating on  instability of the sub-phase A$_3$.

\subsubsection{Which A sub-phase is energetically most preferable? Domains}

If $0<c_1<c_1+c_2+c_3$, then $\xi_{A_3}=\xi_{A_2}<\xi _{A_1}$. Using above done estimates for $d_{A_2}$,  $d_{A_1}$  we have (for $b_2<0$) $d_{A_3}>d_{A_1}^{-}> d_{A_2}^{-}$, and the sub-phase A$_3$ proves to be  energetically favorable compared to the sub-phases A$_1$ and  A$_2$   for all $H$ at $T<T_{\rm cr}^{\rm A}$ under consideration.
Since $\kappa_{2,\rm A_3}>\kappa_{2,\rm A_1}=\kappa_{2,\rm A_2}$,
\begin{align}
H_{c2}^{\rm A_3}=H_{\rm cr}\sqrt{2}\kappa_{2,\rm A_3}=\frac{a}{c_1 |e_{\rm *}|}\,
\end{align}
is higher than $H_{c2}^{\rm A_1}=H_{c2}^{\rm A_2}$ and the sub-phases A$_1$ and A$_2$ are thus destroyed at  a smaller value of the external magnetic field compared to that for the sub-phase A$_3$.

If $0<c_1+c_2+c_3<c_1$, then $\xi _{A_1}<\xi_{A_3}=\xi_{A_2}$,  for $b_2 <0$ we have  $d_{A_1}^{-}<d_{A_3}<d_{A_2}^{-}$, and  the sub-phase A$_1$ is energetically favorable for  low $H$, then with increase of $H$ above the value  $H_{c1}$ the sub-phase A$_2$ might become preferable one and for $H$ near the value $H_{c2}^{\rm A_1}$,  the sub-phase A$_1$ again becomes most favorable.

Assume that  a  domain is in a certain  sub-phase $A_i$, with  $i =1$, either $2$ or 3. Since for $b_2 <0$ each sub-phase $A_i$  is stable to  weak perturbations, in absence of an external force  the domain  remains in the same sub-phase. In presence of the magnetic field or the rotation of the system as the whole, or due to a temperature fluctuation the domain, being in one of sub-phases, after a while  may undergo  transition to another sub-phase.

Thus we demonstrated that even, being in the mean spin-zero phase A, the spin-triplet superconductor has   unconventional properties in the presence of the external magnetic field.

\subsection{Ferromagnetic superconductive phases B and C for $b_2 >0$ in the medium filling half of space}\label{magBC}

\subsubsection{Sub-phases B$_3$ and C$_3$. General consideration}
Above on example of the sub-phase A$_1$ we have demonstrated that for $b_2>0$ the phases A$_i$ are unstable. Let $b_2 >0$, the superconductor fills half-space $x<0$ and as above assume $\vec{H}$ to be directed parallel $z$. To be specific let us focus on the consideration of the  sub-phase B$_3$ (or C$_3$), then $\vec{h}$  is directed parallel or antiparallel $z$.

The gradient contribution to the Gibbs free-energy density  (\ref{GLvector3}) can be rewritten as
\begin{align}
G_{\rm grad}^{\rm ch}&= c_1 |D_i \psi_j|^2 + \frac{c_2 +c_3}{2}[|D_i \psi_i|^2 + (D_i \psi_j)^* D_j \psi_i]\nonumber\\
&-\frac{c_3 -c_2}{2}[|D_i \psi_i|^2 - (D_i \psi_j)^* D_j \psi_i]\label{gradrewritten}
\,,
\end{align}
cf. Ref. \cite{MineevSamokhin}.
Integrating by parts the gradient term in the Gibbs free-energy, using the commutator (\ref{commut}), and retaining only the volume part of the free energy we get
\begin{align}
&\int d^3 x (G_{\rm grad}^{\rm ch}+G_{\rm hom}^{\rm ch})\label{gradc1}\\
=&\int d^3 x \left[-\frac{2c_1+c_2+c_3}{2} \tilde{\psi}^{*}(D_1^2 +D_2^2)\tilde{\psi}\right]\nonumber\\
&+\int d^3 x \left[ e_{\rm *}\frac{c_3-c_2}{2}\vec{n}_3 \vec{h}|\tilde{\psi}|^2\right]
\nonumber\\
+&\int d^3 x \left[\frac{(\vec{h}-\vec{H})^2}{8\pi}-(a+C\vec{\cal{M}}\vec{h})|\tilde{\psi}|^2 +b_1 |\tilde{\psi}|^4 +\gamma |\tilde{\psi}|^6\right]\,,\nonumber
\end{align}
where as above we have chosen simplest form of the 6-th order term and used that  $\widetilde{\psi}$ does not depend on $z$. To be specific we took  $\psi_1 =-\psi_2$ for the B$_3$ and C$_3$ sub-phases  in (\ref{psi21}).  The gradient term is positive due to the stability conditions (\ref{stabgrB}). Presence of the term $\propto i[D_1,D_2]_{-}$ in the gradient contribution to the Gibbs free-energy  resulted in appearance of the contribution
\begin{align}
\int d^3 x \delta G_{\rm intr,1}=-\int d^3 x \vec{M}_{\rm intr,1}\vec{h}|\tilde{\psi}|^2\,,\label{intrin}
\end{align}
with the quantity ${\vec{M}_{\rm intr,1}}=-\vec{n}_3 \frac{1}{2} e_{*} (c_3 -c_2)$
associated in \cite{MineevSamokhin,VolovikMineev84}
with an intrinsic magnetic moment  of the fermion pair  in the spin-triplet superconductor,    $\vec{n}_3 $ is the unit vector aligned in the $z$-direction.
In  \cite{MineevSamokhin} this contribution was considered as the total contribution to the intrinsic magnetic moment density.
However an extra contribution to the effective magnetic moment of the pair may still appear due to the presence of the terms $\propto (D_1^2 +D_2^2)$ in the Gibbs free-energy.

Varying the Gibbs free energy,
 in $\widetilde{\psi}$
 we obtain equation of motion for the order parameter
 \begin{align}
&-(c_1+\frac{c_2+c_3}{2})(D_1^2 +D_2^2)\widetilde{\psi}\label{eqmotinh}\\
&-\left[a+\vec{M}\vec{ h}\right]\widetilde{\psi}+2b_1 |\widetilde{\psi}|^2\widetilde{\psi}+3\gamma |\widetilde{\psi}|^4\widetilde{\psi} =0\,,\nonumber
\end{align}
where we introduced the quantity
 \begin{align}
\vec{M}=C\vec{\cal{M}}-\vec{n}_3 e_{\rm *}(c_3-c_2)/2\,,\label{Min}
\end{align}
$\vec{n}_3 $ is the unit vector aligned in the $z$-direction. If we used $\psi_1 =+\psi_2$ we would get expression with $-\vec{M}$ instead of $\vec{M}$.
The direction of $\vec{{{M}}}$ (the direction of the spin) is selected  to minimize the energy, cf. Eq. (\ref{eqmotinh11ch}).

We note that, {\em if we artificially suppressed  the gradient term $\propto (D_1^2 +D_2^2)$} in (\ref{gradc1}) and performed variation of the resulting Gibbs free energy in $h$ and $\widetilde{\psi}$, we  would recover (in dependence of the sign of the term $b_1-2\pi M^2$) either Eqs. (\ref{psihom}), (\ref{hhom}), (\ref{GB1}) or
Eqs. (\ref{tildepsi2g}),(\ref{tildeh2g}), (\ref{GB2g}), now with $\vec{M}$
instead of $C\vec{\cal{M}}$.

Equation (\ref{eqmotinh}) is supplemented by the Maxwell equation determining the $A_i$, $h_i$ fields:
 \begin{align}
 \partial_i F_{ik}=-4\pi J_k\,,\label{currrent}
 \end{align}
 where $\vec{J}$ is the corresponding current density, cf. Eq. (\ref{AJ}) for the case of the charged vector field.

 Multiplying (\ref{eqmotinh}) by $\widetilde{\psi}^*$ and replacing result back to the expression for the Gibbs free energy we obtain
 \begin{align}
 \int G^{\rm ch} d^3 x = \int d^3 x \left[-b_1 |\widetilde{\psi}|^4 -2\gamma |\widetilde{\psi}|^6 +\frac{(\vec{h}-\vec{H})^2}{8\pi}\right]\,.\label{freeEnBC}
 \end{align}

From (\ref{eqmotinh}) we can immediately recover the value of the upper critical field $H_{c2}$ taking $\widetilde{\psi}\to 0$. This is valid for the consideration of the B phase where the phase transition is of the second order. Neglecting $O(|\widetilde{\psi}|^2)$ terms in (\ref{eqmotinh})
and setting $\vec{h}=\vec{H}$ we get
 \begin{align}
-(c_1+\frac{c_2+c_3}{2})(D_1^2 +D_2^2)\widetilde{\psi}=E\widetilde{\psi}\,,\label{eqmotinhlin}
\end{align}
with $E= a+M_3  H$, cf. Eq. (\ref{Min}).
Directions of the fields should be chosen such that the value $H_{c2}$ be maximum.
 Eq. (\ref{eqmotinhlin}) can be interpreted as the non-relativistic Schr\"odinger equation in the homogeneous magnetic field $H$ for the particle with the mass $m=1/(2c_1+c_2+c_3)>0$ and the energy $E$. The maximum/minimum magnetic field, when there still exists/appears  the solution, corresponds to $E=E(n=0,p_z=0)=|e_{\rm *}|H_{c2}^{\rm B}/(2m)$. Thus we find
  \begin{align}
  H_{c2}^{\rm B}=-a/M_{\pm}\,.\label{hc2}
\end{align}
Here $M_{+}= C{\cal{M}}_3- e_{\rm *}(c_1+c_3)$ corresponds to $e_{\rm *}>0$, and $M_{-}= C{\cal{M}}_3 -|e_{\rm *}|(c_1+c_2)$ relates to $e_{\rm *}<0$.
For $M_{\pm}<0$ solution with $\psi \neq 0$ exists for $H<H_{c2}^{\rm B}$ at $a>0$ (i.e. for  $T<T_{\rm cr}$). For $M_{\pm}>0$ solution with $\psi \neq 0$ exists for $H>H_{c2}^{\rm B}$ at  $a<0$ (i.e. for $T_{\rm cr}^{\rm BH,CH}>T>T_{\rm cr}$), and for any $H$ at $a>0$ (i.e. for $T<T_{\rm cr}$).

Inverting Eq. (\ref{hc2}) we may find the critical temperature $T_{\rm cr}^{\rm BH}$  as a function of $H$. We see that the value of this critical temperature coincides with that follows from  Eqs. (\ref{newcrtem}) (or (\ref{newcrtem1})), but with $M_{\pm}$ instead of $|C{\cal{M}}|$, provided $M_{\pm}>0$. For $e_{\rm *}>0$ and  $c_1=-c_3$ the mentioned values of the critical temperatures coincide completely.

\subsubsection{  Sub-phases B$_3$ and C$_3$. Abrikosov ansatz}
We did not succeed to solve a general problem. Therefore let us consider the matter far from the boundary and employ the variational approach. Let the probe functions satisfy the so called Abrikosov ansatz, cf. \cite{Chernodub,Olesen},
\begin{align}
D_i\psi_i =0\,.\label{Abric}
\end{align}
As we have seen in Sect. \ref{vectorboson} in the problem of the description of the complex vector boson  fields, the condition (\ref{Abric}), cf.   (\ref{condDmu}) and (\ref{CondA}), was required to recover  correct interpretation of the single-particle problem
for $\eta =e$. Also in Sect. \ref{vectorboson} we have shown that the condition (\ref{Abric}) is fulfilled  for arbitrary  $\eta$ at the consideration of the behavior of the vector field interacting with the static uniform magnetic field at $h\simeq H_{\rm cr 2}$.
Here, in the  problem  of the spin-triplet pairing of charged fermions the fulfilment of the condition (\ref{Abric}) is not necessary even for $\eta =C{\cal{M}}=e^*$ but making use of this condition allows to  develop a variational treatment of the problem.
 Besides that, below we show that solution of Eq. (\ref{Abric}) coincides with exact solution of the problem for the value of the external magnetic field $H=H_{\rm cr 2}$.

From Eq. (\ref{Abric}) in the gauge $\vec{A}=(A_1 (y), A_2 (x),0)$ we obtain
\begin{align}
e_{\rm *} (A_2 -i A_1)=-(\partial_1 +i\partial_2) \ln {\psi}_1\,.\label{Abricsol}
\end{align}
We find
  \begin{align}
  \widetilde{\psi}=e^{-e_{\rm *}\int^x A_2(x')dx' +e_{\rm *}\int^y A_1(x')dx'}F(x+iy)\,,\label{anss2}
   \end{align}
   where $F$ is an arbitrary analytical function. On the other hand, from (\ref{Abricsol})  we find
\begin{align}\frac{1}{2}\epsilon_{ki}\partial_i |\tilde{\psi}|^2 + |\tilde{\psi}|^2\partial_k\chi=e_{*}A_k |\tilde{\psi}|^2\,,\quad i,k=1,2\,,
\end{align}
$\epsilon_{12}=1$, $\epsilon_{21}=-1$, $\epsilon_{11}=\epsilon_{22}=0$, $\tilde{\psi}=|\tilde{\psi}|e^{i\chi}$,
and for simplicity choosing $\chi =0$ we get
\begin{align}
-(\partial_x^2 +\partial_y^2)\ln |\widetilde{\psi}|=e_{\rm *} \vec{h}\vec{n}_{3}\,.\label{anss1}
\end{align}

 Using Eqs. (\ref{commut}) and (\ref{Abric}) we derive a helpful relation
\begin{align}
-(D_1^2 +D_2^2)\tilde{\psi}=e_{\rm *}\vec{h}\vec{n}_{3}\tilde{\psi}\,,\label{Dpsi}
\end{align}
cf. Eq. (\ref{eqmotinh11ch}) in Sect. \ref{vectorboson}.

For the current from (\ref{gradc1}) and (\ref{currrent}) using (\ref{Abric}) we obtain
  \begin{align}
 J_k =-\widetilde{M}_3\epsilon_{ki}\partial_i |\widetilde{\psi}|^2\,.
\end{align}
Here we introduced the effective magnetic moment of the Cooper pair
\begin{align}
\vec{\widetilde{M}} =C\vec{\cal{M}}+\vec{M}_{\rm intr}\,,\label{mut}
\end{align}
where $\vec{M}_{\rm intr}=-\vec{n}_3 e_{\rm *}(c_1 +c_3)$
is an intrinsic magnetic moment of the fermion pair, which however differs from the contribution $\vec{M}_{\rm intr,1}$, cf. (\ref{intrin}).

Replacing (\ref{Dpsi}) in (\ref{eqmotinh}) we find
\begin{align}
\vec{\widetilde{M}}\vec{h}=-a+2b_1 |\tilde{\psi}|^2 +3\gamma |\tilde{\psi}|^4\,.\label{hAbr}
\end{align}
Setting (\ref{Dpsi}) in the gradient term in (\ref{gradc1}) we get
\begin{align}
\int d^3 x G_{\rm grad}^{\rm ch}&=\int d^3 x e_{\rm *}\vec{h}\vec{n}_{3}(c_1 +c_3)|\tilde{\psi}|^2\,\label{gradc}
\end{align}
and
\begin{align}
&\int d^3 x G^{\rm ch}\label{totG}\\&=\int d^3 x \left[-(a+\vec{\widetilde{M}}\vec{{h}})|\tilde{\psi}|^2 +b_1 |\tilde{\psi}|^4 +\frac{(\vec{h}-\vec{H})^2}{8\pi}+\gamma |\tilde{\psi}|^6\right]\,.\nonumber
\end{align}
Since the gradient contribution to the Gibbs free-energy should be non-negative our result is valid only  provided the stability condition $(c_1 +c_3)e_{\rm *}{h}_3\geq 0$ is fulfilled.

Minimizing (\ref{totG}) in $h$ we find  the  solution
\begin{align}
\vec{h}=\vec{H}+4\pi \vec{\widetilde{M}}|\tilde{\psi}|^2\,.\label{hsolut}
\end{align}
Note that in general (for $H\neq H_{\rm cr 2}$) the  ansatz (\ref{Abric}) is incompatible with one of the equations of motion, which follow from the minimization of (\ref{totG}) in the order parameter and the electromagnetic field. Indeed, setting in (\ref{totG})  solution (\ref{eqmotinh}), where we  substitute Eq. (\ref{Dpsi}),  in the limit $\gamma \to 0$ in dependence of the sign of $b_1-2\pi \widetilde{M}^2$ we recover  either Eqs. (\ref{psihom}), (\ref{hhom}), (\ref{GB1}) or
Eqs. (\ref{tildepsi2g}),(\ref{tildeh2g}), (\ref{GB2g}), however now with $\widetilde{M}$ from (\ref{mut}) instead of $C{\cal{M}}$.

{\em Only for $H\simeq H_{c2}$ ansatz (\ref{Abric}) is compatible with the solution (\ref{eqmotinh}).}
An analogy of Eq. (\ref{Dpsi}) with the Schr\"odinger equation in a uniform magnetic field (at  $\vec{h}\simeq \vec{H}$ for $\tilde{\psi}\to 0$) demonstrates that the solutions with appropriate boundary condition $|\tilde{\psi}(x,y\to \infty)|<\infty$
exist   provided $e_{\rm *}\vec{h}\vec{n}_{3}=|e_{\rm *}|H_{\rm cr 2}>0$, i.e. for $e_{\rm *}>0$.
Otherwise Abrikosov ansatz cannot be exploited.

Let us employ the variational procedure.
After substitution of  $\vec{h}$ from (\ref{hsolut}) into (\ref{anss1})
the equation for $\widetilde{\psi}$ gets the form:
\begin{align}
-(\partial_x^2 +\partial_y^2)\ln |\widetilde{\psi}|=e_{\rm *} \vec{H}\vec{n}_3 + e_{\rm *}4\pi \vec{\widetilde{M}}\vec{n}_3|\tilde{\psi}|^2\,.\label{anss11}
\end{align}
For example in case $H=0$, the solution of this equation  with periodic boundary conditions is given by the Weierstrass doubly periodic function $\zeta$, cf. \cite{Olesen},
\begin{align}
&&|\tilde{\psi}|=\frac{|\zeta{'}(x+iy)|}{\sqrt{\pi |\widetilde{M} e_{\rm *}| (e_2 -e_3) (e_3 -e_1)}}\nonumber\\
&&\times\frac{(e_3 -e_1)(e_2 -e_3)}{(e_3 -e_1)(e_2 -e_3)+|\zeta(x+iy)-e_3|^2}\,,\label{Weier}
\end{align}
$e_i$ are the roots of equation
\begin{align}
4t^3 -g_2 t -g_3 =0\,,
\end{align}
where the quantities $g_2$, $g_3$ are defined in the standard presentations of the  Weierstrass p-function. We assume
that  these roots are real (that requires $g^3_2 - 27g^2_3 > 0$) and $e_2 > e_3 > e_1$. Other forms of the
solution can be found in \cite{Akerbloom,Doorselai}. If in Eq. (\ref{Weier})
periods of $\zeta$ are $2a$, $2ib$, then $|\tilde{\psi}|$ is periodic function with periods $a$, $ib$.

Now we substitute solution (\ref{hsolut}) in (\ref{totG}). With the solution of Eq. (\ref{anss1})   presented in the form $\tilde{\psi}=\psi_0 \nu (\vec{r})$ we get
\begin{align}
&\overline{G^{\rm ch}}
=-(a+\vec{\widetilde{M}} \vec{H})\overline{\nu^2}|\psi_0|^2 \nonumber\\
&+(b_1\overline{\nu^4} -2\pi \widetilde{M}^2(\overline{\nu^2})^2)|\psi_0|^4 +\gamma \overline{\nu^6}|\psi_0|^6\,.\label{Gtotvar}
\end{align}
Here spatial averaging, $\overline{G^{\rm ch}}=\int d^3 x G^{\rm ch}/\int d^3 x$, is performed with the probed function satisfying Eq. (\ref{anss1}). For $H=0$ we may use solution (\ref{Weier}).
Variational parameter $\psi_0$ is found by minimization of  (\ref{Gtotvar}). We obtain
\begin{align}
&|\psi_0^2| =\frac{(2\pi \widetilde{M}^2(\overline{\nu^2})^2-b_1\overline{\nu^4})}{3\gamma \overline{\nu^6}}\label{sqrtC}\\
&\pm \frac{1}{3\gamma \overline{\nu^6}}\sqrt{(2\pi \widetilde{M}^2(\overline{\nu^2})^2-b_1\overline{\nu^4})^2+3\gamma \overline{\nu^6}(a+\vec{\widetilde{M}}\vec{H})\overline{\nu^2}}\,.\nonumber
\end{align}
For the  probed function describing the periodic  triangular lattice at ordinary spin zero pairing one has \cite{Tilly-Tilly}
$\widetilde{\beta}=\overline{\nu^4}/(\overline{\nu^2})^2 \simeq 1.16$.

In absence of the external magnetic field  the  system of a large size  may exist in a metastable state, being constructed of domains with different directions of $\vec{h}$ and $\vec{\psi}$ in each domain. Since the ground state of the uniform system corresponds to   $\vec{h}$ aligned in one fixed  direction,  the system may undergo transitions with the flip of the domains until it will reach the  state with the minimal surface energy. Note that the process of the alignment of domains should be compatible with the conservation of the magnetic flux. As we have argued, when considered the  B$_i$, C$_i$ phases in neutral superfluids, the spin and the $\vec{h}$ flips   require an energy.
 In the presence of the external magnetic field or for the rotating system a required extra energy can be taken from the energy of the external magnetic and rotation fields. Also flips of the domains are possible via thermal fluctuations.

\subsubsection{ Sub-phase B$_3$. Averaged Gibbs free energy}
Let us focus on the sub- phase B$_3$. In case $H=0$ results are valid also for sub-phases B$_{1,2}$.
Within the variational problem, the B-phase arises provided
\begin{align}
b_1 \tilde{\beta} -2\pi \widetilde{M}^2 >0\,,\label{condBcha}
\end{align}
where as above $\tilde{\beta}=\overline{\nu^4}/(\overline{\nu^2})^2$.
We may for simplicity put $\gamma =0$.

Assume
$(a+\vec{\widetilde{M}}\vec{H})>0$. From (\ref{sqrtC}) we find
\begin{align}
|\psi_0|^2 =\frac{a+\vec{\widetilde{M}} \vec{H}}{2\overline{\nu^2}(b_1 \tilde{\beta} -2\pi \widetilde{M}^2)}\theta \left(a+\vec{\widetilde{M}} \vec{H}\right)\,,\label{psi0av}
\end{align}
\begin{align}
\overline{G}^{\rm ch}_{\rm B_3H} =-\frac{(a+\vec{\widetilde{M}} \vec{H})^2}{4(b_1 \tilde{\beta} -2\pi \widetilde{M}^2)}<0\,,
\end{align}
with $h$ from (\ref{hsolut}). Energetically favorable is the direction of the vector $\vec{\widetilde{M}}$  parallel $\vec{H}$. Thereby we may replace $\vec{\widetilde{M}} \vec{H}$ to $|\vec{\widetilde{M}} \vec{H}|$.  The sub-phase B$_3$ appears for $H=0$ by the second-order phase transition at $T=T_{\rm cr}$  and continues to exist in a certain interval of temperatures above $T_{\rm cr}$ for $H\neq 0$. The value of the new critical temperature $T_{\rm cr}^{\rm B_3H}$ is found from  Eqs. (\ref{newcrtem}), (\ref{newcrtem1}), but with $\widetilde{M}$ from (\ref{mut}) instead of $C{\cal{M}}$. For example, with the parametrization $a=\alpha_0 t$ we get
\begin{align}\label{newcrtemch}
{T}_{\rm cr} ^{\rm B_3H}=T_{\rm cr} (1+|\widetilde{M} H|/\alpha_0)\,.
\end{align}

\subsubsection{ Sub-phase C$_3$. Averaged Gibbs free energy}
Consider sub-phase C$_3$. For $H=0$ results are also valid  for sub-phases C$_{1,2}$.
Now we set
\begin{align}
b_1 \widetilde{\beta} -2\pi \widetilde{M}^2 <0\,.
\end{align}

To get stable solutions we should retain $\gamma\neq 0$ term in (\ref{Gtotvar}). As above, simplifying consideration we assume $\gamma$ to be  positive and  small. Then from (\ref{sqrtC}) in
analogy with (\ref{tildepsi2g}), (\ref{GB2g})  we obtain
\begin{align}
{\psi}^2_0  \simeq\frac{2(2\pi \widetilde{M}^2 -b_1\widetilde{\beta})}{3\gamma\widetilde{\beta}_1}
+
\frac{(a+\vec{\widetilde{M}}\vec{H})\widetilde{\beta}_2}{2\widetilde{\beta}_1(2\pi \widetilde{M}^2 -b_1\widetilde{\beta})}>0\,,\label{tildepsi2gch}\end{align}
\begin{align}
\overline{G}^{\rm ch}_{\rm C_3H} & \simeq -\frac{4}{27\gamma^2\widetilde{\beta}_2^2}(2\pi \widetilde{M}^2 -b_1\widetilde{\beta})^3\,.\label{GB2gch}
\end{align}
Here $\widetilde{\beta}_1 =\overline{\nu^6}/(\overline{\nu^2})^2$, $\widetilde{\beta}_2 =\overline{\nu^6}/(\overline{\nu^2})^3$.
Expansion in the parameter $\gamma$ is valid for
\begin{align}
0<\gamma \ll \frac{(2\pi \widetilde{M}^2 -b_1\widetilde{\beta})^2}{\widetilde{\beta}_2 (a+|{\widetilde{M}} H|)}\, .
\label{gamm}
\end{align}
The own magnetic field is found with the help of  Eqs. (\ref{hsolut}), (\ref{tildepsi2gch}). The new phase appears by the first-order phase transition.

The new critical temperature is determined (for $a=\alpha_0 t$) by setting zero the square root in (\ref{sqrtC}):
\begin{align}
{T}_{\rm cr}^{\rm C_3H} =T_{\rm cr} \left[1+\frac{(2\pi \widetilde{M}^2 -b_1\widetilde{\beta})^2}{3\gamma \widetilde{\beta}_2 \alpha_0}+\frac{|\widetilde{M}H|}{\alpha_0}\right]\,,\label{tcC1}
\end{align}
with ${T}_{\rm cr}^{\rm C_3H}>{T}_{\rm cr}^{\rm C_3}>T_{\rm cr}$, where now
$${T}_{\rm cr}^{\rm C_3} =T_{\rm cr} \left[1+\frac{(2\pi \widetilde{M}^2 -b_1\widetilde{\beta})^2}{3\gamma \widetilde{\beta}_2 \alpha_0}\right]\,.$$

\section{3P$_2$ $nn$ and $pp$ pairings in neutron star interiors}\label{nnpairing}

\subsection{Gibbs free-energy density}
So far we considered the spin-triplet paring in systems with negligible spin-orbital interactions, so that both orbital momentum and spin were assumed to be appropriate quantum numbers and we assumed that orbital momentum and spin can rotate independently. In nuclear matter the spin-orbital interaction is strong and the state of a Cooper pair is described by the total angular momentum $J$ and its projections $m_J$.  The  3P$_2$ phase shift for identical nucleons ($nn$ and $pp$) is the largest among others for the momenta $p>1.3$ fm$^{-1}$.  Thereby, cf. ~\cite{HGRR1970},  for $n \gsim n_0$
 neutrons  in the neutron matter as well as in the  beta-equilibrium matter prove to be paired in the 3P$_2$ state  with $J=2$. Protons might be paired in this channel at a higher density, if their fraction becomes rather high.

The pairing gap of the 3P$_2$ state can be written as
$ \hat{\Delta}=  i \sigma_i \sigma_2 A_{ij}{{n}_j}$, where  $\sigma_{1,2,3}$ are the Pauli spin matrices,  $\vec{n}$ is the unity vector in the direction of the pairing momentum. The matrix $\widehat{A}$ is symmetric and traceless for this type of paring and is determined by the expression~\cite{SS1978} (here presented in another normalization, a more convenient one to compare with results of previous sections)
\begin{align}
\widehat{A}= \left[
\begin{array}{ccc}
\frac{a_{-2}}{2}-\frac{a_{0}}{\sqrt{6}}+\frac{a_{2}}{2}
 & \frac{i}{2}  (a_{2}-a_{-2}) & \frac{1}{2} (a_{-1}-a_{1}) \\
 \frac{i}{2}  (a_{2}-a_{-2}) &
 - \frac{a_{-2}}{2}- \frac{ a_{0}}{\sqrt{6}}- \frac{a_{2}}{2} & -\frac{i}{2}  (a_{-1}+a_{1}) \\
 \frac{1}{2} (a_{-1}-a_{1}) & -\frac{i}{2} (a_{-1}+a_{1}) & \sqrt{\frac{2}{3}} a_{0}
\end{array}
\right]\,.\label{Amatrix}
\end{align}
The Ginzburg-Landau free-energy density functional for the uniform matter has the form
\begin{align}
\label{FA}
F[\widehat{A}] &=
-\bar{\alpha}\, {\rm Tr}(\widehat{A}\,\widehat{A}^*)
+ \bar{\beta}_1\, {\rm Tr}(\widehat{A}\,\widehat{A})\, {\rm Tr}(\widehat{A}^*\,\widehat{A}^*)
\nonumber\\
&+ \bar{\beta}_2\, {\rm Tr}(\widehat{A}\,\widehat{A}^*)\, {\rm Tr}(\widehat{A}\,\widehat{A}^*) +
\bar{\beta}_3\, {\rm Tr}(\widehat{A}\,\widehat{A}\,\widehat{A}^*\,\widehat{A}^*)
\nonumber\\
&+\{\bar{\gamma} \widehat{A}^6\}\,.
\end{align}
The last term, $\{\bar{\gamma} \widehat{A}^6\}$, represents symbolically all terms of the sixth order in $A$. Below we put $\bar{\gamma} =0$, when it does not contradict to the stability condition of the phase. Values $\bar{\alpha}$, $\bar{\beta}_i$ are phenomenological parameters of the model. Assuming (for $\gamma =0$) a second-order phase transition to the paired state, in absence of the external fields one may use $\bar{\alpha}=\bar{\alpha}_0\,  t$ for $|t|\lsim 1$, cf. Eq. (\ref{aLand}). As we have mentioned, being computed in BCS approximation, the $\gamma_6 \widehat{A}^6$ term proves to be  negative \cite{SS1978} that implies necessity to continue the Ginzburg-Landau expansion up to $\gamma_8 \widehat{A}^8$ positive contribution \cite{Yasui:2019unp}. Simplifying consideration, as in previous sections, we will employ the simplest form of the $\{\bar{\gamma} \widehat{A}^6\}$ interaction with
$\gamma >0$.

To consider systems of a finite size we should add the gradient contribution to the free-energy density.
 The generalization to the hypothetical 3P$_2$ $pp$ pairing, which  may be possible for $n\gg n_0$ in neutron star matter, is performed with the help of the replacement of the ordinary derivatives by the long derivatives, i.e. $\partial_i \to D_i =\partial_i +ie_{\rm *} A_i +m_{\rm *}v_i$, $A_i =(A_x, A_y, A_z)$, $e_{\rm *}$ is the charge of the fermion pair, for moving systems $\vec{v}$ is the velocity of the system, $m_{\rm *}$ is the effective mass of the pair.
 Therefore to include the effects associated with the spatial  non-uniformity one should add the gradient terms
\begin{align}
&F_{\rm grad}=c_1 D_i A_{\nu k} D_i^* A^{*}_{\nu k}+c_2 D_i A_{\nu i} D_j^* A^{*}_{\nu j}\\ &+c_3 D_i A_{\nu j} D_j^* A^{*}_{\nu i}\,.\nonumber
\end{align}

 To include interaction of spins of the Cooper pair  with the own magnetic field $\vec{h}$
we add to Eq.~(\ref{FA}) the Zeeman term \cite{MineevSamokhin,AM1973}, $F_{\rm Zeeman}= -\vec{\eta}\vec{h}=-i\eta h_i\epsilon_{ijk} A_{lj}\,A_{lk}^*$.  Also the proper magnetic free-energy density contribution  should be added. To be specific we further assume $\vec{h}=(0,0,h)$, $\vec{h}\parallel \vec{H}$, $\vec{h}\parallel \vec{{\eta}}$ (for $\eta >0$) or $\vec{h}\parallel -\vec{{{\eta}}}$ (for $\eta <0$). Other possibilities can be considered similarly to that we did in Sect. \ref{vectorboson}.
Thus the resulting expression for the Gibbs free-energy density becomes
\begin{align}
G &= F_{\rm grad}[{A}_{ij}, h, \omega] +F[{A}_{ij}]+G_{H},
\label{FA1}\\
G_{H} &= -i\,\eta\, h_{i}
\epsilon_{ijk} A_{lj}\,A_{lk}^* + {\textstyle \frac{1}{8\pi}} (h-H)^2
\nonumber\\
 &= {\textstyle \frac{1}{2}}\,\eta\, h
\big(2\, |{a}_{-2}|^2 + |{a}_{-1}|^2 - |{a}_{1}|^2 - 2\, |{a}_{2}|^2\big)
\nonumber\\
&+ {\textstyle \frac{1}{8\pi}} (h-H)^2 \,.\nonumber
\end{align}
As above, we for simplicity  disregard small polarization terms $\propto h^2$, cf. \cite{MerminStar}.

{\em If we retain only one $m_J$-component} among possible combinations $m_J=0$,  $-1$,  $-2$, $+1$ or $+2$
in matrix (\ref{Amatrix}), the Gibbs free-energy
densities for these states become (for $m_J=0, \pm 1, \pm 2$):
\begin{align}
G_0 = &-  \bar{\alpha}\,|a_{0}|^2 +
(\bar{\beta}_1+\bar{\beta}_2 +\half \bar{\beta}_3)\, |a_{0}|^4\label{G012}
\\
&+ {\textstyle \frac{1}{8\pi}} (h-H)^2
+F_0^{\rm grad}\,,
\nonumber\\
G_{\pm 1} = &- \bar{\alpha} \,|a_{\pm 1}|^2 +(\bar{\beta}_2 +{\textstyle\frac14}
\bar{\beta}_3)|a_{\pm 1}|^4
\mp \half\,\eta\, h  |a_{\pm 1}|^2\nonumber\\
&+ {\textstyle \frac{1}{8\pi}} (h-H)^2
+F_{\pm 1}^{\rm grad}\,,
\nonumber\\
G_{\pm 2} = &- \bar{\alpha}\,|a_{\pm 2}|^2
+ \bar{\beta}_2\, |a_{\pm 2}|^4\mp\,\eta\, h |a_{\pm 2}|^2\nonumber\\
&+ {\textstyle \frac{1}{8\pi}} (h-H)^2
+F_{\pm 2}^{\rm grad}\,.\nonumber
\end{align}

If  one assumes the symmetry among all $a_{m}$ and $a_{-m}$ amplitudes and takes into account the relations  $a_{\pm 2}=\pm\tilde{a}_2\,e^{\pm i\chi_2}$ and $a_{\pm 1}=\pm\tilde{a}_1\,e^{\pm i\chi_1}$ with real amplitudes $\tilde{a}_2$ and $\tilde{a}_1$,
the Gibbs functional $G$ in such a symmetric sub-phase simplifies as
\begin{align}
G_{\rm sym} &= -\bar{\alpha}\,[\tilde{a}_0^2+2(\tilde{a}_{1}^2+\tilde{a}_{2}^2)]
\label{Gsym}\\
&+(\bar{\beta}_1 + \bar{\beta}_2+\half \bar{\beta}_3)
\big[\tilde{a}_0^2+2 (\tilde{a}_1^2+\tilde{a}_2^2)\big]^2\nonumber\\
&+ {\textstyle \frac{1}{8\pi}} (h-H)^2
+F_{\rm sym}^{\rm grad}\,,
\end{align}
yielding in the case of the uniform matter  the same value in the minimum as for the $G_0$, cf. Eq. (\ref{Gsym1}) below.

Note that the critical temperatures for the symmetric sub-phase and the sub-phases $m_J=0$, $m_J=\pm 1$ and $m_J=\pm 2$, respectively, might be different. However according to  \cite{ChenLi} the difference proves to be very small. Thereby, simplifying consideration  we suppose, as we have used it in previous sections, that values $T_{\rm cr}$ are the same for all the sub-phases.

Assume
\begin{align}
\bar{\beta}_1+\bar{\beta}_2+\frac{1}{2}\bar{\beta}_3 >0, \,\,\, \bar{\beta}_2+\frac{1}{4}\bar{\beta}_3 >0,\,\,\,
\bar{\beta}_2 >0 \,,\label{stc}
\end{align}
that is required for the stability of the symmetric and $m_J=0$ sub-phases, the sub-phases with $m_J=\pm 1$ and the sub-phases with $m_J=\pm 2$, respectively. If we put $H=0$ and disregard $h $-dependent terms for a moment, then for the uniform matter we find that for $\bar{\beta}_1 + \frac{1}{2} \bar{\beta}_3 <0$ the
symmetric sub-phase (and the sub-phase with $m_J=0$) is energetically preferable compared to the
sub-phases with $m_J=\pm 2$. For $\bar{\beta}_1 + \frac{1}{4} \bar{\beta}_3
<0$ the former sub-phases are favorable compared to the sub-phase with $m_J=\pm 1$.
For $\bar{\beta}_1 + \frac{1}{2} \bar{\beta}_3 >0$,  the sub-phase with $m_J =\pm 2$ is energetically preferable compared to the symmetric and $m_J=0$ sub-phases and compared to the $m_J=\pm 1$ sub-phases provided
simultaneously $ \bar{\beta}_3>0$, whereas for $ \bar{\beta}_3<0$ the   $m_J=\pm 1$ sub-phases are favorable. In the BCS weak-coupling approximation \cite{RS1976,SS1978} one has $\bar{\beta}_1=0$, $\bar{\beta}_2 =-\bar{\beta}_3 >0$. In this case the symmetric and $m_J=0$ sub-phases prove to be energetically favorable.

To consider finite systems we should include contributions $F^{\rm grad}$. With taking into account these terms degeneracy of the sub-phases 3P$_2$(0) and 3P$_2(\rm sym)$  disappears.
For the matter filling the semi-infinite space $x<0$ in the gauge where ${h}_3=\partial_1 A_2$, $h_1=h_2=0$, for the 3P$_2$(0) sub-phase we obtain
\begin{align}
F^{\rm grad}_0 = (c_1 +\frac{c_2+c_3}{6})[|\partial_1 a_0|^2 +e_{\rm *}^2 A_2^2 |a_0|^2]\,,\label{grad0}
\end{align}
cf. (\ref{GgradAch}), (\ref{GgradAch2}), (\ref{GgradAch3}),
\begin{align}
F^{\rm grad}_{\pm 1} &= (c_1+\frac{c_2+c_3}{4})[|\partial_1 a_{\pm 1}|^2 +e_{\rm *}^2 A_2^2 |a_{\pm 1}|^2]\nonumber\\
&\pm \frac{c_2-c_3}{2}e_{\rm *} A_2 \partial_1 |a_{\pm 1}|^2\,,\label{grad1}
\end{align}
\begin{align}
F^{\rm grad}_{\pm 2} &= (c_1 +\frac{c_2+c_3}{2})[|\partial_1 a_{\pm 2}|^2 +e_{\rm *}^2 A_2^2 |a_{\pm 2}|^2]\nonumber\\
&\pm \frac{c_2-c_3}{2}e_{\rm *} A_2 \partial_1 |a_{\pm 2}|^2\,,\label{grad2}
\end{align}
cf. (\ref{gradc1}).

Difference in the volume and surface  energies for various sub-phases leads to a possibility of domains, see in Sect. \ref{vectorboson}.

\subsection{Sub-phases 3P$_2$(0) and 3P$_2(\rm sym)$ of  $nn$ pairing in  external uniform static magnetic field}\label{3p2A}

Expressions for the Gibbs free-energy densities for the symmetric sub-phase and the $m_J=0$ sub-phase are similar to those for the phase A at the pairing of the neutral fermions considered above in Sect.~\ref{DF}.

Let  $T<T_{\rm cr}$ and $\bar{\beta}_1+\bar{\beta}_2+\frac{1}{2}\bar{\beta}_3 >0\,.$ The order parameter in sub-phases 3P$_2$(0) and 3P$_2(\rm sym)$ of $nn$ pairing decouples with the magnetic field. Thereby  we get $\vec{h}=\vec{H}$.
The order parameters $\tilde{a}_0^2+2(\tilde{a}_{1}^2+\tilde{a}_{2}^2)$ and $|a_0|^2$  are found by the minimization of $G_{\rm sym}$ and $G_0$, respectively. In case of the infinite matter in the minimum we get
\begin{align}
|a_0|^2=\tilde{a}_0^2+2(\tilde{a}_{1}^2+\tilde{a}_{2}^2)=\frac{\bar{\alpha}\theta(t)}{2(\bar{\beta}_1 + \bar{\beta}_2+\frac{1}{2} \bar{\beta}_3)}\,,\label{Aa3P2}
\end{align}
 and
\begin{align}
G_{0}^{\rm hom}=G_{\rm sym}^{\rm hom}=-\frac{\bar{\alpha}^2\theta(t)}{4(\bar{\beta}_1 + \bar{\beta}_2+\frac{1}{2} \bar{\beta}_3)}\,.
\label{Gsym1}
\end{align}
Thus for  the uniform matter these sub-phases   prove to be  degenerate. With a decreasing temperature they appear at $T=T_{\rm cr}$ by the second order phase transition and exist for $T<T_{\rm cr}$.
The order parameter and external magnetic field decouple. The sub-phases are stable respectively transitions to $m_J=\pm 2$ sub-phases provided $\bar{\beta}_1 + \frac{1}{2} \bar{\beta}_3 <0$ and to $m_J=\pm 1$ sub-phases for $\bar{\beta}_1 + \frac{1}{4} \bar{\beta}_3 <0$.

\subsection{Sub-phases 3P$_2$ $(\pm 2)$-B and 3P$_2$ $(\pm 2)$-C of $nn$ pairing in external uniform magnetic field}

The problem is reduced to that considered above  in Sect.~\ref{DF} on example of the phases B and C for the vector order parameter, provided one puts now $\widetilde{\psi} =a_{+2}$ or $\widetilde{\psi} =a_{-2}$.
 We   label the phase ``B" provided $\bar{\beta}_2 -2\pi{\eta}^2 >0$ and ``C", if $\bar{\beta}_2 -2\pi{\eta}^2 <0$.

 \subsubsection{ Sub-phases 3P$_2$ $(\pm 2)$-B$_3$}
 Let us focus on the  3P$_2$ $(\pm 2)$-B sub-phases.
  For $\vec{h}$ and $\vec{\eta}$ directed parallel or antiparallel to $\vec{H}$ we deal with the sub-phase B$_3$. Consider case of the infinite matter. Minimization of Eq.~(\ref{G012}) yields  in case of the phase $3P_2(\pm 2)$-B$_3$:
\begin{align}\label{ha2}
{h}=H\mp 4\pi \eta|a_{\pm 2}|^2
\,,
\end{align}
cf. (\ref{hhom}), and
\begin{align}
&&|a_{\pm 2}^{\rm B_3H}|^2=\frac{(\bar{\alpha}\pm{\eta}{H})\theta(\bar{\alpha}\pm{\eta}{H})}{2(\bar{\beta}_2 -2\pi{\eta}^2)}\,,\label{3p2ord}\\
\,\,
&&G^{\rm B_3H}_{\pm 2} =-\frac{(\bar{\alpha}\pm {\eta}{H})^2\theta(\bar{\alpha}\pm{\eta}{H})}{4(\bar{\beta}_2 -2\pi{\eta}^2)}\,,
\label{G23P2B}
\end{align}
for $\bar{\alpha}\pm {\eta}{H} >0$,
and for $\bar{\gamma} \to 0$, cf. (\ref{psihom}), (\ref{GB1}). Thus even for $H=0$  in this sub-phase there appears the internal magnetic field $h(H=0)$.

The critical temperature found from the condition $\bar{\alpha}\pm {\eta}{H}=0$ is shifted up in presence of the external field $H$ and the new critical temperature equals to
\begin{align}
{T}_{\rm cr}^{\rm B_3H} =T_{\rm cr} (1\pm \eta H/\bar{\alpha}_0)\,,\label{TcB}
 \end{align}
provided $\bar{\alpha}=\bar{\alpha}_0 t$, cf. Eq. (\ref{newcrtem}). For $\eta >0$ the state $m_J=+2$ is profitable and for $\eta <0$ the state $m_J=-2$.

At $T<T_{\rm cr}$ for $\eta >0$ solutions exist at
  arbitrary $H$ for the state $m_J =+2$ and at  $H<H_{\rm cr 2}=-\bar{\alpha}/\eta$ they exist provided  $\eta <0$. For $\eta <0$    solutions exist at arbitrary $H$ for $m_J=-2$ and they exist for  $H<H_{\rm cr 2}=\bar{\alpha}/\eta$ provided $\eta >0$.

  For ${T}_{\rm cr}^{\rm B_3H}>T>T_{\rm cr}$ solutions exist for $\eta >0$ at $H>H_{\rm cr 2}=-\bar{\alpha}/\eta >0$ for the state $m_J =2$  and at   $H>H_{\rm cr 2}=\bar{\alpha}/\eta >0$ for $\eta <0$ for the state $m_J=-2$.

\subsubsection{ Sub-phases 3P$_2$ $(\pm 2)$-C$_3$}
In the sub-phase 3P$_2(\pm 2)$-C$_3$  for small $\bar{\gamma}>0$
we find
\begin{align}\label{aB1}
|a_{\pm 2}^{\rm C_3H}|^2\simeq\frac{2(2\pi{\eta}^2-\bar{\beta}_2)}{3\bar{\gamma}}+
\frac{\bar{\alpha}\pm \eta H}{2 (2\pi{\eta}^2-\bar{\beta}_2)}\,,
 \end{align}
\begin{align}
G_{\pm 2}^{\rm C_3H} \simeq -\frac{4(2\pi{\eta}^2-\bar{\beta}_2)^3}{27\bar{\gamma}^2}\,,
 \end{align}
 cf. Eq. (\ref{GB2g}).
The critical temperature is increased in presence of the external magnetic field $H$ and the new critical temperature is as follows:
\begin{align}
{T}_{\rm cr}^{\rm C_3H} =T_{\rm cr} \left(1 +\frac{(2\pi\eta^2-\bar{\beta}_2)^2}{3\bar{\gamma}\bar{\alpha}_0 }+\frac{|\eta| H}{\bar{\alpha}_0}\right)\,,
\end{align}
provided $\bar{\alpha}=\bar{\alpha}_0 t$, cf. Eq. (\ref{TcC}).

\subsubsection{Sub-phases 3P$_2(\pm 1)$-B and 3P$_2(\pm 1)$-C of $nn$ pairing }
Expressions for $m_J=\pm 1$ can be found from those for $m_J=\pm 2$ with the help of the replacements $\bar{\beta}_2 \to \bar{\beta}_2 +\frac{1}{4}\bar{\beta}_3$, $\eta\to\frac{1}{2}\eta$, cf. Eqs. (\ref{G012}).

\subsection{Sub-phases of 3P$_2$  $pp$ pairing} As above consider medium filling half-space $x<0$ under the action of the external uniform magnetic field $\vec{H}\parallel z$.
Our consideration is completely similar to that performed in Sect. \ref{charged}.

\subsubsection{Sub-phases 3P$_2$(0) and 3P$_2(\rm sym)$}
Penetration of the external static  magnetic  field in case of the 3P$_2$(0) and 3P$_2(\rm sym)$  sub-phases is described similar to that for the A phase  in the  superconducting matter described by the vector order parameter in Sect.~\ref{Ach}.
Using (\ref{G012}), (\ref{grad0}) for simplicity at $\bar{\gamma}\to 0$
we obtain
\begin{align}
&(c_1+\frac{c_2+c_3}{6})[\partial_1^2 a_0 -e_{\rm *}^2 A_2^2 a_0] +\bar{\alpha}a_0  \nonumber\\ &-2(\bar{\beta}_1+\bar{\beta}_2 +\frac{1}{2} \bar{\beta}_3) |a_0|^2 a_0 =0\,,
\end{align}
\begin{align}
\partial_1^2 A_2 -8\pi e_{\rm *}^2 (c_1 +\frac{c_2+c_3}{6}) A_2 |a_0|^2 =0\,.
\end{align}
Thus for  $m_J=0$ sub-phase at low $H$ there  appears Meissner effect and for $\kappa =\sqrt{\frac{\bar{\beta}_1+\bar{\beta}_2 +\frac{1}{2} \bar{\beta}_3}{4\pi e_{*}^2}}>1/\sqrt{2}$ with increasing $H$ for $H_{\rm cr 1}<H<H_{\rm cr 2}$ there exists the Abrikosov mixed state. The question about stability of the sub-phase and a coupling between various sub-phases can be considered, as it  has been done in Sect. \ref{charged}.

\subsubsection{  B and C phases of  $pp$ pairing. The $m\pm 1, \pm 2$ sub-phases}

Using (\ref{G012}), (\ref{grad0})
we obtain
\begin{align}
&(c_1+\frac{c_2+c_3}{4})[\partial_1^2 a_{\pm 1} -e_{\rm *}^2 A_2^2 a_{\pm 1}]
\nonumber\\&\pm (\frac{c_2-c_3}{2}e_{\rm *} +\frac{\eta}{2} )h a_{\pm 1}
\nonumber\\& +\bar{\alpha}a_{\pm 1} -(\bar{\beta}_2
+\frac{1}{4}\bar{\beta}_3) |a_{\pm 1}|^2 a_{\pm 1} =0\,,
\end{align}
\begin{align}
&\partial_1^2 A_2 -8\pi e_{\rm 8}^2 (c_1 +\frac{c_2+c_3}{4}) |a_{\pm 1}|^2 A_2\nonumber\\
&\mp 4\pi (\eta +e_{\rm *} \frac{c_2-c_3}{2})\partial_1 |a_{\pm 1}|^2=0\,,
\end{align}
and
\begin{align}
&(c_1+\frac{c_2+c_3}{2})[\partial_1^2 a_{\pm 2} -e_{\rm *}^2 A_2^2 a_{\pm 2}]
\nonumber\\&\pm (\frac{c_2-c_3}{2}e_{\rm *} +\eta )h a_{\pm 2}+\bar{\alpha}a_{\pm 2} -2\bar{\beta}_2  |a_{\pm 2}|^2 a_{\pm 2} =0\,,
\end{align}
\begin{align}
&\partial_1^2 A_2 -8\pi e_{\rm 8}^2 (c_1 +\frac{c_2+c_3}{2}) |a_{\pm 2}|^2 A_2\nonumber\\
&\mp 4\pi (\eta +e_{\rm *} \frac{c_2-c_3}{2})\partial_1 |a_{\pm 2}|^2=0\,,
\,
\end{align}
cf. (\ref{eqmotinh}).

Instead of solving exact equations of motion let us consider the variational problem.
For that we  employ the Abrikosov ansatz (\ref{Abric}), which for our case of the 3P$_2$ pairing reads as
\begin{align}
(\partial_i +ie_{\rm *} A_i )A_{\nu i}=0\,.
\end{align}
Expressions for the averaged Gibbs free-energy densities for the $m_J =\pm 1$ and $m_J =\pm 2$ sub-phases are similar to those for the sub-phase B$_3$, cf. Sect.~\ref{charged}. We deal with 3P$_2$ $(\pm 1)$-B$_3$ sub-phases provided
$$(\bar{\beta}_2+\frac{1}{4}\bar{\beta}_3)\widetilde{\beta} -2\pi\widetilde{\eta}_{\pm 1}^2>0\,,$$
cf. (\ref{condBcha}), where now
\begin{align}
\tilde{\eta}_{+1} =\frac{\eta}{2}- e_{\rm *}({c_1+c_2})\,,\quad \tilde{\eta}_{-1} =\frac{\eta}{2}- e_{\rm *}({c_1+c_3})\nonumber
\end{align}
and
with 3P$_2$ $(\pm 2)$-B$_3$  sub-phases, if
\begin{align}\bar{\beta}_2\widetilde{\beta} -2\pi\widetilde{\eta}_{\pm 2}^2>0\,,\label{BCcond}\end{align}
where
\begin{align}
\tilde{\eta}_{+2} ={\eta}-e_{\rm *}({c_1+c_2})\,, \quad \tilde{\eta}_{-2} ={\eta}-e_{\rm *}({c_1+c_3}).\label{tildeeta2}
\end{align}
If opposite inequalities are fulfilled, we deal with the corresponding C$_3$ sub-phases.

{\bf Sub-phases $3P_2(\pm 2)$-B$_3$.} 
We employ Eq. (\ref{Dpsi}) in the gauge $\vec{A}=(0, A_2 (x),0)$.
The minimization of the Gibbs free energy ${\cal{G}}=\int d^3 x G$, cf. Eq.~(\ref{G012}), yields for $\bar{\gamma}\to 0$:
\begin{align}
&|a_{\pm 2, {\rm B_3H}}|^2=\frac{\bar{\alpha}\pm {\widetilde{\eta}}_{\pm 2}{H}}{2\overline{\nu^2}(\bar{\beta}_2\widetilde{\beta} -2\pi\widetilde{\eta}^2)}\,,\label{3p2ordch}\\
\,\,
&\overline{G}_{\pm 2, \rm B_3H} =-\frac{(\bar{\alpha}\pm{\widetilde{\eta}}_{\pm 2}{H})^2}{4(\bar{\beta}_2\widetilde{\beta}  -2\pi\widetilde{\eta}^2)}\,.
\label{G23P2Bch}
\end{align}
For $T<T_{\rm cr}$ at $e_* =0$ and $\eta >0$ the phase $m_J=+2$ is energetically preferable whereas
for $e_* =0$ and $\eta <0$ wins the phase $m_J =-2$.
 Equality
 $$\bar{\alpha}\pm {\widetilde{\eta}}_{\pm 2}\vec{H} =0$$
 determines the critical point for the second order phase transition,
\begin{align}
\vec{h}=\vec{H}\pm 4\pi {\widetilde{\eta}}_{\pm 2}|a_{\pm 2,{\rm B}_3H}|^2
\,,\label{ha2in}
\end{align}
Even for $H=0$ in this phase there exists an own magnetic  field $h(H=0)$.

The critical temperature is shifted up in the presence of the external magnetic field $H$ and the new critical temperature becomes (for $\bar{\alpha}=\bar{\alpha}_0 t$):
\begin{align}\label{TcB}
{T}_{\rm cr}^{{\rm B}_3H} =T_{\rm cr} (1\pm \widetilde{\eta}_{\pm 2} H/\bar{\alpha}_0)\,.
 \end{align}
 Upper sign is for $m_J=2$ and $\widetilde{\eta}_2 >0$ and lower sign is for $m_J=-2$ and $\widetilde{\eta}_2 <0$.

{\bf Sub-phases $3P_2(\pm 2)$-C$_3$.} We deal with  $3P_2(\pm 2)$-C$_3$ sub-phase provided
\begin{align}
\bar{\beta}_2\widetilde{\beta} -2\pi\widetilde{\eta}_{\pm 2}^2<0\,.\nonumber
\end{align}
For $\bar{\gamma}\leq 0$ the ground state is unstable.
For $\bar{\gamma}>0$ we deal with the first-order phase transition.
For a small $\bar{\gamma}>0$ we find
\begin{align}
|a_{\pm 2,{\rm C}_3H}|^2\simeq\frac{2(2\pi\widetilde{\eta}_{\pm 2}^2-\bar{\beta}_2\widetilde{\beta})}
{3\bar{\gamma}\widetilde{\beta}_1}+
\frac{(\bar{\alpha}\pm {\widetilde{\eta}_{\pm 2}}{H})\widetilde{\beta}_2}{2\widetilde{\beta}_1 (2\pi\widetilde{\eta}_{\pm 2}^2-\bar{\beta}_2\widetilde{\beta})}\,,\label{aB1}
 \end{align}
\begin{align}
\overline{G}_{\pm 2,{\rm C}_3H} \simeq -\frac{4(2\pi\widetilde{\eta}_{\pm 2}^2-\bar{\beta}_2\widetilde{\beta})^3}
{27\bar{\gamma}^2\widetilde{\beta}_2^2}\,,
 \end{align}
and the own magnetic field is determined by Eqs. (\ref{ha2}).

The critical temperature is shifted up in presence of the external magnetic field $H$, and the new critical temperature  is given by
\begin{align}
\frac{{T}_{\rm cr}^{\rm C_3H}}{T_{\rm cr}}= 1 +\frac{(2\pi\widetilde{\eta}_{\pm 2}^2-\bar{\beta}_2\widetilde{\beta})^2}{3\bar{\gamma}\bar{\alpha}_0 \widetilde{\beta}_2}\pm\frac{ \widetilde{\eta}_{\pm 2} H}{\bar{\alpha}_0}\,.\label{TcrHOm3P2}
\end{align}

\section{ Numerical evaluations: BCS approximation and beyond }\label{BCS}
\def\BCS{{\rm BCS}}

As we have mentioned, existing in the literature estimates of the typical value of $T_{\rm cr}$ for the 3P$_2$ $nn$-pairing  in a dense neutron star matter are  controversial. Following BCS estimates \cite{HGRR1970}, typical value of $T_{\rm cr}$ for the 3P$_2$ $nn$-pairing is $T_{\rm cr}^{\BCS} \sim 0.1\mbox{--}1$MeV, cf.~\cite{Takatsuka:2004zq}.
Contrary, Ref. ~\cite{Schwenk:2003bc} with taking into account of the polarization effects estimated $T_{\rm cr}\lsim 10$ keV for the 3P$_2$ $nn$-pairing.
 Values of the Fermi liquid parameters for the isospin-asymmetric nuclear matter in the pairing channel  at $n\neq n_0$, as well as  their density dependence, are poorly known. Only rough estimates were performed, cf.  ~\cite{VS}.
Bearing this in mind, in our estimates we consider $T_{\rm cr}$ as a free parameter, which  we vary in the range  $T_{\rm cr}^{\BCS} \sim (0.01\mbox{--}1)$~MeV.

Values of the parameters used in Eq.~(\ref{FA}) were calculated in the weak coupling limit (BCS)~\cite{RS1976,SS1978,SaulsSS82}:
\begin{align}
& \bar{\alpha}_0^{\BCS}=N(0)/3\,,\qquad \bar{\beta}_1^{\BCS} =0,
\nonumber\\
& \bar{\beta}_2^{\BCS} =-\bar{\beta}_3^{\BCS}=4|\beta|=\frac{7\zeta(3)N(0)}{60 \pi^2 T_{\rm cr}^2}\,.
\label{BCSparam}
\end{align}
Here $N(0) =m_{\rm F}^* p_{\rm F}/(2\pi^2)$ is the density of states with $m_{\rm F}^*$ standing for the effective fermion  mass and $p_{\rm F}$ is the   Fermi momentum,
$\zeta(x)$ is the Riemann function and $\zeta(3)\approx 1.202$. In the approximation
of a symmetry of the particles and holes on the Fermi surface
the coefficients
$c_1$, $c_2$ and $c_3$ approximately coincide \cite{VolovikMineev84}
\begin{align}
c_1^{\BCS}\simeq c_2^{\BCS} \simeq c_3^{\BCS} \simeq \frac{7\zeta (3)N(0)}{120\pi^2}\frac{\epsilon_{\rm F}}{ m_{\rm F}^{*} T_{\rm cr}^2}\equiv c\,,
\end{align}
 where $\epsilon_\rmF$ is the  Fermi energy, $\epsilon_\rmF=p_\rmF^2/2m_{\rm F}^*$.
Exploiting presence of a slight asymmetry
of the particles and holes near the Fermi surface  ($|c_2 -c_3|/c_2 \ll 1$) Refs. \cite{VolovikMineev84,SamokhinWalker}
estimated
\begin{align}
c_2-c_3\simeq c (T_{\rm cr}/\epsilon_{\rm F})^2\ln (\epsilon_{\rm F}/T_{\rm cr})\,.
\end{align}

As for $T$ in the vicinity of $T_{\rm cr}$ as for $T\ll T_{\rm cr}$, with a logarithmic accuracy \cite{AM1973,MineevSamokhin} we obtain
\begin{align}
\eta_{\pm 2}^\BCS =\frac13\mu_{\rm pair} N'(0)\ln \frac{\epsilon_\rmF}{T_{\rm cr}}
 \,.
\label{BCSparam1}
\end{align}
We used that $\Delta^2 =2|a_{\pm 2}|^2 /3$ for the pairing in $m_J=\pm 2$ states, $\Delta$ is the pairing gap.
The quantity $N'(0)$ is the derivative of the density of states with respect to the energy,
$N'(0)=N(0)/2 \epsilon_{\rm F}$, $\mu_{\rm pair}$ is  the magnetic moment of the Cooper pair.

Following an estimate \cite{SS1978}, in the BCS theory $\bar{\gamma}^{\rm
BCS} <0$ for the $m_J=0$ phase. The coefficients in the $\{A^6\}$ contribution to the Gibbs free energy are $\delta G_6 (m_J=0) =\bar{\gamma}_1 ({\rm Tr}A^2 )^3 + \bar{\gamma}_2 {\rm Tr}(A^6 )$, and
\begin{align}
\bar{\gamma}_2^{\BCS} =\frac{4}{5}\bar{\gamma}_1^{\BCS}
=-\frac{31}{16}\frac{\zeta (5)}{35}\frac{N(0)}{\pi^4 T_{\rm cr}^4}\,,
\label{gamma2}
\end{align}
where $\zeta(5)\approx 1.0369$, that forces to keep $A^8$ term in expansion of $G$.
Reference \cite{Yasui:2019unp} uses a more complicated structure of the $A^6$ term and keeps $A^8$ term in expansion of $G$.
To avoid these complications in our rough numerical estimates, e.g. for $m_J=-1$ and $-2$,  we  take
 $\delta G_6 (m_J=-1) =\bar{\gamma}^{\BCS}(m_J=-1)|a_{-1}|^6$, $\delta G_6 (m_J=-2) =\bar{\gamma}^{\BCS}(m_J=-2)|a_{-2}|^6$ with
 $\bar{\gamma}^{\BCS}_i =|\bar{\gamma}_2^{\BCS}|$ taken in modulus. Certainly, within such a simplified analysis we disregard a possibility of existence of some other  phases, which may appear in a sequence of the first-order phase transitions. Using (\ref{gamm}) we find that $|\bar{\gamma}^{\BCS}|\ll 1$  for $|t|\ll 1$. Then, dealing with the phase B we may use  expressions for $|\bar{\gamma}^{\BCS}|\to 0$.

With the BCS parameters (\ref{BCSparam}), stability conditions (\ref{stc}) are fulfilled in a case of  a weak external magnetic field $H$ for all the sub-phases 3P$_2(0)$, 3P$_2({\rm sym})$, 3P$_2(\pm 1)$, 3P$_2(\pm 2)$ considered above. Also, in the BCS approximation the value $T_{\rm cr}$ is the same for all these phases.

Actual values of the parameters of  the Ginzburg-Landau functional  in the strong coupling theory are poorly known. Only rough estimations have been performed \cite{VS}. Existing estimates of the gradient terms are controversial. Reference \cite{Rosenstein:2015iza} calculated for the triplet
 superconductivity in 3D Dirac semimetals $c_3 =[u_L -u_T]/4$, $c_1=u_T/4$, $c_2 =0$,   $u_L =u_T/32$,  $u_T= \frac{7\zeta (3)N(0)v^2_{\rm F}}{15\pi^2 T_{\rm cr}^2}$,  i.e. $c_1\simeq -c_3$, $c_2=0$, and   derived values  $b_1 = \frac{7\zeta(3)N(0)}{640 \pi^2 T_{\rm cr}^2}$ and $b_2=-b_1/3$, $\Delta_i =\psi_i/2$,
 $v_{\rmF}$ is the  Fermi velocity. As one of the choices (E$_2$ model) Ref. \cite{SaulsAdv} employs   $c_2=c_3\ll c_1\sim N(0)v_{\rm F}^2/(\pi^2 T_{\rm cr}^2)$ that does not contain a small numerical pre-factor appeared in estimate \cite{VolovikMineev84,SamokhinWalker}.  On the other hand the heat capacity measurements  performed for UPt$_3$
by several groups give $b_2/b_1 =(0.2-0.5)$, cf. \cite{SaulsAdv,Hasselbach}.  We remind that neglecting $u_L$ contribution one recovers the relation $c_1 =-c_3$, as follows from   the microscopical consideration   of the $W$ boson fields \cite{Olesen}.

If there were $\bar{\beta}_3 > 0$,  with the same BSC estimates for other parameters  the phase B would be preferable even for $H=0$ and for $T<T_{\rm cr}$.

\subsection{3P$_2$ $nn$ pairing in neutron stars}

For \emph{the $nn$ pairing in the sub-phase 3P$_2(0)$-A} following (\ref{Aa3P2}), (\ref{Gsym1}) with values of the parameters estimated within the BCS approximation  (\ref{BCSparam}) we find
 \begin{align}
|a_0^{\BCS}|^2 =\frac{20 \pi^2 }{7\zeta(3)}T_{\rm cr}^2 t\,,
\quad
G_{\rm 0}^{\BCS} =-\frac{10\pi^2  N(0) }{21\zeta(3)}T_{\rm cr}^2t^2\,.
\end{align}

For \emph{the $nn$ pairing in the sub-phase 3P$_2(0)$-B}
 using $\mu_{\rm pair} =\mu_{nn}=-3.83\mu_N$,  $\mu_N\approx 4.5\cdot 10^{-5}/{\rm MeV}$,
for $n\sim n_0$ and $T_{\rm cr}\sim (0.01-1)$MeV following (\ref{BCSparam1}) we  estimate $\eta^\BCS\sim \mp (10^{-2}-10^{-1})$, $\mp \eta_{\pm 2}^\BCS H/\bar{\alpha}\sim \pm 3\cdot (10^{-2}-10^{-1}) H/m_\pi^2$, and we obtain $(T_{\rm cr}^{\rm B_3H}-T_{\rm cr})/T_{\rm cr}\lsim 3\cdot 10^{-1} $  for $H\lsim m_\pi^2$ for all relevant values $T_{\rm cr}$.
From
Eqs. ~(\ref{3p2ord}), (\ref{G23P2B})
we obtain (for $\bar{\beta}_2>2\pi\tilde{\eta}^2$ which is indeed fulfilled),
\begin{align}
&|a_{-2,\rm{\BCS}}^{\rm B_3H}|^2 =
\frac{\frac12|a_0^{\BCS}|^2\nu_{-2}}{1-\frac{T_{\rm cr}^2}{T_{\mu,-2}^2}}  \,,\label{BCSB3field}
\end{align}
\begin{align}
G_{\rm -2,\BCS}^{\rm B_3H}=
\frac{\frac12 G_{\rm 0}^{\BCS}\nu_{-2}^2}{1-\frac{T_{\rm cr}^2}{T_{\mu,\rm -2}^2}}
  \,,\label{BCSB3GE}
\end{align}
cf. also
Eqs.~(\ref{3p2ordch}), (\ref{G23P2Bch}) at $e^*=0$. We used that
\begin{align}
\nu_{-2} =1-{\eta H}/{\bar{\alpha}}
\end{align}
is positive for relevant values $H\lsim m_\pi^2$
and that
\begin{align}
T_{\mu,\rm -2}=\Big(\frac{21\zeta(3)}{20\,\pi}\Big)^{1/2} \frac{v_{\rmF}^{3/2}}{|\mu_{nn}|\ln\frac{\epsilon_\rmF}{T_{\rm cr}}}
\,,
\end{align}
 $v_{\rmF}=p_\rmF/m^*_n$.  We estimate
$T_{\mu,-2}\simeq
2\cdot 10^3 (v_\rmF^{3/2}/\ln\frac{\epsilon_\rmF}{T_{\rm cr}})$~MeV. At  $n\sim n_0\simeq 0.16/{\rm fm}^3$, we have  $v_\rmF\sim 0.4$, and $T_{\mu,-2}\simeq
500/\ln\frac{\epsilon_\rmF}{T_{\rm cr}} $~MeV $\gg T_{\rm cr}$. Thereby $\bar{\beta}_2>2\pi\tilde{\eta}^2$, as we have used deriving (\ref{BCSB3field}) and (\ref{BCSB3GE})  and we indeed deal with the B phase rather than with the C phase.

We see that
$
G_{-2,\rm B_3}^{\BCS}\simeq \frac{1}{2}G_{\rm 0}$.
Similarly $G_{-1,\rm B}^{\BCS}\simeq \frac{2}{3}G_{\rm 0}$. Therefore, if the BCS estimates (\ref{BCSparam}) were correct, the sub-phases   3P$_2 (- 2)$-B and 3P$_2 (- 1)$-B of the $nn$ pairing would not be realized in the neutron stars for
$T<T_{\rm cr}$ till $\nu_{-2} >1/\sqrt{2}$ and $\nu_{-1} >\sqrt{2/3}$. However in the temperature interval $T_{\rm cr}<T< T_{\rm cr}^{\rm B_3H}$, where the A phase is impossible, the 3P$_2 (- 1)$-B$_3$ sub-phase is realized in any case.

Using the relation  $b_2 =-b_1/3$ derived in \cite{Rosenstein:2015iza} for the description of the superconductivity in  3D semimetals, that corresponds to the relation $\bar{\beta}_3 =-\bar{\beta}_2/3$ in the functional (\ref{FA1}), with  $\bar{\beta}_1 =0$
we evaluate $G_{-1,\rm B}^{\BCS}\simeq \frac{11}{12}G_{\rm 0}$, i.e. $G_{-1,\rm B}^{\BCS}$ is only slightly larger than  $G_{\rm 0}$.

On the other hand, with $\bar{\beta}_3 /\bar{\beta}_2 >0$, as follows from experiments on UPt$_3$, for $T<T_{\rm cr}$ and in the temperature interval $T_{\rm cr}<T< T_{\rm cr}^{\rm B_3H}$ the 3P$_2 (- 1,-2)$-B$_3$ sub-phases are energetically favorable compared to the A phase.

As we have mentioned, the heat capacity measurements  performed for UPt$_3$
by several groups give $b_2/b_1 =(0.2-0.5)$, cf. \cite{SaulsAdv,Hasselbach}.
Choosing  estimate of $b_2=b_1/2$, that corresponds to $\bar{\beta}_3 =+\bar{\beta}_2/2$,
we find that $G_{-2,\rm B_3}^{\BCS}\simeq \frac{5}{4}G_{\rm 0}$ and $G_{-1,\rm B}^{\BCS}\simeq \frac{9}{8}G_{\rm 0}$. With these estimates the sub-phase   3P$_2 (- 2)$-B  of the $nn$ pairing would  be realized in the neutron stars for $0<T< T_{\rm cr}^{\rm B_3H}$.

With the help of Eqs. (\ref{ha2}), (\ref{G23P2B}) and making use of the  estimate (\ref{BCSB3field}) we find
\begin{align}
h_{-2}^{\rm B_3} = \frac{\epsilon_\rmF}{|\mu_{nn}|}
\frac{2|t|\frac{T_{\rm cr}^2}{T_{\mu,-2}^2 \ln \frac{\epsilon_{\rm F}}{T_{\rm cr}}}  }{(1-\frac{T_{\rm cr}^2}
{T_{\mu,-2}^2})}
\,.
\label{hbcsb}\end{align}
We have put  $\nu_{-2} \simeq 1$, $\epsilon_\rmF/|\mu_{nn}|\approx 5.7\cdot 10^3 [\epsilon_\rmF/{\rm MeV}]\,{\rm MeV}^2\approx 8.3\cdot 10^{16}[\epsilon_\rmF/{\rm MeV}]$~Gs. Thus for ${T_{\rm cr}^2}/
{T_{\mu,-2}^2}\ll 1$ we estimate $h_{-2}^{\rm B_3}\sim  10^{11} |t|\left(\frac{ T_{\rm cr}}{{\rm MeV}}\right)^2\frac{\epsilon_\rmF}{\rm MeV}$~Gs for $n=n_0$. For $T_{\rm cr} \sim 1$ MeV we estimate $h \sim 10^{13}$ Gs. Note  that, as we have shown, $h(H=0)\propto |a_{-2}^{\rm B_3}(x)|^2$ and thereby it vanishes at the superfluid -- normal matter boundary. If by some reason the field $h$ had a magnetic-dipole component outside the superfluid star interior, the neutron star would substantially  diminish its rotation during first $\sim (10^3 - 10^4)$ years of its evolution. At least millisecond pulsars should not have such a strong  magnetic-dipole fields. For $T_{\rm cr} \lsim 10$ keV, cf. \cite{Schwenk:2003bc},  we estimate $h\lsim 10^{9}$ Gs, which value any case does not contradict to the data on millisecond pulsars.

\subsection{ Estimates for  hypothetical  3P$_2$ $pp$ pairing in neutron stars}

Since $e_* (c_1+c_3) \sim 4\cdot 10^{-5}(m_\pi/T_{\rm cr})^2$
 for $n\sim n_0$, for
$T_{\rm cr}\sim (0.01-1)$MeV using (\ref{tildeeta2}) we estimate    $\widetilde{\eta}_{\pm 1}\sim \widetilde{\eta}_{\pm 2}\sim -(1-10^4)$, being valid   for the 3P$_2(2)$-B and 3P$_2(1)$-B sub-phases of the $pp$ pairing.
 Making use of this estimate and (\ref{BCSparam}) we see that the condition (\ref{BCcond}) is fulfilled for all values $T_{\rm cr}$ of our interest. Thus C phase is not realized.
The value $T_{\rm cr}^{\rm B_3H}$ proves to be significantly shifted up for  $T_{\rm cr}\sim 0.01$MeV already for $H\gsim 10^{14}$Gs.  For a higher value  $H$ the B$_3$ sub-phase becomes energetically profitable compared to the A phase, as it follows from   Eqs.  (\ref{G23P2Bch}) and (\ref{Gsym1}). With the help of  (\ref{3p2ordch}), (\ref{ha2in}) we roughly estimate  the value of the own magnetic field    as $h\sim 10^{16}$Gs. Recall that at the surfaces of the magnetars the strength of the magnetic field reaches values $h\lsim 10^{16}$Gs. For $T_{\rm cr}\sim 1$MeV the value $T_{\rm cr}^{\rm B_3H}$ is significantly shifted up respectively $T_{\rm cr}$ only for $H\gsim 10^{18}$Gs.

Reference \cite{Schulze} expressed an idea about a possibility of the triplet 3PF$_2$ $pp$ pairing in the hyperon enriched dense region. Then one should study a coexistence of the considered above phases of the $nn$ pairing and those available for the $pp$ pairing.

\subsection{ Estimates for  3S$_1$ $np$ pairing in isospin-symmetrical systems}
The 3S$_1$ channel provides the largest  attractive interaction  for the triplet $np$ pairing in the  isospin-symmetrical matter for $n\lsim n_0$. With increasing density the 3D$_2$ channel becomes most attractive, cf.  \cite{Sedrakian:2018ydt}. The BCS calculations for the symmetric matter with polarization effects included \cite{Guo:2018lna} predict the $np$ pairing gaps  $\sim {\rm (several -10)}$ MeV.   As for the 3P$_2$ $pp$ pairing, the own magnetic field in the B$_3$ sub-phase is estimated   as $h\sim 10^{16}$ Gs.  In this phase the nucleon matter is spin-polarized that might be checked experimentally.
 For example, in peripheral heavy-ion collisions of approximately isospin-symmetric nuclei, where the temperature is rather low, the spin-triplet $np$  pairing in the 3S$_1$ channel  can be formed.
Moreover, in peripheral heavy-ion collisions the external magnetic field reaches values $10^{17}-10^{19}$Gs, cf. \cite{VA1980,IST}. In such strong fields the value $T_{\rm cr}^{\rm B_3}$
might be significantly larger than $T_{\rm cr}$, favoring $np$ pairing in the 3S$_1$ channel.
Also, the $np$ pairing in the 3SD$_1$ state is possible in  the nuclei \cite{Bertsch:2009xz,Guo:2018lna}.

\section{Conclusion}\label{conclusion}
This paper studies effects of the vector boson condensation and spin-triplet superfluidity and superconductivity, such as ferromagnetic superfluidity,  as well as the effects of the 3P$_2$ $nn$ and $pp$ pairing in the neutron-star matter and the 3S$_1$ $np$ pairing in the isospin-symmetrical matter in absence and in presence of the external static uniform magnetic field. Possible effects of the self-rotation and  response of the system on  ``external'' rotation  were for simplicity disregarded and will be considered elsewhere.

We started in Sect. \ref{preliminary}  with   the description  of the condensation of the complex scalar field characterized by a negative squared effective mass inside a half-space medium $x<0$, placed in  an external static uniform magnetic field. Next, we considered a role of the Zeeman coupling for neutral fermions and discussed a possibility of the existence of the ferromagnetic state in the fermion matter (e.g., in the neutron star matter).

In Sect. \ref{vectorboson}  focus was made  on the study of the complex neutral and charged vector boson fields with negative and positive  squared effective mass. A possibility of existence of the A, B and C  phases was found.  In the phase A  the mean spin density is zero and in the phase B  spins are aligned in one direction. The simplest choice to describe the phase A  is to chose only one Lorentz component of the complex vector field to be non-zero.
The C phase is not realized provided the hadron-hadron coupling constant $\Lambda \gg e^2$.

The behavior of the  charge-neutral complex  vector boson field  inside the half-space medium, $x<0$ was studied in presence of the uniform static external magnetic field.  Two   A sub-phases are then permitted for $m_{\rm sc}^2<0$: A$_2$ provided the $y$ component of the vector boson field is non-zero and A$_3$, provided $z$ component is non-zero. The  vector boson field, and the magnetic field  decouple and the Gibbs free energies in the sub-phases are the same.  Thus the  A phase of the neutral vector bosons is non-magnetic. For $m_{\rm sc}^2>0$ there is no  condensate.

In the  phase B, which is described by two non-zero complex components of the neutral vector boson  field,  the  system behaves  as a ferromagnetic superfluid. In the condensate region there appears an own static magnetic field. We considered the matter filling half-space $x<0$  in presence of the external uniform static magnetic field  either directed parallel to the system boundary or perpendicular to it.

In the sub-phase B$_2$ for $\vec{H}\parallel y$ (i.e. parallel to the system boundary and to the direction of the spin) and in the sub-phase B$_3 $ for $\vec{H}\parallel z$ and spin parallel $z$
 the condensate amplitude grows with $H$. At $H>H_{\rm cr }^{\rm neut}$, cf. Eq. (\ref{Hneut}), the superfluid condensate exists not only for $m_{\rm sc}^2<0$ but also  for $m_{\rm sc}^2>0$.
Which phase A or B is energetically favorable depends on the form of the self-interaction term in the Lagrangian. For the very same values $m_{\rm sc}^2$, with the self-interaction  in the form (\ref{phiphisimp}) for $\xi_1 =0$ the phase B proves to be energetically preferable in comparison with the phase A.

We demonstrated that the difference  in the volume and surface energies for  the sub-phases   motivates a possibility of the existence of domains with different directions of the magnetic moment in each domain. Domains may merge in presence of the external  fields.

Then we studied the behavior of   {\em the charged complex vector field}  interacting with the electromagnetic field by the minimal  and the Zeeman couplings.  As for neutral vector bosons, we first considered charged complex vector field with the negative squared effective mass, $m^2_{\rm ef}<0$, in the half-space $x<0$ under the action of  the external static uniform magnetic field $\vec{H}$.

For the state with zero spin density (A-phase) for $\vec{H}$  parallel to the system boundary ($\vec{H}\parallel y$) the  sub-phase  A$_2$ demonstrates superdiamagnetic response on a weak external magnetic field, as
 for
 the charged scalar boson  field, and for $\vec{H}$ parallel to the system boundary ($\vec{H}\parallel z$) the sub-phase  A$_2$  is nonmagnetic, as  for a neutral complex vector boson field. The phase A$_3$ demonstrates superdiamagnetic response for a weak external magnetic field
 $\vec{H}\parallel z$ and it is nonmagnetic for $\vec{H}\parallel x$.  The  Gibbs free energies for  the  sub-phase  A$_2$ at $\vec{H}\parallel z$ and for  the  sub-phase  A$_3$ at $\vec{H}\parallel x$ are equal  and they are
 lower than those for the  sub-phase A$_2$ at $\vec{H}\parallel y$ and for the sub-phase A$_3$ at $\vec{H}\parallel z$.
 There are no solutions in case of  the  charged complex vector field in the phase A at $m^2_{\rm ef}>0$.

 Then we found  solution for the sub-phase B$_3$ at $\vec{H}\parallel z$. In this case for $H<H_{\rm cr 1}$ there exists ordinary  Meissner effect. However, for increasing $H$, for $m^2_{\rm ef,0}<0$, $\eta <0$, $e<0$ we did not find  a solution with $H=H_{cr 2}$, such that  the condensate vanishes for $H\to H_{\rm cr 2}$ from below. For $m^2_{\rm ef}>0$ the superconductive condensate appears for $H>H_{\rm cr 2}=-m^2_{\rm ef}/\eta>0.$ For rather low $H\neq 0$, $m^2_{\rm ef,0}<0$, the nonmagnetic A$_2$ sub-phase for $\vec{H}\parallel z$ and A$_3$ sub-phase for $\vec{H}\parallel x$  are more energetically preferable compared to the B$_3$ sub-phase for $\vec{H}\parallel z$, whereas for $H\to 0$  the sub-phase B$_3$ wins due to a smaller surface energy, if the system occupies a finite size layer.

In Sects. \ref{vector}, \ref{charged},  \ref{nnpairing} the focus is made on the description of the spin-triplet pairing of neutral and charged  fermions coupled with
the magnetic  field by the Zeeman coupling. First, in Sects. \ref{vector}, \ref{charged} we considered the case, when the spin of the pair can be treated as a conserved quantity. This is the case for a negligibly small  spin-orbit interaction (as for 3S$_1$ $np$ pairing in isospin-symmetrical nuclear matter). Then  the order parameter is  a vector with complex components and the description is similar to that  for the spin 1 vector bosons considered in Sect. \ref{vectorboson}: the vector order parameter is characterized by the two complex vectors of different amplitudes.

 In Sect. \ref{vector} we consider {\em  triplet pairing of neutral fermions.}
In  the p-wave triplet phase with zero projection of the spin of the pair on a quantization axis (the phase A) the two unit vectors $\vec{n}$ and $\vec{m}$ characterizing the vector order parameter are co-linear. The A-phase appears for $b_1+b_2 >0$
by the second-order phase transition for the temperature $T<T_{\rm cr}$ ($b_1$ and $b_2$ are coefficients at the $\psi^4$ terms in the free energy, cf. Eq. (\ref{GLvector})). In the absence of the external magnetic field (for $b_2 +2\pi C^2 {\cal{M}}^2 <0$, where $C{\cal{M}}$ is the appropriately normalized effective magnetic moment of the fermion pair, cf. Eq. (\ref{GLvector})) the A phase proves to be stable.  In difference with the case
of the vector bosons considered in Sect. \ref{vectorboson}, where the A$_1$ sub-phase is not realized, for the triplet pairing of fermions all three sub-phases can be realized, with  the same volume contribution to the energy.  The surface energies in sub-phases A$_2$ and A$_3$ are the same, whereas the  surface energy in the A$_1$ sub-phase is another. The vector $\vec{n}\parallel \vec{m}$ may change the direction depending on the spatial point, since the surface contributions to the Gibbs free energy depend on the direction of the vector order parameter respectively to the surface boundary. Owing to this property there may appear  domains with different directions of $\vec{n}$ in each domain.
In presence of the domains the system remains for a while in a metastable state. The system may transform to the uniform state under the action of the external  magnetic field and in presence of the external rotation, or the energetic barrier can be overcame by  a heating of the system.

We have shown that with an increase of the external magnetic field the system from the phase A transforms to another phase (labeled as AH), such that there appears an angle between vectors $\vec{n}$ and $\vec{m}$ growing with increase of $H$. The critical temperature of the phase transition also is increased with the growth of $H$.
 For $T<T_{\rm cr}$ not all spins of Cooper pairs are aligned in the direction parallel $\vec{H}$, and in the temperature interval  $T_{\rm cr}<T<T_{\rm cr}^{\rm AH}$ all spins prove to be aligned in the direction parallel $\vec{H}$. The AH phase exists at $H\leq H_{\rm cr}^{\rm AH}$ for $T<T_{\rm cr}$, cf. Eq. (\ref{HcrAH}), and at $H\geq H_{\rm cr}^{\rm AH}$ for
 $T_{\rm cr}<T<T_{\rm cr}^{\rm AH}$, cf. Eq. (\ref{HcrAH1}).

Besides the A-phase,   we found a possibility of the ferromagnetic
superfluid  phases B and C in neutral superfluids characterized even for $H=0$ by the  $+1$ or $-1$ projections of the spin of the Cooper pair on the quantization axis. Here, the vector order parameter is the sum of two perpendicular vectors, i.e. here   $\vec{n}\perp \vec{m}$.
The phase B  appears, if  $b_2 > 2\pi C^2{\cal M}^2$ and  $b_1 >2\pi C^2{\cal M}^2$, cf. Eq. (\ref{conda}),
and the C phase occurs,  if $b_2 > 2\pi C^2{\cal M}^2$ but $b_1 < 2\pi C^2{\cal M}^2$, cf. Eq. (\ref{condC}).
The A and B phases  arise by the second-order phase transitions, whereas the C phase appears by the first-order phase transition. For simplicity we   put $T_{\rm cr}^{\rm A}=T_{\rm cr}^{\rm B}=T_{\rm cr}$, whereas $T_{\rm cr}^{\rm C}\neq T_{\rm cr}$,  since the phase transition to the phase C proves to be of the first order.
 The B and C phases  of neutral superfluids are characterized by an own uniform
 magnetic field.
 For some values of parameters at $T<T_{\rm cr}$
  the B or C phases  win a competition with the phase A, for other values of parameters the A phase  wins. For $T<T_{\rm cr}$ the condensate amplitude grows with increasing $H$.
  The sub-phases B and C may exist for $T_{\rm cr}<T<T_{\rm cr}^{\rm B_3H}, T_{\rm cr}^{\rm C_3H}$, where $T_{\rm cr}^{\rm B_3H}, T_{\rm cr}^{\rm C_3H} >T_{\rm cr}$, cf. Eqs. (\ref{newcrtem}), (\ref{TcC}) provided $H>H_{\rm cr 2}^{\rm BH}$, cf. Eq. (\ref{HcrBH1}).

Then in Sect. \ref{charged} we studied {\em the spin-triplet pairing of charged fermions.} Here, as in case of neutral superfluids, there may exist the  A, B, and C phases. In the A$_3$ sub-phase the spin-triplet superconductor, occupying half-space $x<0$, placed in uniform external magnetic field
 parallel $z$ behaves as an ordinary second-order supeconductor characterized by the Ginzburg-Landau parameter $\kappa_{\rm A_3}$ (we considered the case $\kappa_{\rm A_3}\gg 1$). The sub-phases A$_1$ and A$_2$ have some peculiarities. The  critical values of the  magnetic field, $H_{\rm cr 1}^{\rm A_1}$, $H_{\rm cr 1}^{\rm A_2}$ and $H_{\rm cr 2}^{\rm A_1}$, $H_{\rm cr 2}^{\rm A_2}$, are characterized by the two Ginzburg-Landau parameters  $\kappa_{1, \rm  A_1}\gg 1$, $\kappa_{1, \rm  A_2}\gg 1$ and $\kappa_{2, \rm  A_1}$,
$\kappa_{2, \rm  A_2}$ in each case.

Then focus was made on the description of  the B$_3$ sub-phase. We solved the variational problem using the Abrikosov ansatz (\ref{Abric}) for the probe functions. It was demonstrated that the value $\vec{\widetilde{M}}=C\vec{\cal{M}}-\vec{n}_3 e_{*} (c_1+c_3)$, cf. (\ref{mut}),  gets sense of the effective magnetic moment, $e_{*}$ is the effective charge of the  fermion pair, $c_1$ and $c_3$ are the coefficients at the gradient contributions to the free energy, $\vec{n}_3$ is the unit vector parallel $z$. For $T<T_{\rm cr}$ and $\widetilde{M}>0$ the condensate exists for any value of $H$ and  for $T_{\rm cr}<T<T_{\rm cr}^{\rm BH}$ the condensate exists also for $H>H_{\rm cr 2}^{\rm B}$, cf. Eq. (\ref{hc2}) and (\ref{mut}). For $\widetilde{M}<0$ the condensate exists only for $T<T_{\rm cr}$ and $H<H_{\rm cr 2}^{\rm B}$. Similarly the sub-phase C$_3$ may esist in a certain  temperature interval above $T_{\rm cr}$.

Then in Sect. \ref{nnpairing} we studied {\em the 3P$_2$ pairing in nuclear systems.} Due to a strong spin-orbit $nn$ interaction   3P$_2$ phase of the  $nn$ pairing is supposed to exist in the baryon density interval $0.8 n_0\lsim n\lsim (3\div 4)n_0$ in the neutron star interiors.    We focused on cases when the projection of the total angular momentum on quantization axis is fixed as $m_J=2,1,0, -1,$ or $-2$,
and also we considered a symmetric phase. It was demonstrated that  the sub-phase $m_J=0$ and symmetric sub-phase (labeled 3P$_2$(0)-A and 3P$_2$(sym)-A) are described similarly to the sub-phases of the phase A. The sub-phases $m_J=1$ and $2$ (and $-1$ and $-2$) are described similarly to the sub-phase B$_3$, and then we label them as  $3P_2(\pm 1)$-B$_3$, $3P_2(\pm 2)$-B$_3$, or to the sub-phase C$_3$. For the $nn$ pairing in the mentioned sub-phases the description is similar to that for the neutral complex vector boson field and for the $pp$ pairing it is similar to the case of the charged complex vector boson field.

 The values of  the parameters of the
Gibbs free energy functional for strongly interacting systems are  unknown because of absence
of sound microscopic calculations with inclusion of  the polarization effects.
However these parameters can be easily evaluated  in the BCS weak coupling
approximation exploiting the bare pairing potentials. In Sect. \ref{BCS} within the BCS approximation and beyond it  we performed some estimates relevant for the $3P_2$ $nn$ and $pp$ pairings in  the neutron star matter and for the 3S$_1$ $np$ pairing in the isospin-symmetric nuclear matter. We found at which conditions the ferromagnetic superfluid phases characterized by  own  magnetic  field prove to be  energetically favorable.

A lot of work remains to be done. Let me list some problems related to the spin-triplet superfluidity in nuclear systems.
 In the paper body  only simplest available phases of the 3P$_2$ $nn$ pairing were studied, whereas some other phases may  also exist.   Calculations of parameters of the Ginzburg-Landau functional are very desirable. A possibility of the  ferromagnetic color superconductivity in hybrid stars should be studied. Gluons become massive in the hot quark-gluon plasma and  may form vector field condensates at some conditions. Question about a possibility of a self-rotation in ferromagnetic superfluids was not considered, as well as the response of the spin-triplet superfluid sub-system on the rotation of the normal component.  Another interesting issue is the problem of the neutron star cooling with taking into account a possibility of the ferromagnetic superfluidity and superconductivity including effects on the cooling of millisecond pulsars, cf. \cite{KV2015},  and strong magnetic fields for  magnetars. If the 3P$_2$ $nn$ pairing were realized in the $m_J\neq 0$ state,  the neutron specific heat and the neutrino emissivity of the nucleon involved processes would decrease with decrease of the temperature as a power of the temperature rather than exponentially, since the gap  vanishes  at the Fermi sphere poles. This was noticed in \cite{Voskresensky:1987hm} and in \cite{Migdal:1990vm}, and then considered in a more detail in  \cite{Yak}. However all presently existing neutron star cooling scenarios explored $3P_2$ $nn$ pairing in $m_J=0$ state, since mechanisms for the formation of the $nn$ pairs in the $m_J\neq 0$ states were not yet explored, cf. \cite{Migdal:1990vm,Grigorian:2016leu,Grigorian:2018bvg,Yak}. Possibilities of the $3P_2$ $pp$, hyperon-hyperon and   $\Delta$ isobar -- $\Delta$ isobar pairings in interiors of sufficiently  massive neutron stars   should be additionally investigated.
 Triplet pairing in non-equilibrium systems should be studied. Spin polarization effects owing to the possibility of a feasible  3S$_1$ $np$ pairing in peripheral heavy-ion collisions were not yet considered.
 Presence of magnetic fields of the order of $(10^{17}-10^{19})$Gs, cf. \cite{VA1980,IST}, and of high   angular momenta in peripheral heavy-ion collisions may act in favor of the spin-triplet pairing. Novel spin-triplet sub-phases can be formed during very low energy collisions of normal and superfluid nuclei and in the rotating nuclei. Energetically favorable transitions from one phase to another one may result in an increase of the duration of the process of the collision of nuclei. In neutron star interiors the magnetic field may reach values $\sim 10^{18}$ Gs. At such conditions the charged $\rho$ meson  condensates  may appear, may be forming a ferromagnetic superfluid.
A further more detailed quantitative study is welcome to answer these and other  intriguing  questions.

\acknowledgments I thank E.E. Kolomeitsev  for the help in the  beginning stage of this work started  several years ago and for valuable advices.
The work  was  supported by   the Ministry of
 Science and High Education of the Russian Federation within the state assignment,
project No 3.6062.2017/6.7.


\end{document}